\documentclass[a4paper,11pt]{article}

\usepackage{jheppub} 
\usepackage[table]{xcolor}
\usepackage[english]{babel}
\usepackage[utf8]{inputenc}
\usepackage{amsmath}
\usepackage{graphicx}
\usepackage{amsmath}
\usepackage{amsmath}
\usepackage{amsfonts}
\usepackage{amssymb}
\usepackage{indentfirst}
\usepackage{color}
\usepackage{amssymb}
\usepackage{array} 
\usepackage{float}
\usepackage{comment}

\newcolumntype{M}[1]{>{\centering\arraybackslash}m{#1}}
\newcolumntype{N}{@{}m{0pt}@{}}

\def\be{\begin{equation}}
\def\ee{\end{equation}}
\def\bea{\begin{eqnarray}}
\def\eea{\end{eqnarray}}

\newcommand{\mc}{\mathcal}

\def\bel#1{\begin{equation} \label{#1}}
\def\eel{\end{equation}}

\def\ltap{\ \raise.3ex\hbox{$<$\kern-.75em\lower1ex\hbox{$\sim$}}\ }
\def\gtap{\ \raise.3ex\hbox{$>$\kern-.75em\lower1ex\hbox{$\sim$}}\ }
\def\gl{\ \raise.5ex\hbox{$>$}\kern-.8em\lower.5ex\hbox{$<$}\ }
\def\roughly#1{\raise.3ex\hbox{$#1$\kern-.75em\lower1ex\hbox{$\sim$}}}

\def\nn{\nonumber}

\def\pref#1{(\ref{#1})}

\def\del{\partial}

\newcommand{\comments}[1]{}

\newcommand{\ben}{\begin{enumerate}}
\newcommand{\een}{\end{enumerate}}
\newcommand{\bi}{\begin{itemize}}
\newcommand{\ei}{\end{itemize}}
\newcommand{\ba}{\begin{align}}
\newcommand{\ea}{\end{align}}

\def\beq{\begin{equation}}
\def\eeq{\end{equation}}

\newcommand{\dd}{\text{d}}
\newcommand{\conj}[1]{\overline{#1}}

\def\M{{\scriptscriptstyle \rm M}}
\def\D3{{\scriptscriptstyle \rm D3}}

\title{Flux Vacua with Approximate Flat Directions}

\author[a,b]{Michele Cicoli,}
\author[a,b]{Matteo Licheri,}
\author[b]{Ratul Mahanta,}
\author[c]{Anshuman Maharana}

\affiliation[a]{\footnotesize Dipartimento di Fisica e Astronomia, Universit\`a di Bologna, via Irnerio 46, 40126 Bologna, Italy}
\affiliation[b]{\footnotesize INFN, Sezione di Bologna, viale Berti Pichat 6/2, 40127 Bologna, Italy}
\affiliation[c]{\footnotesize Harish-Chandra Research Institute, A CI of Homi Bhabha National Institute, Allahabad 211019, India}

\emailAdd{michele.cicoli@unibo.it}
\emailAdd{matteo.licheri@unibo.it}
\emailAdd{mahanta@bo.infn.it}
\emailAdd{anshumanmaharana@hri.res.in}

\abstract{We present a novel method to obtain type IIB flux vacua with flat directions at tree level. We perform appropriate choices of flux quanta that induce relations between the flux superpotential and its derivatives. This method is implemented in toroidal and Calabi-Yau compactifications in the large complex structure limit. Explicit solutions are obtained and classified on the basis of duality equivalences. In the toroidal case we present solutions with $N=1$ and $N=2$ supersymmetry and arbitrarily weak coupling. In Calabi-Yaus we find novel perturbatively flat vacua, as well as solutions with non-zero flux superpotential and an axionic flat direction which represent a promising starting point for de Sitter constructions from non-zero F-terms in the complex structure sector. The higher order (perturbative and non-perturbative) effects that can lift these flat directions are discussed. We also outline applications in a wide variety of settings involving the classical Regge growth conjecture, inflation and quintessence, supersymmetry breaking and F-term de Sitter uplifting.}

\begin{document} 

\maketitle
\flushbottom

\section{Introduction}
\label{Intro}

Potentials for moduli fields play a central role in string phenomenology. The simplest way to generate these potentials is to consider solutions with background fluxes, see e.g. \cite{Michelson:1996pn, Dasgupta:1999ss,Gukov:1999ya, Curio:2000sc, Giddings:2001yu}. In the type IIB setting the effect of fluxes is to stabilise the complex structure moduli and the axio-dilaton \cite{Giddings:2001yu}. This is encoded in the Gukov-Vafa-Witten (GVW) superpotential \cite{Gukov:1999ya}: 
\be
W = \int_X G_3 \wedge \Omega \,,
\label{gvw}
\ee 
where $G_3 = F_3 - \phi H_3$ is the complexified $3$-form flux, $\phi$ is the axio-dilaton\footnote{In this paper we do not follow the standard convention to denote the axio-dilaton as $\tau$ to avoid confusion with the period matrix $\tau^{ij}$ of the toroidal case.} and $\Omega$ the holomorphic $3$-form of the (orientifolded) Calabi-Yau (CY) $X$ on which the theory is compactified. The $3$-form fluxes thread $3$-cycles of the CY with their integrals over the cycles satisfying Dirac quantisation conditions. Depending on the choice of fluxes, minima of the associated potential can be isolated or have flat directions. Once the axio-dilaton and the complex structure moduli are integrated out, the effect of fluxes is captured by a constant superpotential:
\be
W_0 \equiv   \,\langle \int_X G_3 \wedge \Omega \rangle\,.
\ee
The value of $W_0$ is a key input for phenomenology.\footnote{For a general discussion on $W_0$ in the context of moduli stabilisation and its role in phenomenology, see \cite{Cicoli:2013swa} and references therein.} Again, this is determined by the choice of flux quanta. Various recent studies have shown an interesting interplay between the existence of (approximate) flat directions and a low value of $W_0$ \cite{Demirtas:2019sip, Demirtas:2021ote, Demirtas:2021nlu, Demirtas:2020ffz, Alvarez-Garcia:2020pxd}. Ref. \cite{Burgess:2020qsc} argued that these perturbative flat directions are associated to pseudo-Goldstone bosons of a $2$-parameter family of scale invariance of the classical $10$-dimensional theory. One of these $2$ symmetries is the scale transformation included in $SL(2,\mathbb{R})$, while the other transforms the metric. Both of them are spontaneously broken by the fact that both the metric and the dilaton acquire a vacuum expectation value. When combined with axionic shift symmetries, this is reflected in the $4$-dimensional effective theory in the fact that both the axio-dilaton and the overall K\"ahler modulus are flat directions at classical level. However, in the $10$-dimensional theory also $G_3$ transforms with a non-zero weight. When compactifying, $3$-form fluxes take quantised background values, and so act as explicit breaking parameters which lift the axio-dilaton. In \cite{Burgess:2020qsc} the explicit breaking parameter was identified in $W_0$ (promoted to a spurion), arguing that $W_0=0$ implies the existence of a flat direction, in agreement with the findings of \cite{Demirtas:2019sip, Demirtas:2021ote, Demirtas:2021nlu, Demirtas:2020ffz, Alvarez-Garcia:2020pxd}. 

Notice that the condition to have a flat axio-dilaton is that $W_0=0$ after the complex structure moduli have been integrated out. In fact, in this case the $4$-dimensional action does not see any explicit scale breaking parameter since the flux quanta do not contribute to the scalar potential. On the other hand, $W_0=0$ can clearly be compatible with a stable axio-dilaton at classical level when $W_0$ has an appropriate dependence on $\phi$ after complex structure moduli stabilisation, even if the generic case would be characterised by $W_0\neq 0$. 

Let us stress that these flat directions are only approximate since they are expected to be lifted by a combination of non-perturbative and perturbative effects. Nevertheless they can have a wide variety of interesting phenomenological applications. The first is in the context of K\"ahler moduli stabilisation. A low value of $W_0$ is an essential ingredient for the KKLT scenario \cite{Kachru:2003aw} for moduli stabilisation. A method to construct vacua with low $W_0$ has been put forward in \cite{Demirtas:2019sip}.\footnote{For earlier work on obtaining low values of $W_0$ see \cite{Giryavets:2003vd, Cole:2019enn}, while for challenges in implementing moduli stabilisation and obtaining dS vacua in this setting see \cite{ Lust:2022lfc, Blumenhagen:2022dbo}.} This is in the large complex structure limit of the underlying CY compactification. Flux quanta are so chosen that they yield a GVW superpotential which, when computed using the perturbative part of the prepotential, is a degree-$2$ homogeneous polynomial. The homogeneity property and the request of a vanishing flux superpotential for non-zero values of the moduli guarantees the presence of a flat direction. This flat direction is lifted when non-perturbative corrections to the prepotential are incorporated. Hence $W_0$ acquires an exponentially small value (at weak string coupling). Working with a CY orientifold obtained by considering a degree-$18$ hypersurface in $\mathbb{CP}_{[1,1,1,6,9]}$, \cite{Demirtas:2019sip} presented an explicit choice of flux quanta corresponding to $W_0 \sim 10^{-8}$. Using the same method, an example with $W_0$ as low as $10^{-95}$ was constructed in \cite{Demirtas:2021ote, Demirtas:2021nlu}. Further studies of this setup have been carried out in \cite{Demirtas:2020ffz,Alvarez-Garcia:2020pxd, Broeckel:2021uty, Honma:2021klo, Grimm:2021ckh, Carta:2021kpk}.

Low values of $W_0$ might also be important in the context of LVS models \cite{Balasubramanian:2005zx}. Recently explicit LVS realisations of the Standard Model have been carried out by considering D3-branes at an orientifolded dP$_5$ singularity \cite{Cicoli:2021dhg}. Here the cancellation of all D7-charges and Freed-Witten anomalies forces the presence of a hidden D7 sector with non-zero gauge fluxes which induce a T-brane background suitable for de Sitter (dS) uplifting \cite{Cicoli:2015ylx}. The T-brane contribution can give a leading order Minkowski vacuum if the value of the $W_0$ is exponentially small in the string coupling, i.e. it is precisely of the form described above. A dS minimum with soft terms above the TeV scale requires $W_0$ as small as $10^{-13}$. Let us point out that, contrary to KKLT, in LVS an exponentially small value of $W_0$ is just a model-dependent condition since \cite{Cicoli:2017shd} presented a chiral global D3-brane model at an orientifolded dP$_0$ singularity which can allow for dS moduli stabilisation with T-brane uplifting for $W_0\sim \mc{O}(1)$. 

Approximate flat directions are naturally interesting also in the context of cosmology. In fact, the idea of focusing on degree-$2$ homogeneous superpotentials so as to obtain flat direction(s) was first used in \cite{Hebecker:2017lxm} to enhance the inflaton field range. More recently, flat directions in the type IIB flux superpotential have been used to construct models of sequestered inflation \cite{Kallosh:2021vcf}. Interestingly, the predictions of the models carry signatures of the moduli space geometry. Moreover, leading order flat directions can be promising candidates to realise quintessence models in order to avoid any destabilisation problem due to the inflationary energy contribution and to reproduce the correct tiny value of the cosmological constant scale \cite{Cicoli:2021fsd, Cicoli:2021skd}. 

Regarding supersymmetry breaking, type IIB models are characterised by a no-scale relation which implies that generically the main contribution to supersymmetry breaking comes from the K\"ahler moduli sector. In fact, typically at semi-classical level the complex structure moduli and the axio-dilaton are fixed by setting their F-terms to zero with $W_0 \neq 0$ which induces instead non-zero F-terms for the K\"ahler moduli (that are still flat at this level of approximation). However in scenarios with $W_0 = 0$ and a flat axio-dilaton, all F-terms are zero at leading order and the effective field theory after integrating out the complex structure moduli has to include both $\phi$ and the K\"ahler moduli \cite{Choi:2005ge}. Therefore the F-term of the axio-dilaton can also play an important role in supersymmetry breaking, especially in sequestered models with D3-branes where gaugino masses are controlled by the F-term of $\phi$ \cite{Blumenhagen:2009gk, Aparicio:2014wxa}.

Moreover, flux vacua with a leading order axionic flat direction and $W_0\neq 0$ have been shown in \cite{Hebecker:2020ejb} to be very promising to obtain a dS uplifting contribution from non-zero F-terms of the complex structure moduli, so providing an explicit realisation of the idea proposed in \cite{Saltman:2004sn,Gallego:2017dvd} without however the assumption of continuous $3$-form fluxes. More precisely, at perturbative level all F-terms of the complex structure moduli are zero with $W_0\neq 0$ and a flat axion. Instanton corrections to the superpotential lift the axion and shift all the remaining moduli, so that the corresponding F-terms become non-zero and can act as a dS uplifting source by an appropriate tuning of background fluxes \cite{Hebecker:2020ejb}.

From a more theoretical point of view, developments under the name `the tadpole problem' \cite{Bena:2020xrh, Marchesano:2021gyv, Plauschinn:2021hkp, Lust:2021xds, Tsagkaris:2022apo} seem to suggest that flat directions of the GVW superpotential might be a generic feature when the number of complex structure moduli is large. Thus classifying and studying the precise nature of flat directions, together with finding mechanisms for lifting them, are needed to develop a comprehensive understanding of the string landscape.

Finally let us make a few comments in the context of the statistical approach to string phenomenology (see e.g. \cite{Douglas:2003um, Ashok:2003gk, Denef:2004ze, Denef:2004dm, Denef:2004cf, Broeckel:2020fdz,Broeckel:2021dpz, Halverson:2019cmy, Susskind:2004uv, Douglas:2004qg, Dine:2004is, Arkani-Hamed:2005zuc, Kallosh:2004yh, Dine:2005yq, Sun:2022xdl}). Given the rich phenomenological applications of vacua with exponentially small $W_0$ and (approximate) flat directions in the complex structure and axio-dilaton moduli space, it is important to study how they fit in the full ensemble of type IIB vacua and develop an understanding of their statistical significance. Preliminary steps in this direction were taken in \cite{Broeckel:2021uty}. The analysis indicated that the class of vacua obtained in \cite{Demirtas:2019sip} occupy a small fraction of the full set of vacua at low $W_0$ as computed by the statistical methods in \cite{Denef:2004ze}. Given this, it is important to look for novel classes so as to enrich our knowledge of vacua at low $W_0$.

In this paper we will present a novel and more general method to find supersymmetric solutions with approximate flat directions in type IIB flux compactifications. We now give a qualitative description of our method to obtain the solutions and we provide a summary of the key results.
  
\subsection{Summary of results}
\label{sec:qual}
 
The superpotential in type IIB compactifications is given by the sum of the GVW superpotential \pref{gvw} and non-perturbative corrections. We will work with the GVW term (the non-perturbative terms are small corrections in the large radius limit) and search for supersymmetric minima with flat directions. At this level the conditions for supersymmetry are $D_\phi W = \del_\phi W + W \del_\phi K = 0$ and $D_{U_\alpha} W = \del_{U_\alpha} W+ W \del_{U_\alpha} K = 0$ where $U_\alpha$ $(\alpha = 1,\dots,h^{2,1}_-)$ are the complex structure moduli \cite{Giddings:2001yu}.\footnote{For toroidal compactifications primitivity of the fluxes has to be imposed as an additional requirement. This is due to the presence of holomorphic $1$-forms on tori \cite{Giddings:2001yu}. In our study of a toroidal case we impose this condition at the very end, after having obtained solutions to the F-flatness conditions.} Given that (\ref{gvw}) does not depend on the K\"ahler moduli $T_i$ $(i = 1,\dots,h^{1,1}_+)$, the F-flatness conditions for these modes is $W = 0$ since $D_{T_i} W = W \del_{T_i} K$ with $\del_{T_i} K \neq 0$ for finite field values. Thus supersymmetry at classical level requires $W=\del_\phi W = \del_{U_\alpha} W = 0$. Notice that flat directions can clearly exist also for $W\neq 0$ where supersymmetry is definitely broken by the K\"ahler moduli (and potentially by the axio-dilaton and the complex structure moduli as well). Despite being interesting for phenomenological and cosmological applications, these solutions would typically be characterised by large values of $W_0$ which are incompatible with the KKLT scenario (and some LVS models with T-brane uplifting). 

Our analysis will be for toroidal orientifolds and CY compactifications in the large complex structure limit where the superpotential is a polynomial (after dropping exponentially small terms in the large complex structure limit). Thus the F-flatness conditions $W=\del_\phi W = \del_{U_\alpha} W = 0$ are $n + 1= h^{2,1}_-+2$ polynomial equations in $n$ complex variables which in general do not have a solution since the system is overdetermined. In addition, we are interested in solutions with $p\geq 1$ flat directions which can exist if the number of linearly independent equation is reduced from $(n+1)$ to $(n-p)$ by an appropriate flux choice. Thus the first step is to understand which choice of flux quanta can yield solutions with flat directions. At present the answer to this in full generality is unknown, and so we have to resort to a well motivated ansatz.

Before discussing our ansatz, let us recapitulate the basic idea in \cite{Hebecker:2017lxm, Demirtas:2019sip}. Flux quanta were chosen so that $W$ was 
a degree-$2$ homogeneous polynomial. For such superpotentials:
\be
2 W = \phi \,\del_\phi W + U_\alpha \,\del_{U_\alpha} W\,,
\label{degreetwo}
\ee
holds as a functional relation (i.e. on all points on the moduli space). This implies that the $W =0$ equation is automatically satisfied once the derivatives of $W$ vanish. Furthermore the scaling behaviour of $W$ implies that, if $(\hat\phi,\hat{U}_\alpha)$ is a solution, $\phi = \lambda \hat\phi$, $U_\alpha = \lambda \hat{U}_\alpha$ remains a solution, signaling the existence of a flat direction parametrised by $\lambda$.\footnote{Unless $\hat\phi=0$ and $\hat{U}_\alpha=0$ $\forall \alpha$ which is however a situation that we do not consider since it would lead to a breakdown of the effective field theory.} This implies that, on top of (\ref{degreetwo}), $\del_\phi W$ can be expressed as a linear combination of the derivatives of $W$ with respect to the complex structure moduli. This can be easily seen in the $h^{1,2}_-=1$ case where, setting $c\equiv W_{\phi\phi} W_{UU} - W_{\phi U} W_{U\phi}$, one has:
\be
\begin{cases}
W_{\phi\phi} W_U = W_{U\phi} W_\phi + c\, U \\
2 W_{\phi\phi} W = W_\phi^2 + c\, U^2
\end{cases}
\qquad \underrightarrow{c=0} \qquad 
\begin{cases}
W_U = \left(\frac{W_{U\phi}}{W_{\phi\phi}}\right) W_\phi \\
2 W_{\phi\phi} W = W_\phi^2
\end{cases}
\ee
showing that the flux choice $c=0$ (or $W_{\phi\phi} W_{UU} = W_{\phi U} W_{U\phi}$) guarantees that $W=0$ for $U\neq 0$ and the fact that $W=\del_U W=0$ is an automatic consequence of $\del_\phi W=0$, signaling the presence of a flat direction. 

The lesson to take from the above is that superpotentials where there are functional relations between $W$ and its derivatives, such that the vanishing of some implies the vanishing of other(s), are particularly suited for obtaining solutions with flat directions. In this paper we will focus on the more general case where $W$ is not necessarily a homogeneous function but its derivatives are linearly dependent:
\be
\lambda_\phi \,\del_\phi W + \lambda_\alpha \,\del_{U_\alpha} W = 0\,,
\label{linrel}
\ee
where $\lambda_\phi$ and $\lambda_\alpha$ are constants with no moduli dependence. Our strategy is as follows:
\begin{enumerate}
\item Given a toroidal orientifold or an orientifolded CY in the large complex structure limit, we compute the superpotential in full generality as a function of the flux vectors and moduli. 

\item We impose that a condition of the form \pref{linrel} holds as a functional relation, and determine the constraints that this sets on the fluxes. At this stage $\lambda_\phi$ and $\lambda_\alpha$ are to be thought of as parameters in the ansatz for the fluxes. Thus the constrained fluxes are allowed to depend on them. This in general reduces the number of independent equations from $n+1$ to $n$.

\item Taking the fluxes obtained in the previous step, we impose the F-flatness conditions and the requirement to have at least $1$ flat direction. Unlike the case of a degree-$2$ homogeneous superpotential, a flat direction is not guaranteed if just a condition of the form \pref{linrel} holds. When possible, the existence of a flat direction is obtained by an appropriate choice of $\lambda_\phi$ and $\lambda_\alpha$ which reduces further the number of independent equation from $n$ to $n-p$ with $p\geq 1$. Thus the requirement of a flat direction can further constrain the fluxes.\footnote{In effect, we adjust fluxes to ensure the following. We have $n$ independent equations in $n$ variables: $f^{l}(U_k)=0,\ U_k=\phi,U^a$, after step 2. For cases with flat directions,  $\det\left(\partial_k f^l\right)$ vanishes at the solution.}

\item The end result of step $3$ are solutions to the F-flatness conditions with at least $1$ flat direction and flux vectors parametrised by $\lambda_\phi$ and $\lambda_\alpha$. Of these we isolate the subset of flux vectors that satisfy the integrality and the D3 tadpole condition. We also impose physical restrictions such as the positivity of ${\rm Im}\,\phi$ which sets the string coupling (${\rm Im}\,\phi=g_s^{-1}$).

\item For toroidal examples we finally impose also the primitivity of $G_3$ to have a supersymmetric solution.
\end{enumerate}
A few comments are in order. Implementing the procedure working with the general form of the linear relation is rather cumbersome. It is easier to work case by case with the linear relations being classified by which of the $\lambda_\phi$ and $\lambda_\alpha$ are non-vanishing. We have included the checks for the solutions being physical in step $4$ of the procedure. In practice, it is easier to check for these conditions at every stage and discard any candidate solution as soon as it becomes clear that it is unphysical.

Let us highlight our key results. An explicit implementation of the algorithm has been carried out for the $T^6/ \mathbb{Z}_2$ orientifold \cite{Kachru:2002he, Frey:2002hf}, an orientifold of the CY obtained by considering a degree-$18$ hypersurface in $\mathbb{CP}_{[1,1,1,6,9]}$ (first studied in the context of mirror symmetry in \cite{Candelas:1994hw} and also the example studied in \cite{Demirtas:2019sip}), and an orientifold of the CY discussed in \cite{Cicoli:2013cha}.
\begin{itemize}
\item For the $T^6/\mathbb{Z}_2$ orientifold, we find solutions with $1$ and $2$ flat directions (and no more). The solutions fall into various families (classified according to the nature of the linear relation that holds). In all solutions the residual moduli space contains regions in which the string coupling is arbitrarily small.

\item For the $T^6/\mathbb{Z}_2$ orientifold, there are solutions which preserve $N=2$ supersymmetry in $4$ dimensions. Being novel solutions with extended supersymmetry, they are interesting in their own right.

\item For the $\mathbb{CP}_{[1,1,1,6,9]}[18]$ case, we find essentially $1$ family of fluxes which lead to solutions with $1$ flat direction corresponding to the axio-dilaton. One can ensure that the moduli take on values in the large complex structure limit (as is required for the consistency of our analysis) when the string coupling is taken to arbitrarily small values.

\item We find $68$ distinct solutions in the $\mathbb{CP}_{[1,1,1,6,9]}[18]$ case, $15$ of which are entirely novel since the superpotential is a non-homogeneous polynomial. The remaining $53$ solutions can instead be mapped by duality to the case when the superpotential is a degree-$2$ homogeneous polynomial. However only $2$ out of these $53$ solutions lie at weak string coupling and in a regime where the large complex structure limit is definitely under control, reproducing the old vacua already found in \cite{Demirtas:2019sip, Broeckel:2021uty, Carta:2021kpk}.

\item For the $\mathbb{CP}_{[1,1,1,6,9]}[18]$ case, we find also solutions with $1$ axionic flat direction and $W\neq 0$ which represent promising starting points for an explicit CY realisation of winding dS uplift \cite{Hebecker:2020ejb}. In this case, $W$ is still a polynomial of degree $2$ but not a homogeneous function.

\item For the CY studied in \cite{Cicoli:2013cha}, which features effectively $1$ complex structure modulus more than the $\mathbb{CP}_{[1,1,1,6,9]}[18]$ example, we present a preliminary analysis where we find solutions with $2$ flat directions. Again, $W$ is a polynomial of degree $2$ that is always non-homogeneous when $W\neq 0$, while it can become a homogeneous function for some flux quanta only when $W=0$ at the minimum (as in the $\mathbb{CP}_{[1,1,1,6,9]}[18]$ case, there are $W=0$ cases where $W$ cannot be made non-homogeneous by duality).
\end{itemize}

Before closing the introduction we would like to mention that, while we provide a systematic classification of the solutions into families, we do not carry out an exhaustive search determining all solutions in each family. For the toroidal case, we isolate the family that contains all solutions up to duality equivalences. We provide a large class of representative examples for both toroidal and CY cases, leaving exhaustive tabulations for future work.

This article is structured as follows. Sec. \ref{sec:tor} is on the $T^6/ \mathbb{Z}_2$ orientifold. Here, after reviewing some background material, we provide a classifications of the solutions with explicit examples. In particular, Sec. \ref{n2} is devoted to solutions with $N=2$ supersymmetry. Sec. \ref{sec:cy} is instead on orientifolded CYs in the large complex structure limit. After reviewing the basics and a general discussion, we give a detailed treatment of the $\mathbb{CP}_{[1,1,1,6,9]}[18]$ example and a preliminary analysis of the CY studied in \cite{Cicoli:2013cha} which features effectively $3$ complex structure moduli. Sec. \ref{sec:pheno} gives a general discussion of how the flat directions can get lifted and potential phenomenological implications. We conclude in Sec. \ref{sec:conc}.

\section{Flat directions in the $T^6/\mathbb{Z}_2$ orientifold}
\label{sec:tor}

In this section we will study classical supersymmetric solutions with flat directions that can arise in the $T^6/\mathbb{Z}_2$ orientifold. This is the setting where some of the first explicit computations of the flux potential in type IIB were carried out \cite{Kachru:2002he, Frey:2002hf}. Flux vacua in the toroidal setting have been studied in much detail (see e.g. \cite{Aldazabal:2011yz,Yang:2005fa,Grana:2005jc,Lust:2005dy,Antoniadis:2005nu,Cvetic:2005bn,Frey:2003sd,DAuria:2003nhg,Blumenhagen:2003vr, Kobayashi:2020hoc} for related studies). We will follow the conventions of \cite{Kachru:2002he} in our treatment.

\subsection{Type IIB toroidal flux compactifications}

In this section we review some basic ingredients of type IIB compactifications on the $T^6/\mathbb{Z}_2$ orientifold with non-trivial $3$-form fluxes turned on. This will also help to set our notation. The type IIB supergravity action in Einstein frame is:
\bea
S_{\rm IIB}&=&\frac{1}{2\kappa_{10}^2}\int \dd^{10}x\sqrt{-g}\left( R-\frac{\partial_\M\phi\partial^\M\phi}{2\left({\rm Im}\phi\right)^2}-\frac{G_3\cdot\bar{G}_3}{2\cdot 3!\,{\rm Im}\phi}-\frac{\tilde{F}_5^2}{4\cdot 5!} \right) \nn \\
&+&\frac{1}{2\kappa_{10}^2}\int \dd^{10}x\,\frac{C_4\wedge G_3\wedge \bar{G}_3}{4{\rm i}\,{\rm Im}\phi}\ +S_{\rm local}\,,
\eea
where:
\bea
\phi&=&C_0+i/g_s\,,\qquad F_3 = \dd C_2\,,\qquad H_3= \dd B_2\,, \nn \\
G_3 &=& F_3-\phi H_3\,, \qquad \tilde{F}_5 = F_5 - \frac12 C_2 \wedge H_3 + \frac12 F_3\wedge B_2\,,\qquad *\tilde{F}_5 = \tilde{F}_5\,.
\eea
Upon compactifying on the $T^6/\mathbb{Z}_2$ orientifold with spacetime-filling D3-branes, the D3 tadpole condition is (setting $2 \pi\sqrt{\alpha'}=1$):
\be
\frac12\, N_{\rm flux}+N_\D3 -16=0\,,\qquad N_{\rm flux}\equiv \int_{T^6} H_3 \wedge F_3\,,
\label{Nflux}
\ee
where $N_\D3$ is the number of D3-branes. The flux contribution can been shown to be positive semi-definite. The negative contribution arises from the $2^6$ O3-planes. Clearly this condition implies $0< N_{\rm flux} \leq 32$.\footnote{We do not consider the $N_{\rm flux}=0$ case since, due to the imaginary self-duality condition on the fluxes, it corresponds to either $g_s\to \infty$ or trivial flux quanta.}

The geometry of the torus will be parametrised as follows. The $6$ real periodic coordinates on $T^6$ are denoted as $x^i$, $y^i$, $i=1,2,3$ with $x^i\sim x^i+1$, $y^i\sim y^i+1$. The holomorphic $1$-forms are taken to be $\dd z^i=\dd x^i+\tau^{ij}\dd y^j$, where $\tau^{ij}$ is the period matrix. The choice of orientation is:
\be
\int \dd x^1 \wedge \dd x^2 \wedge \dd x^3 \wedge \dd y^1 \wedge \dd y^2 \wedge \dd y^3 = 1 \,.
\label{choice}
\ee  
We will make use of the following orthonormal basis $\{\alpha_0,\alpha_{ij},\beta^{ij},\beta^0\}$ for $H^3(T^6,\mathbb{Z})$:
\bea
\alpha_0 &=& \dd x^1 \wedge \dd x^2 \wedge \dd x^3\,, \qquad \alpha_{ij} =\frac12 \epsilon_{ilm} \dd x^l \wedge \dd x^m \wedge \dd y^j\,, \nn \\
\beta^{ij} &=&-\frac12 \epsilon_{jlm} \dd y^l \wedge \dd y^m \wedge \dd x^i\,,\qquad \beta^0 = \dd y^1 \wedge \dd y^2 \wedge \dd y^3 \,,\quad i,j=1,2,3\,,
\label{threebasis}
\eea
with:
\be
\int \alpha_I\wedge\beta^J = \delta_I^J\,.
\ee
Finally the holomorphic $3$-form is taken to be $\Omega = \dd z^1 \wedge \dd z^2 \wedge \dd z^3$. The NSNS and RR fluxes can be expanded in terms of the orthonormal basis as:
\bea
F_3 &=& a^0\alpha_0+a^{ij}\alpha_{ij}+b_{ij}\beta^{ij}+b_0\beta^0\,, \nn \\
H_3 &=& c^0\alpha_0+c^{ij}\alpha_{ij}+d_{ij}\beta^{ij}+d_0\beta^0\,,
\eea
where the Dirac quantisation condition requires $(a^0,a^{ij},b_{ij},b_0,c^0,c^{ij},d_{ij},d_0)$ to be integers. We will restrict them to be even integers so as to avoid the need for any discrete flux on the orientifold planes.\footnote{See \cite{Kachru:2002he,Frey:2002hf} for a discussion on this point.} The flux contribution to the D3 tadpole (\ref{Nflux}) takes the form:
\be
N_{\rm flux} = (c^0b_0-a^0d_0)+(c^{ij}b_{ij}-a^{ij}d_{ij})\,,
\ee
while the GVW superpotential (\ref{gvw}) becomes: 
\be
W =(a^0-\phi c^0)\det \tau-(a^{ij}-\phi c^{ij})({\rm cof}\,\tau)_{ij}-(b_{ij}-\phi d_{ij})\,\tau^{ij}-(b_0-\phi d_0)\,.
\label{gvwtorus}
\ee
Supersymmetry is preserved when the F-flatness conditions of this superpotential are satisfied, together with $W=0$ (which can be thought of as the F-flatness condition for the K\"ahler moduli) and the requirement of primitivity of $G_3$ (i.e. the existence of a K\"ahler form such that $J \wedge G_3 = 0$). We will use the method described in Sec. \ref{sec:qual} to obtain supersymmetric solutions with at least $1$ flat direction. The F-flatness and $W=0$ conditions are equivalent to:
\bea
\label{torusF}
f^{(1)} &\equiv& a^0 \det \tau - a^{ij} ({\rm cof}\,\tau )_{ij} - b_{ij} \tau^{ij} - b_0 = 0\,, \nn \\
f^{(2)} &\equiv& c^0 \det \tau - c^{ij} ({\rm cof}\,\tau )_{ij} - d_{ij} \tau^{ij} - d_0 = 0\,, \nn \\
f^{(3)}_{kl} &\equiv& (a^0 - \phi c^0 )({\rm cof}\,\tau )_{kl} - (a^{ij} - \phi c^{ij} ) \epsilon_{kim} \epsilon_{ljn} \tau^{mn} - (b_{ij} - \phi d_{ij} ) \delta_k^i \delta_l^j = 0\,.
\eea
The primitivity of $G_3$ will be imposed as a final condition. We will see that a suitable choice of the K\"ahler form satisfying the primitivity condition can be found for all cases.

\subsection{Supersymmetric solutions with $W=0$}

In this section we present explicit solutions for the $T^6/\mathbb{Z}_2$ orientifold. We will consider the class in which the flux vectors are diagonal, i.e.:
\be
a^{ij}={\rm diag}\{a_1,a_2,a_3\},\,\, b_{ij}={\rm diag}\{b_1,b_2,b_3\},\,\, c^{ij}={\rm diag}\{c_1,c_2,c_3\},\,\, d_{ij} = {\rm diag}\{d_1,d_2,d_3\}\,,
\label{diagflux}
\ee
which lead to:
\be
N_{\rm flux} = (b_0c_0 - a_0d_0) + (b_1c_1 - a_1d_1) + (b_2c_2 - a_2d_2) + (b_3c_3 - a_3d_3)\,.
\ee
Given that the structure of (\ref{torusF}) implies a diagonal form of the period matrix, we take:
\be
\tau^{ij}={\rm diag}\{\tau_1,\tau_2,\tau_3\}\,.
\ee
Note that this corresponds to a $T^2 \times T^2 \times T^2$ factorisation of the $T^6$ with $\tau_\alpha$ $(\alpha=1,2,3)$ as the complex structure moduli of the $3$ $2$-tori. For notational convenience we introduce:
\be
(U_1,U_2,U_3,U_4) \equiv (\tau_1,\tau_2,\tau_3,\phi)\,. 
\ee   
With this, (\ref{gvwtorus}) takes the form:
\bea
W &=& (a^0-U_4 c^0)U_1U_2U_3-(a_1-U_4 c_1)U_2U_3-(a_2-U_4 c_2)U_1U_3-(a_3-U_4 c_3)U_1U_2 \nn \\
 &-& (b_1-U_4 d_1)U_1-(b_2-U_4 d_2)U_2-(b_3-U_4 d_3)U_3-(b_0-U_4 d_0)\,,
\label{eq:Wexpres}
\eea
and the system of equations (\ref{torusF}) reduces to:
\begin{align}
  &a^0 U_1U_2 U_3 - (a_1U_2 U_3+a_2U_1U_3+a_3U_1U_2) - (b_1U_1+b_2U_2+b_3U_3) - b_0 = 0\,,\label{eq:sup3}\\
  &c^0 U_1U_2 U_3 - (c_1U_2 U_3+c_2 U_1U_3+c_3 U_1U_2) - (d_1U_1+d_2 U_2+d_3 U_3) - d_0 = 0\,,\\
  &(a^0 - U_4 c^0 ) U_2 U_3 - ((a_2 U_3+a_3 U_2) - U_4 (c_2 U_3+c_3 U_2) ) - (b_1 - U_4 d_1 ) = 0\,,\\
  &(a^0 - U_4 c^0 ) U_1U_3 - ((a_1 U_3+a_3 U_1) - U_4 (c_1U_3+c_3U_1) ) - (b_2 - U_4 d_2 ) = 0\,,\\
  &(a^0 - U_4 c^0 )U_1U_2 - ((a_1U_2+a_2U_1) - U_4(c_1U_2+c_2 U_1) ) - (b_3 - U_4 d_3 ) = 0\,.\label{eq:sup4}
\end{align}
In the next sections we present different families of solutions to these F-flatness and $W=0$ conditions. We start with an example without any flat direction and we then provide our classification of the solutions with flat directions.\footnote{Let us point out that this is not a full classification of the solutions since we obtain only those which satisfy the linear dependence ansatz \pref{linrel}.} Representative examples of flux vectors satisfying the integrality condition are provided for all the families that arise in the classification. We first present solutions with $1$ flat direction and then solutions with $2$ flat directions (our ansatz does not lead to any solutions with higher number of flat directions). As mentioned earlier, the solutions will be classified according to the nature of the linear relation that the derivatives of the superpotential satisfy. This leads to $3$ different cases (all compatible with $N_{\rm flux}\neq 0$) for which \eqref{eq:sup3}-\eqref{eq:sup4} admit complex solutions, i.e. ${\rm Im}\,U_a \neq 0$ $ \forall a$, with $1$ or $2$ flat directions:
\begin{enumerate}
\item Linear relation among all derivatives: $\lambda_1\,\partial_1 W +\lambda_2\,\partial_2 W + \lambda_3 \partial_3 W + \partial_4 W = 0$ with $\lambda_\alpha\neq 0$ $\forall \alpha=1,2,3$ which can allow for solutions with $W=0$ and either $1$ or $2$ flat directions;

\item Linear relation among the derivatives of $W$ with respect to the axio-dilaton and $1$ complex structure modulus: $\lambda_\alpha\,\partial_\alpha W + \partial_4 W = 0$ (no sum over $\alpha$) with $\alpha = 1,2,3$ which can feature solutions with $W=0$ and $2$ flat directions;

\item Linear relation among the derivatives of $W$ with respect to $2$ different complex structure moduli: $\partial_\alpha\, W = \lambda_\beta\, \partial_\beta W$ (no sum over $\beta$) with $\alpha\neq \beta$ and $\alpha,\beta=1,2,3$ which can give solutions with $W=0$ and $2$ flat directions.
\end{enumerate}

\subsubsection*{Solutions without flat directions}

In this section we review a solution presented in \cite{Kachru:2002he} which has $W=0$ but no linearity relation among the superpotential and its derivatives. Hence it does not feature any flat direction since it can be shown that the solution is not part of a continuous family. In this case the fluxes are taken to be proportional to identity:
\be
(a^{ij},b_{ij},c^{ij},d_{ij}) = (a,b,c,d)\,\delta_{ij}\,,\quad a_0=b_0=c_0= -c=-d=2\,,\quad a= b =0 \,,\quad  d_0 = -4\,, \nn
\ee
and an explicit solution to \eqref{torusF} is given by:
\be
\tau^{ij} = \tau\,\delta^{ij}\,,\qquad \tau = \phi = e^{{\rm i}\frac{2\pi}{3}}\,.
\label{eq:exKST}
\ee
For a set of fluxes to find whether a given solution is isolated or part of a continuous family, we will use linearised perturbation theory.\footnote{This technique is not limited to diagonal fluxes. For a generic choice of fluxes, even if a given solution has diagonal $\tau^{ij}$, that may be a part of a continuous family with non-zero off-diagonal terms. As a result, in general we must deal with a $11\times10$ matrix, as shown below. However only $\tau^{ij}=\tau\delta^{ij}$ can satisfy \eqref{torusF} for fluxes proportional to the identity. As a result, it is possible to work with a matrix with lower dimensions.} For this let us write abstractly the system of equations (\ref{torusF}) as:
\be
f^{(I)}(U_a) = 0\,,
\ee
where $I$ runs over the $11$ equations and $U_a$ runs over the $10$ variables $(\tau^{ij},\phi)$. Then if the solution $\hat{U}_a$ is part of a continuous family, the following linear system for $\delta U_a$ must have a solution:
\be
\partial_{U_a}f^{(I)}\bigg\vert_{\hat{U}_a}\delta U_a = 0\,,
\ee
i.e. the rank of the matrix $\partial_{U_a}f^{(I)}(\hat{U}_a)$ should be less than $10$. Now, the matrix elements are given by:\footnote{Here we use $\partial_{\tau^{ij}} \det\tau = \frac12\epsilon_{ikl}\epsilon_{jmn}\tau^{km}\tau^{ln},\ \partial_{\tau^{ij}} ({\rm cof}\,\tau)_{ab} =  \epsilon_{ial}\epsilon_{jbn}\tau^{ln}$, and repeated indices are summed.}
\begin{align}
  &\partial_{\tau^{ij}}f^{(1)} = \frac12 a^0\epsilon_{ikl}\epsilon_{jmn}\tau^{km}\tau^{ln} - a^{km}\epsilon_{ikl}\epsilon_{jmn}\tau^{ln}-b_{ij}\,,\\
  &\partial_{\tau^{ij}}f^{(2)} = \frac12 c^0\epsilon_{ikl}\epsilon_{jmn}\tau^{km}\tau^{ln} - c^{km}\epsilon_{ikl}\epsilon_{jmn}\tau^{ln}-d_{ij}\,,\\
  &\partial_{\tau^{ij}}f^{(3)}_{kl} = (a^0 -\phi c^0) \epsilon_{ikm}\epsilon_{jln}\tau^{mn} - (a^{mn}-\phi c^{mn})\epsilon_{ikm}\epsilon_{ljn}\,,\\
  &\partial_{\phi}f^{(1)} = \partial_{\phi}f^{(2)} = 0\,,\\
  &\partial_{\phi}f^{(3)}_{kl} = - c^0({\rm cof}\,\tau)_{kl} + c^{ij}\epsilon_{ikm}\epsilon_{jln}\tau^{mn} + d_{kl}\,.
\end{align}
For fluxes proportional to the identity, these matrix elements evaluated at $(\tau\delta^{ij},\phi)$ become:
\begin{align}
  &\partial_{\tau^{ij}}f^{(1)} = (a^0\tau^2 - 2 a \tau-b)\delta_{ij}\,, \\
  &\partial_{\tau^{ij}}f^{(2)} = (c^0\tau^2 - 2 c \tau-d)\delta_{ij}\,, \\
  &\partial_{\tau^{ij}}f^{(3)}_{kl} =[(a^0-\phi c^0)\tau - (a-\phi c)](\delta_{ij}\delta_{kl}-\delta_{il}\delta_{jk})\,, \\
  &\partial_{\phi}f^{(1)} = \partial_{\phi}f^{(2)} = 0\,, \\
  &\partial_{\phi}f^{(3)}_{kl} =-(c^0\tau^2 - 2 c \tau-d)\delta_{kl}\,.
\end{align}
The above matrix has rank $10$ at \eqref{eq:exKST}, implying that it is a solution with no flat directions.

\subsubsection*{Solutions with $1$ flat direction}

Solutions with $1$ flat direction are all in $1$ family. The linear relation satisfied in this family is:
\be
\lambda_1\,\partial_1 W + \lambda_2\,\partial_2 W + \lambda_3\,\partial_3 W + \partial_4 W = 0\,,
\ee 
with $\lambda_\alpha\neq0$ $\forall \alpha=1,2,3$. The flux quanta (introduced in \pref{diagflux}) take the form:
\bea
\{a_0, a_1, a_2, a_3\}&=& \{0,\frac{d_3}{\lambda_2}+\frac{d_2}{\lambda_3},-\frac{d_2\lambda_2}{\lambda_1\lambda_3},-\frac{d_3\lambda_3}{\lambda_1\lambda_2}\}\,,\quad \{c_0, c_1, c_2, c_3\} = \{0,0,0,0\}\,, \nn \\
\{b_0, b_1, b_2, b_3\}&=&\{b_0,\frac{d_0-b_2\lambda_2-b_3\lambda_3}{\lambda_1},b_2,b_3\}\,,\quad \{d_0, d_1, d_2, d_3\}=\{d_0,-\frac{d_2\lambda_2+d_3\lambda_3}{\lambda_1},d_2,d_3\}\,, \nn
\label{eq:fluxesoneflat}
\eea
with the condition $d_2,d_3,d_2\lambda_2+d_3\lambda_3\neq 0$. With this choice of fluxes $N_{\rm flux}$ becomes:
\be
N_{\rm flux} = \frac{2}{\lambda_1\lambda_2\lambda_3 } \left(\lambda_2 ^2 d_2^2+\lambda_2  \lambda_3  d_2 d_3 +\lambda_3^2 d_3^2\right)\,,
\label{eq:Nfluxcond111}
\ee
and the GVW superpotential reduces to:
\bea
W &=& \frac{1}{\lambda_1} \left[(b_2 \lambda_2 +b_3 \lambda_3-d_0) U_1 -\lambda_1 (b_2 U_2 + b_3 U_3 - d_0 U_4+ b_0 )\right] \nn \\
&+& \frac{d_2}{\lambda_1 \lambda_3} (U_3-\lambda_3  U_4) (\lambda_2  U_1-\beta  U_2)+\frac{d_3}{\lambda_1\lambda_2} (U_2-\lambda_2 U_4) (\lambda_3  U_1-\lambda_1 U_3)\,.
\eea
Demanding that the derivatives of the superpotential vanish implies that the $3$ complex structure moduli $U_\alpha$, $\alpha=1,2,3$, are related to the axio-dilaton $U_4$ as follows:
\bea
U_1 &=& -\frac{\lambda_1(b_3d_2+b_2d_3)}{2d_2d_3}+\frac{\lambda_1 d_0(\lambda_2 d_2+\lambda_3 d_3)}{2\lambda_2\lambda_3 d_2d_3} +\lambda_1\, U_4\,, \nn\\
U_2 &=& -\frac{\lambda_2 b_3}{d_3}+\frac{\lambda_2^2 (b_3d_2-b_2d_3)}{2d_3 (\lambda_2 d_2+\lambda_3 d_3)}+\frac{\lambda_2 d_0}{2\lambda_3 d_3}+\lambda_2\, U_4\,, \nn \\
U_3 &=& -\frac{\lambda_3 b_2}{2d_2}-\frac{\lambda_3(\lambda_2 b_2+\lambda_3 b_3)}{2(\lambda_2 d_2+\lambda_3 d_3)}+\frac{\lambda_3  d_0}{2 \lambda_2  d_2}+\lambda_3\,  U_4\,.
\eea
The $W=0$ condition instead implies:
\be
(\lambda_2 d_2+\lambda_3  d_3) \left[4 b_0 d_2 d_3 -2 \lambda_2  \lambda_3  d_0 (b_2d_3+b_3 d_2) +d_0^2 (\lambda_2 d_2+\lambda_3 d_3)\right] 
+\lambda_2 \lambda_3  (b_3 d_2-b_2 d_3)^2 = 0\,.
\label{eq:diophantine}
\ee
Note that this can be thought of as a relation between the parameters $\lambda_2$ and $\lambda_3$. Hence the flux quanta are essentially parametrised by $2$ parameters and some integers. We could have presented the flux vectors as functions of $2$ parameters from the very beginning. In this case the $W=0$ condition would have been automatically satisfied. We did not do so to avoid cluttering the notation.

In summary, the solutions are obtained by choosing the even integers $b_0$, $b_2$, $b_3$, $d_0$, $d_2$, $d_3$ and the parameters $\lambda_\alpha$ $\alpha=1,2,3$ such that all flux quanta in \pref{eq:fluxesoneflat} are even, the $W=0$ condition \pref{eq:diophantine} is met and the D3 tadpole condition $N_{\rm flux} \leq 32$ (with $N_{\rm flux}$ given in \pref{eq:Nfluxcond111}) is satisfied. Furthermore, physical consistency conditions such as ${\rm{Im}}(U_4) > 0$ must be satisfied. It is easy to find explicit examples. For instance:
\be
\lambda_1=\lambda_2=\lambda_3=1\,, \qquad b_2=b_3=0\,,\qquad d_2= d_3=2\,,
\ee
and:
\be
b_0=-4p^2\,,\qquad d_0=4p\,,\quad p\in\mathbb{Z}\,,
\ee
yields a family of solutions parametrised by $p\in\mathbb{Z}$. The corresponding flux quanta are:
\bea
\{a_0, a_1, a_2, a_3\} &=& \{0, 4, -2, -2\}\,,\qquad \{b_0, b_1, b_2, b_3\}=\{-4 p^2, 4 p, 0, 0\}\,, \nn \\
\{c_0, c_1, c_2, c_3\} &=& \{0,0,0,0\}\,,\qquad \{d_0, d_1, d_2, d_3\}=\{4 p, -4, 2, 2\}\,.
\label{eq:fluxesEx1flat}
\eea
It follows that $N_{\rm flux}=24$ and the superpotential can be written as:
\be
W = 2 \left(2 p^2+2 p (U_4-U_1)+U_1 (U_2+U_3-2 U_4)+U_4 (U_2+U_3)-2 U_2 U_3\right),
\ee
which satisfies:
\be
W \propto (\partial_2W - \partial_3W)^2 + 4 (\partial_2W + \partial_3W) \partial_4W + 4 (\partial_4W)^2\,.
\ee
Due to above relation, solving $\partial_a W=0$, $a=1,\dots,4$ automatically sets $W=0$, although $W$ does not have any scaling property when $p\neq0$.\footnote{By scaling property of a function $g(U_1\dots,U_n)$, we mean that there exists a set of numbers $\lambda_1,\dots,\lambda_n$ not all zeros, such that $$g(\lambda^{w_1}U_1,\dots,\lambda^{w_n}U_n)=\lambda^{w(w_1,\dots,w_n)}g(U_1,\dots,U_n),\quad w(w_1,\dots,w_n)\neq0\,.$$} For $p=0$, $W$ is a degree-$2$ homogeneous function. Let us mention that we are unable to find even integer fluxes \eqref{eq:fluxesoneflat} subject to \eqref{eq:diophantine} and $0<N_{\rm flux}\le32$, for which $N_{\rm flux}$ is other than $24$. At the F-flatness locus the moduli take the values:
\be
(U_1,U_2,U_3,U_4) = \left(U_4 +2p, U_4 +p, U_4+p, U_4\right).
\label{Sol1flat}
\ee
This might seem as giving an infinite number of solutions. However, one needs to check if the solutions are physically distinct or related by duality transformations. We give a summary of the relevant duality transformations in App. \ref{AppB}. Applying these we find that the infinite class actually corresponds to just $1$ distinct solution with representative the case $p=0$.

\subsubsection*{Solutions with $2$ flat directions}
\label{twoflat}

In this section we discuss solutions with $2$ flat directions. Classified according to the nature of the linear relations satisfied by the derivatives of the superpotential, these solutions fall into $3$ families. 

\vspace{0.5cm}
\noindent \underline{\bf{Family $\mc{A}$}}: For this family the linear relation involves the derivatives of $W$ with respect to all moduli and looks like:
\be
\lambda_1\,\partial_1W+\lambda_2\,\partial_2W+\lambda_3\,\partial_3W+\partial_4W = 0\,,\qquad \lambda_1,\lambda_2,\lambda_3\neq 0\,.
\label{eq:linrelA}
\ee
With this, the allowed flux quanta fall into $3$ subfamilies. We will refer to them as $\mc{A}_1$, $\mc{A}_2$ and $\mc{A}_3$.

\vspace{0.5cm}
\noindent  \underline{Subfamily $\mc{A}_1$}: Here the flux quanta are characterised by $d_3\neq0$ and take the form:
\bea
\{a_0, a_1, a_2, a_3\}&=&\{0,\frac{d_3}{\lambda_2},0,-\frac{d_3\lambda_3}{\lambda_1\lambda_2}\}\,,\quad \{b_0, b_1, b_2, b_3\}=\{\frac{b_3d_0}{d_3},-\frac{b_3\lambda_3}{\lambda_1},\frac{d_0}{\lambda_2},b_3\}\,, \nn \\
\{c_0, c_1, c_2, c_3\} &=& \{0,0,0,0\}\,,\quad \{d_0, d_1, d_2, d_3\}=\{d_0,-\frac{d_3\lambda_3}{\lambda_1},0,d_3\}\,.
\label{a1flux}
\eea
With this choice $N_{\rm flux}$ and $W$ become:
\be
N_{\rm flux} = \frac{2d_3^2\lambda_3 }{\lambda_1\lambda_2}\,,\qquad 
W = \left(d_3 U_4-\frac{d_3}{\lambda_2} U_2-b_3 \right)\left( U_3-\frac{\lambda_3}{\lambda_1}  U_1+ \frac{d_0}{d_3}\right).
\ee
The superpotential and its derivatives vanish when the moduli take the values: 
\be
(U_1,U_2,U_3,U_4) = \left(\frac{\lambda_1}{\lambda_3}\left(U_3+\frac{d_0}{d_3}\right),\lambda_2\left(U_4-\frac{b_3}{d_3}\right),U_3,U_4\right) .
\label{eq:solgenA1}
\ee
Note that the residual moduli space is $2$-dimensional and parametrised by $U_3$ and $U_4$. Let us present an explicit solution. For $\lambda_\alpha=1$, $\forall \alpha=1,2,3$ and $d_3=2$, $N_{\rm flux}=8$ and the fluxes in (\ref{a1flux}) become:
\bea
\{a_0, a_1, a_2, a_3\} &=& \{0, 2, 0, -2\}\,,\quad \{b_0, b_1, b_2, b_3\}=\{\frac{b_3 d_0}{2},-b_3,d_0,b_3\}\,, \nn \\
\{c_0, c_1, c_2, c_3\} &=& \{0,0,0,0\}\,,\quad \{d_0, d_1, d_2, d_3\}=\{d_0, -2, 0, 2\}\,.
\eea
Clearly $b_3=2p$ and $d_0=2q$ with $p,q\in\mathbb{Z}$ retain all fluxes even. With these choices we get a quadratic superpotential:
\be
W = -2 \left(U_2-U_4+p\right) \left(U_3-U_1+q\right)\,,
\label{WA1}
\ee
and the solution to $W=\partial_a W=0$ $\forall a=1,\dots,4$ is given by:
\be
(U_1,U_2,U_3,U_4) = \left(U_3 +q,U_4 -p,U_3,U_4\right).
\ee
All the solutions in this class, parametrised by a pair of integers $(p,q)$, are shown to be dual to $1$ physically distinct solution with representative $p=q=0$ in App. \ref{AppB}.

\vspace{0.5cm}
\noindent \underline{Subfamily $\mc{A}_2$}: In this case $d_3\neq0$ again but the fluxes take the form:
\bea
\{a_0, a_1, a_2, a_3\} &=& \{0,0,\frac{d_3}{\lambda_1},-\frac{d_3\lambda_3}{\lambda_1\lambda_2}\}\,,\quad 
\{b_0, b_1, b_2, b_3\}=\{\frac{b_3d_0}{d_3},\frac{d_0}{\lambda_1},-\frac{b_3\lambda_3}{\lambda_2},b_3\}\,, \nn \\
\{c_0, c_1, c_2, c_3\} &=& \{0,0,0,0\}\,,\quad \{d_0, d_1, d_2, d_3\}=\{d_0,0,-\frac{d_3\lambda_3}{\lambda_2},d_3\}\,.
\label{eq:fluxesA2}
\eea
With this choice $N_{\rm flux}$ and $W$ become:
\be
N_{\rm flux} = \frac{2d_3^2\lambda_3 }{\lambda_1\lambda_2}\,,\qquad
W = \left(d_3 U_4-\frac{d_3}{\lambda_1} U_1-b_3\right) \left( U_3 -\frac{\lambda_3}{\lambda_2}  U_2+\frac{d_0}{d_3}\right).
\ee
The superpotential and its derivatives vanish when the moduli take the values:
\be
(U_1,U_2,U_3,U_4) = \left(\lambda_1\left(U_4-\frac{b_3}{d_3}\right),\frac{\lambda_2}{\lambda_3}\left(U_3+\frac{d_0}{d_3}\right),U_3,U_4\right) .
\label{eq:solgenA2}
\ee
Note that the residual moduli space is $2$-dimensional and parametrised by $U_3$ and $U_4$. Let us present an explicit example. For $\lambda_\alpha=1$, $\forall \alpha=1,2,3$ and $d_3=4$, $N_{\rm flux}=32$ and the fluxes in (\ref{eq:fluxesA2}) become:
\bea
\{a_0, a_1, a_2, a_3\} &=& \{0, 0, 4, -4\}\,,\quad \{b_0, b_1, b_2, b_3\}=\{\frac{b_3 d_0}{4},d_0,-b_3,b_3\}\,, \nn \\
\{c_0, c_1, c_2, c_3\} &=& \{0,0,0,0\}\,,\quad \{d_0, d_1, d_2, d_3\}=\{d_0, 0, -4, 4\}\,.
\eea
Clearly, $b_3=2p$ and $d_0=4q$ with $p,q\in\mathbb{Z}$ retain all fluxes even. With these choices we get a quadratic superpotential:
\be
W = -4 \left( U_1- U_4+\frac{p}{2}\right) \left(U_3-U_2+q\right)\,,
\label{WA2}
\ee
and the solution to $W=\partial_a W=0$ $\forall a=1,\dots,4$ is given by:
\be
(U_1,U_2,U_3,U_4) = \left(U_4-\frac{p}{2},U_3 +q,U_3,U_4\right) .
\ee
All the solutions in this class, parametrised by a pair of integers $(p,q)$, are shown to be dual to $2$ physically distinct solutions with representatives $p=q=0$ and $p=1$, $q=0$ in App. \ref{AppB}.

\vspace{0.5cm}
\noindent \underline{Subfamily $\mc{A}_3$}: Here $d_2\neq0$ and the fluxes look like:
\bea
\{a_0, a_1, a_2, a_3\} &=& \{0,\frac{d_2}{\lambda_3},-\frac{d_2\lambda_2}{\lambda_1\lambda_3},0\}\,,\quad 
\{b_0, b_1, b_2, b_3\}=\{\frac{b_2d_0}{d_2},-\frac{b_2\lambda_2}{\lambda_1},b_2,\frac{d_0}{\lambda_3}\}\,, \nn \\
\{c_0, c_1, c_2, c_3\} &=& \{0,0,0,0\}\,,\quad \{d_0, d_1, d_2, d_3\}=\{d_0,-\frac{d_2\lambda_2}{\lambda_1},d_2,0\}\,,
\label{eq:fluxesA3}
\eea
With this choice $N_{\rm flux}$ and the superpotential become:
\be
N_{\rm flux} = \frac{2d_2^2\lambda_2}{\lambda_1\lambda_3}\,,\qquad 
W = \left( d_2 U_4 -\frac{d_2}{\lambda_3} U_3-b_2 \right) \left( U_2-\frac{\lambda_2}{\lambda_1}   U_1+\frac{d_0}{d_2}\right).
\ee
$W$ and its derivatives vanish if the moduli take the values:
\be
(U_1,U_2,U_3,U_4) = \left(\frac{\lambda_1}{\lambda_2}\left(U_2+\frac{d_0}{d_2}\right),U_2,\lambda_3\left(U_4-\frac{b_2}{d_2}\right),U_4\right) .
\label{eq:solgenA3}
\ee
Note that the residual moduli space is $2$-dimensional and parametrised by $U_2$ and $U_4$. Let us present an explicit example. For $\lambda_1=\lambda_3= 1$ and $\lambda_2= d_2=2$, $N_{\rm flux} = 16$ and the fluxes in (\ref{eq:fluxesA3}) become:
\bea
\{a_0, a_1, a_2, a_3\} &=& \{0, 2, -4, 0\}\,,\quad \{b_0, b_1, b_2, b_3\}=\{\frac12 b_2 d_0,-2 b_2,b_2,d_0\}\,,\nn \\
\{c_0, c_1, c_2, c_3\} &=& \{0,0,0,0\}\,,\quad \{d_0, d_1, d_2, d_3\}=\{d_0, -4, 2, 0\}\,.
\eea
Clearly $b_2=2p$ and $d_0=2q$ with $p,q\in\mathbb{Z}$ retain all fluxes even. With these choices we get a quadratic superpotential:
\be
W = -2 \left(U_3-U_4+p\right) \left(U_2-2 U_1+q\right),
\label{WA3}
\ee
and the solution to $W=\partial_a W=0$ $\forall a=1,\dots,4$ is given by:
\be
(U_1,U_2,U_3,U_4) = \left(\frac{U_2+q}{2},U_2, U_4-p,U_4\right) .
\ee
All the solutions in this class, parametrised by a pair of integers $(p,q)$, are shown to be dual to $2$ physically distinct solutions with representatives $p=q=0$ and $p=0$, $q=1$ in App. \ref{AppB}.

\vspace{0.5cm}
\noindent \underline{\bf{Family $\mc{B}$}}: For this family the linear relation involves derivatives of $W$ with respect to the dilaton and $1$ complex structure modulus and reads:
\be
\lambda_3\, \partial_3 W + \partial_4 W = 0\,,\qquad \lambda_3\neq0\,.
\label{eq:linrelB}
\ee
Similar solutions exist for linear relations of the form $\lambda_\alpha \partial_\alpha W + \partial_4 W = 0$ with $\lambda_\alpha \neq 0$ for $\alpha=1,2$, and so we do not list them separately. The allowed flux quanta fall into $2$ subfamilies which we call $\mc{B}_1$ and $\mc{B}_2$.

\vspace{0.5 cm}
\noindent \underline{Subfamily $\mc{B}_1$}: Here $d_2\neq0$, $c_3d_0\neq d_1d_2$ and the fluxes take the form:
\bea
\{a_0, a_1, a_2, a_3\} &=& \{-\frac{c_3}{\lambda_3},\frac{d_2}{\lambda_3},\frac{d_1}{\lambda_3},\frac{b_2c_3}{d_2}\}\,,\quad 
\{b_0, b_1, b_2, b_3\}=\{\frac{b_2d_0}{d_2},\frac{b_2d_1}{d_2},b_2,\frac{d_0}{\lambda_3}\}\,, \nn \\
\{c_0, c_1, c_2, c_3\} &=& \{0,0,0,c_3\}\,,\qquad \{d_0, d_1, d_2, d_3\}=\{d_0,d_1,d_2,0\}\,.
\label{eq:fluxesB1}
\eea
With this choice $N_{\rm flux}$ and the superpotential become:
\be
N_{\rm flux} = \frac{2}{\lambda_3}\left(c_3d_0-d_1d_2\right),\qquad 
W=\left(U_4 -\frac{U_3}{\lambda_3} -\frac{b_2}{d_2}\right) \left(U_2 (c_3 U_1+d_2)+d_1 U_1 +d_0\right).
\ee
$W$ and its derivatives vanish at:
\be
(U_1,U_2,U_3,U_4) = \left(-\frac{d_2U_2+d_0}{c_3 U_2+d_1},U_2,\lambda_3\left(U_4-\frac{b_2}{d_2}\right),U_4\right) .
\label{eq:solgenB1}
\ee
Note that the residual moduli space is $2$-dimensional and parametrised by $U_2$ and $U_4$. Let us present an explicit solution. For $\lambda_3=1$, $c_3=6$, $d_0=d_2=2$ and $d_1=0$, $N_{\rm flux} = 24$ and the fluxes in (\ref{eq:fluxesB1}) become:
\bea
\{a_0, a_1, a_2, a_3\} &=& \{-6, 2, 0, 3 b_2\}\,,\quad \{b_0, b_1, b_2, b_3\}=\{b_2, 0, b_2, 2\}\,, \nn \\
\{c_0, c_1, c_2, c_3\} &=& \{0, 0, 0, 6\}\,,\quad \{d_0, d_1, d_2, d_3\}=\{2, 0, 2, 0\}\,.
\eea
Clearly $b_2=2p$ with $p\in\mathbb{Z}$ retains all fluxes even. With these choices we get a cubic superpotential:
\be
W = -2 \left(3 U_1 U_2+U_2+1\right) \left(U_3-U_4+p\right),
\ee
and the solution to $W=\partial_a W=0$ $\forall a=1,\dots,4$ is given by:
\be
(U_1,U_2,U_3,U_4) = \left(-\frac{U_2+1}{3 U_2},U_2,U_4-p,U_4\right) .
\ee
All the solutions in this class, parametrised by an integer $p$, are shown to be dual to $1$ physically distinct solution with representative $p=0$ in App. \ref{AppB}.

\vspace{0.5cm}
\noindent \underline{Subfamily $\mc{B}_2$}: In this case $c_3\neq0$, $c_3d_0\neq d_1d_2$ and the fluxes read:
\bea
\{a_0, a_1, a_2, a_3\} &=& \{-\frac{c_3}{\lambda_3},\frac{d_2}{\lambda_3},\frac{d_1}{\lambda_3},a_3\}\,,\quad 
\{b_0, b_1, b_2, b_3\}=\{\frac{a_3d_0}{c_3},\frac{a_3d_1}{c_3},\frac{a_3d_2}{c_3},\frac{d_0}{\lambda_3}\}\,, \nn \\
\{c_0, c_1, c_2, c_3\} &=& \{0,0,0,c_3\}\,,\quad \{d_0, d_1, d_2, d_3\}=\{d_0,d_1,d_2,0\}\,.
\label{eq:fluxesB2}
\eea
This choice induce a flux contribution to the D3 tadpole and a superpotential of the form:
\be
N_{\rm flux} = \frac{2}{\lambda_3}\left(c_3d_0-d_1d_2\right),\qquad
W = \left(U_4 - \frac{U_3}{\lambda_3}-\frac{a_3}{c_3}\right) \left(U_2 (c_3 U_1+d_2)+ d_1 U_1+d_0\right).
\ee
The superpotential and its derivatives vanish if the moduli take the values:
\be
(U_1,U_2,U_3,U_4) = \left(-\frac{d_2U_2+d_0}{c_3U_2+d_1},U_2,\lambda_3\left(U_4-\frac{a_3}{c_3}\right),U_4\right).
\label{eq:solgenB2}
\ee
Note that the residual moduli space is $2$-dimensional and parametrised by $U_2$ and $U_4$. Let us present an explicit example. For $\lambda_3=1$, $c_3=d_1 = d_2 = 2$ and $d_0=4$, $N_{\rm flux}=8$ and the fluxes in (\ref{eq:fluxesB2}) take the form:
\bea
\{a_0, a_1, a_2, a_3\} &=& \{-2, 2, 2, a_3\}\,,\quad \{b_0, b_1, b_2, b_3\}=\{2 a_3, a_3, a_3, 4\}\,, \nn \\
\{c_0, c_1, c_2, c_3\} &=& \{0, 0, 0, 2\}\,,\quad \{d_0, d_1, d_2, d_3\}=\{4, 2, 2, 0\}\,.
\eea
Clearly $a_3=2p$ with $p\in\mathbb{Z}$ retains all fluxes even. With these choices the superpotential is cubic:
\be
W = -2 \left(U_1 U_2+U_1+U_2+2\right) \left(U_3-U_4 +p\right),
\ee
and the solution to $W=\partial_a W=0$ $\forall a=1,\dots,4$ is given by:
\be
(U_1,U_2,U_3,U_4) = \left(-\frac{U_2+2}{U_2+1},U_2,U_4-p,U_4\right).
\ee
All the solutions in this class, parametrised by an integer $p$, are shown to be dual to $1$ physically distinct solution with representative $p=0$ in App. \ref{AppB}.

\vspace{0.5cm}
\noindent \underline{\bf{Family $\mc{C}$}}: For this family the linear relation involves the derivatives of $W$ with respect to $2$ complex structure moduli and takes the form:
\be
\partial_1 W = \lambda_2\partial_2W\,,\qquad \lambda_2\neq0\,.
\label{crel}
\ee
Similar solutions exist with linear relations of the form $\partial_\alpha W= \lambda_3\,\partial_3 W$ with $\alpha=1,2$ and $\lambda_3\neq0$, and so we do not list them separately. With a relation of the form (\ref{crel}) the allowed flux quanta fall into $3$ subfamilies. We will refer to them as $\mc{C}_1$, $\mc{C}_2$ and $\mc{C}_3$.

\vspace{0.5cm}
\noindent  \underline{Subfamily $\mc{C}_1$}: Here $b_2c_2\neq a_2d_2$ and the flux quanta look like:
\bea
\{a_0, a_1, a_2, a_3\} &=& \{0,\frac{a_2}{\lambda_2},a_2,0\}\,,\quad \{b_0, b_1, b_2, b_3\}=\{0,b_2\lambda_2,b_2,0\}\,, \nn \\
\{c_0, c_1, c_2, c_3\} &=& \{0,\frac{c_2}{\lambda_2},c_2,0\}\,,\quad \{d_0, d_1, d_2, d_3\}=\{0,d_2\lambda_2,d_2,0\}\,.
\label{eq:fluxesC1}
\eea
With this choice we have:
\be
N_{\rm flux} = 2\left(b_2c_2-a_2d_2\right),\qquad 
W = \left( U_1+\frac{U_2}{\lambda_2}\right) \left(U_3 (c_2 U_4 -a_2)+\lambda_2  d_2 U_4-b_2 \lambda_2 \right).
\ee
The superpotential and its derivatives vanish at:
\be
(U_1,U_2,U_3,U_4) = \left(-\frac{U_2}{\lambda_2},U_2,-\lambda_2\frac{d_2U_4-b_2}{c_2U_4-a_2},U_4\right).
\label{eq:solgenC1}
\ee
Note that the residual moduli space is $2$-dimensional and parametrised by $U_2$ and $U_4$. Let us present an explicit example. For $a_2=d_2=0$, $b_2=4$ and $c_2=2$, $N_{\rm flux}=16$ and the fluxes in (\ref{eq:fluxesC1}) become:
\bea
\{a_0, a_1, a_2, a_3\} &=& \{0, 0, 0, 0\}\,,\quad \{b_0, b_1, b_2, b_3\}=\{0,4 \lambda_2 ,4,0\}\,, \nn \\
\{c_0, c_1, c_2, c_3\} &=& \{0,\frac{2}{\lambda_2 },2,0\}\,,\quad \{d_0, d_1, d_2, d_3\}=\{0, 0, 0, 0\}\,.
\eea
Clearly $\lambda_2=\pm1,\pm\frac12$ retain all fluxes even. With these choices we get a cubic superpotential:
\be
W = 2 \left(U_1+\frac{U_2}{\lambda_2}\right) \left(U_3 U_4-2 \lambda_2 \right),
\ee
and the solution to $W=\partial_a W=0$ $\forall a=1,\dots,4$ is given by:
\be
(U_1,U_2,U_3,U_4) = \left(-\frac{U_2}{\lambda_2},U_2,\frac{2 \lambda_2}{U_4},U_4\right).
\ee

\vspace{0.5cm}
\noindent \underline{Subfamily $\mc{C}_2$}: In this case $b_2,c_2,d_3\neq0$ and the fluxes look like:
\bea
\{a_0, a_1, a_2, a_3\} &=& \{0,\frac{b_3c_2}{d_3\lambda_2},\frac{b_3c_2}{d_3},0\}\,,\quad 
\{b_0, b_1, b_2, b_3\}=\{\frac{b_2d_3\lambda_2}{c_2},b_2\lambda_2,b_2,b_3\}\,, \nn \\
\{c_0, c_1, c_2, c_3\} &=& \{0,\frac{c_2}{\lambda_2},c_2,0\}\,,\quad \{d_0, d_1, d_2, d_3\}=\{0,0,0,d_3\}\,.
\label{eq:fluxesC2}
\eea
This choices induces:
\be
N_{\rm flux} = 2b_2c_2\,,\qquad 
W= \left(\left(U_1+\frac{U_2}{\lambda_2} \right)+  \frac{d_3}{c_2}\right) \left(c_2 U_3 \left(U_4-\frac{b_3}{d_3}\right) -b_2\lambda_2 \right).
\ee
The superpotential and its derivatives vanish if the moduli take the values:
\be
(U_1,U_2,U_3,U_4)=\left(-\frac{U_2}{\lambda_2}-\frac{d_3}{c_2},U_2,\frac{b_2d_3\lambda_2}{c_2d_3U_4-b_3 c_2},U_4\right).
\label{eq:solgenC2}
\ee
Note that the residual moduli space is $2$-dimensional and parametrised by $U_2$ and $U_4$. Let us present an explicit solution. For $b_2=c_2=4$, $N_{\rm flux}=32$ and the fluxes in (\ref{eq:fluxesC2}) take the form:
\bea
\{a_0, a_1, a_2, a_3\} &=& \{0,\frac{4 b_3}{\lambda_2  d_3},\frac{4 b_3}{d_3},0\}\,,\quad \{b_0, b_1, b_2, b_3\}=\{\lambda_2  d_3,4 \lambda_2 ,4,b_3\}\,, \nn \\
\{c_0, c_1, c_2, c_3\} &=& \{0,\frac{4}{\lambda_2},4,0\}\,,\quad \{d_0, d_1, d_2, d_3\}=\{0, 0, 0, d_3\}\,.
\eea
Clearly $\lambda_2= 1$, $d_3=2p$, $b_3=qd_3$ with $p,q\in\mathbb{Z}$ retain all fluxes even. With these choices we get a cubic superpotential:
\be
W = 4 \left(U_3 U_4-q U_3-1\right)\left(U_1+U_2+\frac{p}{2}\right),
\ee
and the solution to $W=\partial_a W=0$ $\forall a=1,\dots,4$ is given by:
\be
(U_1,U_2,U_3,U_4) = \left(-U_2-\frac{p}{2},U_2,\frac{1}{U_4-q},U_4\right).
\ee
All the solutions in this class, parametrised by a pair of integers $(p,q)$, are shown to be dual to $2$ physically distinct solutions with representatives $p=q=0$ and $p=1$, $q=0$ in App. \ref{AppB}.

\vspace{0.5cm}
\noindent \underline{Subfamily $\mc{C}_3$}: In this case $d_0,d_2\neq0$, $b_2d_3\neq b_3d_2$ and the fluxes take the form:
\bea
\{a_0, a_1, a_2, a_3\} &=& \{0,\frac{b_3d_2}{d_0},\frac{b_3d_2\lambda_2}{d_0},0\}\,,\quad 
\{b_0, b_1, b_2, b_3\}=\{\frac{b_2d_0}{d_2},b_2\lambda_2,b_2,b_3\}\,, \nn \\
\{c_0, c_1, c_2, c_3\} &=& \{0,\frac{d_2d_3}{d_0},\frac{d_2d_3\lambda_2}{d_0},0\}\,,\quad \{d_0, d_1, d_2, d_3\}=\{d_0,d_2\lambda_2,d_2,d_3\}\,.
\label{eq:fluxesC3}
\eea
The expressions for $N_{\rm flux}$ and $W$ become:
\be
N_{\rm flux}=\frac{2 d_2\lambda_2 }{d_0}\left(b_2 d_3-b_3 d_2\right),\quad 
W = \left(d_2 \left(\lambda_2  U_1+U_2\right)+d_0\right) \left(U_4 \left(\frac{d_3}{d_0}U_3+1\right)-\frac{b_3}{d_0} U_3-b_2 d_0\right).
\ee
The superpotential and its derivatives vanish at:
\be
(U_1,U_2,U_3,U_4) = \left(-\frac{1}{\lambda_2}\left(U_2+\frac{d_0}{d_2}\right),U_2,-\frac{d_0\left(d_2U_4-b_2\right)}{d_2\left(d_3U_4-b_3\right)},U_4\right) .
\label{eq:solgenC3}
\ee
Note that the residual moduli space is $2$-dimensional and parametrised by $U_2$ and $U_4$. Let us present an explicit example. For $\lambda_2=1$, $b_2 = 0$, $b_3 = -4 p$, $d_0 = 4 p$, $d_2 = -2$ and $d_3 = 4 p$ with $p\in\mathbb{Z}$, $N_{\rm flux}=8$ and the fluxes in (\ref{eq:fluxesC3}) become:
\bea
\{a_0, a_1, a_2, a_3\} &=& \{0, 2, 2, 0\}\,,\quad \{b_0, b_1, b_2, b_3\}=\{0, 0, 0, -4 p\}\,, \nn \\
\{c_0, c_1, c_2, c_3\} &=& \{0, -2, -2, 0\}\,,\quad \{d_0, d_1, d_2, d_3\}=\{4 p, -2, -2, 4 p\}\,.
\eea
Clearly all fluxes are even. With these choices we get a cubic superpotential:
\be
W = -2 \left(U_3 U_4+U_3+U_4\right) \left(U_1+U_2-2p\right),
\ee
and the solution to $W=\partial_a W=0$ $\forall a=1,\dots,4$ is given by:
\be
(U_1,U_2,U_3,U_4) = \left(2 p - U_2,U_2,\frac{1}{U_4+1}-1,U_4\right) .
\ee
All the solutions in this class, parametrised by an integer $p$, are shown to be dual to $1$ physically distinct solution with representative $p=0$ in App. \ref{AppB}.

\subsubsection*{Dualities among solutions}

The duality relations among the different classes of solutions presented above are analysed in detail in App. \ref{AppB} (for the case where $\lambda_\alpha \in \mathbb{Z}$). Here we just summarise the main results. The solutions of all $3$ subfamilies in family $\mc{A}$ are dual to each other, and subfamily $\mc{A}_1$ features inequivalent solutions. Similarly, all the solutions in $\mc{B}_1$ are dual to solutions in $\mc{B}_2$, and $\mc{B}_1$ has physically different solutions. On the other hand, even if $\mc{C}_2$ and $\mc{C}_3$ are dual to each other, $\mc{C}_1$ is dual only to a subset of $\mc{C}_2$.\footnote{Precisely, $\mc{C}_3$ contains 2 copies of $\mc{C}_2$.} App. \ref{AppB} discusses the classification of inequivalent solutions within $\mc{C}_2$, together with an explicit example of a solution which is in $\mc{C}_2$ but not in $\mc{C}_1$.

An interesting fact is the presence of inter-family dualities despite distinct linear functional relations for the derivatives of the superpotential across the families $\mc{A},\mc{B},\mc{C}$. In App. \ref{AppB} we have found that $\mc{C}_2$ is dual to $\mc{B}_1$ and $\mc{A}_3$ is dual to a subset of $\mc{B}_1$. Hence, $\mc{B}_1$ is the subject of focus, for which physically distinct solutions have been classified in great detail in App. \ref{AppB}.

Notice finally that in family $\mc{A}$ each subfamily gives a quadratic superpotential. Setting $p=q=0$ in \eqref{WA1}, \eqref{WA2} or \eqref{WA3} yields the superpotential discussed in \cite{Hebecker:2017lxm} which can also be reproduced for suitable choices of fluxes in the cases $\mc{B}_1$, $\mc{C}_1$ and $\mc{C}_3$. On the other hand, in the cases $\mc{B}_2$ and $\mc{C}_2$ the superpotential is always cubic. In light of aforesaid dualities, a cubic superpotential can be mapped to a quadratic one in certain cases. However, we can find cases where cubic $W$ can be made quadratic but not a homogeneous function of degree $2$, e.g. setting $\lambda_3 = 1,b_2 = 2,c_3 = 4,d_0 = 4,d_1 = 0,d_2 = 4$ in \eqref{eq:fluxesB1}.

Let us close this section commenting on some general features of the superpotential that we observe in these cases. $W$ is always a product of $2$ factors, each of which depends on $2$ variables among $U_1,\dots,U_4$. They also do not depend on the same $U_a$, and one of them is linear while the other is at most quadratic. Hence $W$ can be written as:
\be
W(U_1,\dots,U_4) = f(U_{p(1)},U_{p(2)})\,g(U_{p(3)},U_{p(4)})\,,
\ee
where $(p(1),\dots,p(4))$ is a permutation of $(1,\dots,4)$, $f$ is linear, $g$ is at most quadratic and the quadratic term in $g$ (if any) is only the cross-term $U_{p(3)}U_{p(4)}$. Clearly:
\be
\partial_{U_{p(1)}}W\propto g\,,\qquad \partial_{U_{p(2)}}W\propto g\,,\qquad
\partial_{U_{p(3)}}W = f\partial_{U_{p(3)}}g\,,\qquad \partial_{U_{p(4)}}W= f\partial_{U_{p(4)}}g\,. 
\ee
Moreover, for $a=3,4$, $\partial_{U_{p(a)}}g$ is of the form $\mu_a U_{p(b\neq a)}+\nu_a$ for some real coefficients $\mu_a$ and $\nu_a$ at least one of which is non-zero. Hence $\partial_{U_{p(a)}}W=0$ with $a=3,4$ sets $f=0$ for complex solutions $\hat{U}_a$. Due to this, solving $f=g=0$ automatically sets
$W = \partial_{U_{p(a)}}W=0$, $\forall a=1,\dots,4$, reducing the number of linearly independent equation to $2$. This shows clearly the existence of $2$ flat direction since the number of moduli is $4$. Notice also that in general neither $f$ nor $g$ (hereby $W$) has any scaling property. However, each class of fluxes presented above includes examples where at least one of the $f$ and $g$, or both, can be made homogeneous in their arguments by suitably setting some fluxes to zero. For example, for a given class, $f(U_{p(1)},U_{p(2)})= \mu_1 U_{p(1)} + \mu_2 U_{p(2)} + \nu$ can be made homogeneous in $U_{p(1)}$, $U_{p(2)}$ by setting $\nu=0$ whenever allowed.

\subsubsection*{Primitivity}

The solutions in the previous sections correspond to situations where the F-terms of the axio-dilaton and the complex structure moduli vanish and $W=0$. Additionally, to be supersymmetric solutions of the $10$-dimensional equations of motion, $G_3$ needs to be primitive. In this section we present a suitable K\"ahler form for all cases so that $G_3$ is primitive. The analysis is along the lines of \cite{Kachru:2002he}.

The fluxes considered are diagonal, and so their expansion in the basis elements defined in \pref{threebasis} is of the form:
\bea
F_3 &=& a_0\alpha_0+a_1\alpha_{11}+a_2\alpha_{22}+a_3\alpha_{33}+b_1\beta^{11}+b_2\beta^{22}+b_3\beta^{33}+b_0\beta^0\,, \nn \\
H_3 &=& c_0\alpha_0+c_1\alpha_{11}+c_2\alpha_{22}+c_3\alpha_{33}+d_1\beta^{11}+d_2\beta^{22}+d_3\beta^{33}+d_0\beta^0\,.
\eea
The period matrix is also diagonal for all the solutions obtained. Thus $\dd z^j=\dd x^j+\tau_j \dd y^j,\ d\bar{z}^j=\dd x^j+\bar{\tau}_j \dd y^j,\ j=1,2,3$. Now, taking the K\"ahler form to be:
\be
J=\sum_{j=1}^3r_j^2\ \dd z^j\wedge d\bar{z}^j=-2{\rm i}\sum_{j=1}^3\text{Im}\left(\tau_j\right)r_j^2\ \dd x^j\wedge \dd y^j\,,
\label{eq:genJ}
\ee
it is easy to see that $J\wedge G_3=0$, i.e. $G_3$ is primitive.

\subsection{Solutions with $N=2$ supersymmetry and flat directions}
\label{n2}

Solutions with extended supersymmetry in $4$ dimensions have been useful laboratories for developing our understanding of string theory. Some of our solutions with $2$ flat directions preserve $N=2$ supersymmetry in $4$ dimensions. Being warped flux Minkowski compactifications with extended supersymmetry where the string coupling can be tuned to arbitrarily small values, they should be of interest for various theoretical studies.
   
The number of supersymmetries that a solution preserves can be determined by examining the decomposition of the $G_3$ flux under $SU(2)_L \times SU(2)_R \times U(1) \subset SO(6)$ (where $SO(6)$ is the group of rotations of the internal torus) \cite{Kachru:2002he}. In the charge convention of \cite{Kachru:2002he}, a general $3$-form decomposes as:
\be
[6 \times 6 \times 6]_A \to (2,2)_0 + (2,2)_0 + (3,0)_2 + (3,0)_{-2} + (0,3)_2 + (0,3)_{-2}\,.
\ee
The requirement of extended supersymmetry is that $G_3$ must take values so that only the $(0,3)_2$ component is present. This implies that when $G_3$ is written as:
\be
G_3 = \omega \wedge \dd z^\alpha\,,
\label{selfd}
\ee
where $z^\alpha$ is the `complex direction' with $U(1)$ charge $2$, then $\omega$ has to be self dual in the remaining $4$ (real) directions, with the orientation choice for Hodge duality which is consistent with (\ref{choice}). We present $2$ explicit solutions which preserve $N =2$ supersymmetry. In all our computations we will consider a metric of the form:
\be
g_{i\bar{\jmath}}  = r_i^2 \,\delta_{i\bar{\jmath}}\,,
\ee
that will ensure primitivity of the solutions.

\vspace{0.5 cm}
\noindent \underline{Example 1}: This solution lies in subfamily $\mc{A}_1$ of Sec. \ref{twoflat}. Choosing $\lambda_1 = \lambda_2 = \lambda_3 = 1$, $b_3=d_0=0$ and $d_3=2$ in the expressions for the flux quanta in \eqref{a1flux} we obtain:
\bea
\{a_0,a_1,a_2,a_3 \} &=& \{0,2,0,-2 \}\,,\qquad \{b_0,b_1,b_2,b_3 \} = \{ 0,0,0,0 \}\,, \nn \\
\{ c_0,c_1,c_2,c_3 \} &=& \{ 0,0,0,0 \}\,,\qquad \{ d_0,d_1,d_2,d_3 \} = \{ 0,-2,0,2\}\,,
\eea
together with $N_{\rm flux}=8$, and so the tadpole bound is satisfied. The superpotential is given by
\be
W=2 (U_1 - U_3) (U_2 - U_4)\,.
\ee
The residual moduli space can be parametrised as:
\be
(U_1,U_2,U_3,U_4) = (U_3,U_4,U_3,U_4)\,,
\ee
where we take $U_3,U_4$ to be in the fundamental domain of the upper half plane modulo modular transformations. Thus, all $U_a$ have positive imaginary parts. The $3$-form fluxes are:
\bea
F_3 &=& -2 \dd x^1 \wedge \dd x^2 \wedge \dd y^3 + 2 \dd x^2 \wedge \dd x^3 \wedge \dd y^1\,, \nn \\
H_3 &=& 2 \dd x^1 \wedge \dd y^2 \wedge \dd y^3 - 2 \dd x^3 \wedge \dd y^1 \wedge \dd y^2\,,
\eea
leading to the complexified $3$-form:
\be
G_3 = -\frac{2}{U_3-\conj{U}_3} \,(\dd z^1 \wedge \dd \conj{z}^3 + \dd z^3 \wedge \dd \conj{z}^1) \wedge \dd z^2 \equiv \omega\wedge \dd z^2\,.
\ee
Identifying $z^2$ as the $U(1)$ coordinate, we see that $G_3$ has hypercharge $+2$. Furthermore, computing the $SO(4) \supset SU(2)_L\times SU(2)_R$ dual we get:
\be
\star_4 \omega = \omega\,.
\ee
Thus the solution preserves $N=2$ supersymmetry. Notice that this corresponds to the case studied in \cite{Hebecker:2017lxm}. 

\vspace{0.5 cm}

\noindent \underline{Example 2}: This solution lies in subfamily $\mc{B}_1$ of Sec. \ref{twoflat}. Choosing $\lambda_3 = 1$, $b_2 = 2$, $d_1=0$ and $ c_3 = d_0 = d_2 =4$ in the expressions for the flux quanta in \eqref{eq:fluxesB1} we obtain:
\bea
\{a_0,a_1,a_2,a_3\} &=& \{-4,4,0,2\}\,,\qquad \{b_0,b_1,b_2,b_3\} = \{2,0,2,4\}\,, \nn \\
\{c_0,c_1,c_2,c_3\} &=& \{0,0,0,4\}\,,\qquad \{d_0,d_1,d_2,d_3\} = \{4,0,4,0\}\,,
\eea
together with $N_{\rm flux} = 8$. The superpotential is given by
\be
W = -2(U_1 U_2+U_2+1)(2 U_3 + U_4+1)\,. 
\ee
This is an example where using dualities the superpotential cannot be brought to a degree-2 homogeneous polynomial. The residual moduli space can be parametrised as:
\be
(U_1,U_2,U_3,U_4) = \left(-\frac{U_2+1}{U_2},U_2,U_4-\frac12,U_4\right),
\ee
where we take $U_2$ and $U_4$ to be in the fundamental domain of the upper half plane modulo modular transformations. Thus, all $U_a$ have positive imaginary parts. The $3$-form fluxes are:
\bea
F_3 &=& -4 \dd x^1 \wedge \dd x^2 \wedge \dd x^3 + 2 \dd x^1 \wedge \dd x^2 \wedge \dd y^3+ 4 \dd x^2 \wedge \dd x^3 \wedge \dd y^1+ 2 \dd x^2 \wedge \dd y^1 \wedge \dd y^3 \nn\\
&&- 4 \dd x^3 \wedge \dd y^1 \wedge \dd y^2+ 2 \dd y^1 \wedge \dd y^2 \wedge \dd y^3\,, \nn \\
H_3 &=& 4 \dd x^1 \wedge \dd x^2 \wedge \dd y^3 + 4 \dd x^2 \wedge \dd y^1 \wedge \dd y^3+ 4 \dd y^1 \wedge \dd y^2 \wedge \dd y^3\,, 
\eea
leading to the complexified $3$-form:
\be
G_3 = -\frac{4}{U_2-\conj{U}_2} \,(U_2\,\dd z^1 \wedge \dd \conj{z}^2 + \conj{U}_2\, \dd z^2 \wedge \dd \conj{z}^1) \wedge \dd z^3 \equiv \omega\wedge \dd z^3\,.
\ee
Identifying $z^3$ as the $U(1)$ coordinate, we see that $G_3$ has hypercharge $+2$. Furthermore, computing the $SO(4) \supset SU(2)_L\times SU(2)_R$ dual we get:
\be
\star_4 \omega=\omega\,.
\ee
Thus the solution preserves $N=2$ supersymmetry.

\section{Flat directions in Calabi-Yau orientifolds}
\label{sec:cy}

In this section we turn to CYs in the large complex structure limit. A detailed study will be carried out using the CY obtained by considering a degree-$18$ hypersurface in $\mathbb{CP}_{[1,1,1,6,9]}$ (first studied in the context of mirror symmetry in \cite{Candelas:1994hw}). Then, we also briefly discuss another CY with more moduli.

\subsection{Type IIB Calabi-Yau flux compactifications at large complex structure}
\label{cyreview}

In this section we first recapitulate some basic material on type IIB flux compactifications in the large complex structure limit\footnote{For detailed studies of flux vacua in the large complex structure limit see e.g. \cite{Dimofte:2008jg, Cicoli:2013cha, Blumenhagen:2014nba, Marsh:2015zoa, Kobayashi:2015aaa, Honma:2017uzn, Grimm:2019ixq, Blanco-Pillado:2020hbw}.} and the $\mathbb{CP}_{[1,1,1,6,9]}$ example.\footnote{We follow the notation and conventions of \cite{Demirtas:2019sip} but with $2$ exceptions: ($i$) in the definition of the GVW superpotential, the paper has an overall factor of $\sqrt{2/\pi}$ which we set equal to unity to be consistent with our earlier discussion. ($ii$) the paper uses $\tau$ to denote the axio-dilaton, while we will continue to use $\phi$.} Given that our discussion shall be quite brief, we refer the reader to \cite{Hosono:1994av, Klemm:2005tw, Giddings:2001yu, Demirtas:2019sip} for further details.

Type IIB flux compactifications have an internal manifold that is conformally an orientifolded CY $X$. To describe these in the language of special geometry, one works with a symplectic basis for $H_3(X,\mathbb{Z})$, $\{A_a, B^a \}$ for $a=0,...,h^{1,2}_-(X)$ with $A_a\cap A_b=0,$ $A_a\cap B^b=\delta^{~b}_a,$ and $B^a\cap B^b=0$, and projective coordinates on the complex structure moduli $U^a$ (in what follows, we will take $U^0 =1$). The central object is the prepotential $\mc{F}$, which is degree-$2$ and homogeneous in the projective coordinates. The period vector is given by:
\be
\Pi=\left( \begin{array}{c} \int_{B^a}\Omega\\ \int_{A_a}\Omega\end{array}\right)=\left(\begin{array}{c} \mc{F}_a\\ U^a \end{array}\right)\,,
\label{per}
\ee
where $\mc{F}_0 = 2\mc{F}-U^a\mc{F}_a$ with $\mc{F}_a\equiv \partial_{U^a}\mc{F}$. Similarly, (integer valued) flux vectors $F$ and $H$ are obtained by performing integrals of the $3$-form field strengths over the $A_a$ and $B_a$ cycles. The flux superpotential, which is classically exact, is given by:
\be
W = \left(F-\phi H\right)^t \cdot\Sigma \cdot\Pi \,,
\label{eq:wis}
\ee
where:
\be
\Sigma = 
 \begin{pmatrix}
 0 & 1 \cr
 -1 & 0
 \end{pmatrix}\,,
\label{SimplMatrix}
\ee
is the symplectic matrix. The tree-level K\"ahler potential (for the complex structure moduli and the axio-dilaton) is:
\be
\mc{K} = -\ln\left(-{\rm i}\Pi^\dagger \cdot\Sigma\cdot \Pi\right)-\ln\left(-{\rm i}(\phi-\bar{\phi})\right)\,.
\label{eq:wis2}
\ee
In the large complex structure limit, the prepotential is a sum of perturbative terms which are at most degree-$3$ and instanton corrections, i.e. $\mc{F}(U)= \mc{F}_{\rm pert}(U)+\mc{F}_{\rm inst}(U)$ with:
\be
\mc{F}_{\rm pert}(U)=-\frac{1}{3!}\, \mc{K}_{abc} U^a U^b U^c + \frac12\, {\bf{a}}_{ab} U^a U^b + b_a U^a +\xi\,,
\label{eq:LCSprepotential}
\ee
where $\mc{K}_{abc}$ are the triple intersection numbers of the mirror CY, ${\bf{a}}_{ab}$ and $b_a$ are rational, and $\xi=-\frac{\zeta(3)\chi}{2(2\pi i)^3}$, with $\chi$ the CY Euler number. The instanton corrections are:
\be
\mc{F}_{\rm inst}(U)=\frac{1}{(2\pi {\rm i})^3}\sum_{\vec{q}} A_{\vec{q}}\,e^{2\pi {\rm i} \vec{q}\cdot \vec{U}}\,,
\ee
where the sum runs over effective curves in the mirror CY. The form of the perturbative part of the prepotential implies that it leads to a superpotential that is at most degree-$3$ polynomial in the complex structure moduli and the fluxes. Thus the search for supersymmetric minima with flat directions can be carried out using  the method we have put forward in Sec. \ref{sec:qual}.

\subsection{Supersymmetric solutions with flat directions for $\mathbb{CP}_{[1,1,1,6,9]}[18]$}
\label{cy11169}

\subsubsection*{The $\mathbb{CP}_{[1,1,1,6,9]}[18]$ example}

In this section we implement our method to find supersymmetric minima with flat directions focusing on the example of the degree-$18$ hypersurface in $\mathbb{CP}_{[1,1,1,6,9]}$. Let us record some basic facts about this CY which has $272$ complex structure moduli and a $\mc{G}=\mathbb{Z}_6\times \mathbb{Z}_{18}$ symmetry. By considering fluxes which are $\mc{G}$-invariant, one stabilises on the $\mc{G}$-symmetric locus (see \cite{Giryavets:2003vd}). Thus the stabilisation problem can be effectively reduced to a $2$-moduli one. For this, the relevant geometric data are:
\be
\mc{K}_{111} = 9\,,\qquad \mc{K}_{112} = 3\,,\qquad \mc{K}_{122} = 1\,,\qquad {\bf{a}}= \frac12 
\begin{pmatrix}
9 & 3 \\
3 & 0
\end{pmatrix}\,,\qquad \vec{b}=\frac14
\begin{pmatrix}
17\\
6
\end{pmatrix}\,,
\label{eq:geodata}
\ee
and the instanton corrections are $(2\pi {\rm i})^3 \mc{F}_{\rm inst}=\mc{F}_1 + \mc{F}_2 + \cdots$ with:
\be
\mc{F}_1 =  - 540\, q_1 - 3\,q_2\,,\qquad
\mc{F}_2 = -\frac{1215}{2}\, q_1^2 +1080\,q_1 q_2 + \frac{45}{8}\,q_2^2\,,
\label{eq:theinst}
\ee
where $q_a=\exp(2\pi {\rm i} U^a)$ with $a=1,2$. We will consider the orientifold described in \cite{Louis:2012nb} with the D7 tadpole cancelled by $4$ D7-branes on top of each O7-plane. This setup yields a D3-charge $Q_\D3=138$. 

Neglecting exponentially small corrections in the prepotential, the F-flatness conditions are a set of $3$ polynomial equations in $3$ variables. We examine both cases, in which the superpotential vanishes or assumes a non-zero value at the minimum. As described earlier, our ansatz will involve looking for solutions where there is a linear relation between the derivatives of the superpotential.

Following the algorithm described in Sec. \ref{sec:qual}, we start by writing the flux vectors as:
\be
F = (f_1\ f_2\ f_3\ f_4\ f_5\ f_6)^t\,, \qquad H=(h_1\ h_2\ h_3\ h_4\ h_5\ h_6)^t\,, \qquad f_i, h_i \in \mathbb{Z}\,.
\ee
For simplicity in this paper we will take $f_4 = h_4 =0$. As a result, the contribution to the superpotential of the term involving the CY Euler number in the prepotential \pref{eq:LCSprepotential} vanishes and the superpotential is polynomial with rational coefficients. This simplifies the search for solutions. Now, defining $\mc{N}_{\rm flux}\equiv\frac12 N_{\rm flux}= -\frac12\,H^t\cdot\Sigma\cdot F$ and denoting $(U^1,U^2,\phi)$ by $(U_1,U_2,U_3)$, we have:
\bea
\mc{N}_{\rm flux} &=& \frac12 (f_2 h_5+f_3 h_6-f_5 h_2-f_6 h_3)\,, \nn \\
W &=& f_1+U_1 (f_2-h_2 U_3)+U_2 (f_3-h_3 U_3) \nn \\
  &+& \frac14 \left(2 (3U_1+U_2)^2 -18 U_1-6 U_2-17\right) (f_5-h_5 U_3) \nn \\
  &+& \frac12 \left(U_1 (3 U_1+2 U_2-3)-3\right) (f_6-h_6 U_3)-h_1 U_3\,.
\label{eq:CYmod2}
\eea

\subsubsection*{Solutions with $W=0$}

Given \eqref{eq:CYmod2}, consider the $4$ polynomial equations in $3$ variables: $W(U_a)=\partial_a W=0$ $\forall a=1,2,3$. The degree of each of these equations is $3$ or less, depending upon the choices of fluxes. For this system of equations to admit a solution, one of them should be dependent on the others. Here, we examine cases when this dependence is linear:\footnote{As the derivatives of the polynomial $W$ are of lower degree, $W$ can never be equal to a linear combination of its derivatives.}
\begin{itemize}
\item[($i$)] When $\partial_1 W=\lambda_2\, \partial_2 W+ \lambda_3\,\partial_3 W$ with at least one of $\lambda_2$ and $\lambda_3$ which is non-zero and subject to $\mc{N}_{\rm flux} \neq0$, we find only $1$ family of fluxes (details are given below) for which $W=\partial_a W=0$ $\forall a=1,2,3$ admit solutions in the large complex structure limit. Here, we do not need to impose any further conditions ensuring the existence of a flat direction since it turns out that we always have a flat direction (parametrised by the axio-dilaton) with the above family of fluxes;

\item[($ii$)] When $\partial_2 W=\lambda_3\,\partial_3 W$, $\lambda_3\neq 0$, the conditions on the fluxes have no solution in keeping with $\mc{N}_{\rm flux}\neq0$.
\end{itemize}

We now provide the aforesaid family of fluxes which are dependent on the $5$ parameters $\lambda_2$, $\lambda_3$, $f_3$, $h_1$ and $h_3$:
\bea
&&\{f_1,f_2,f_3,f_4,f_5,f_6\}= \nn \\
&&\left\{\frac{4 f_3 h_1-\frac{\lambda_3  \left((2 \lambda_2 -3) (2 h_1^2 + 6 h_1 h_3+ h_3^2)-6 \lambda_2 ^2 h_3^2\right)}{(\lambda_2 -3) \lambda_2 }}{4 h_3},\lambda_2  f_3-\lambda_3  h_1-\frac{3 \lambda_3  h_3}{2},f_3,0,\frac{(2 \lambda_2 -3) \lambda_3  h_3}{(\lambda_2 -3) \lambda_2 },\frac{(\lambda_2 -3) \lambda_3  h_3}{\lambda_2 }\right\}, \nn \\
&&\{h_1,h_2,h_3,h_4,h_5,h_6\}=\{h_1,\lambda_2  h_3,h_3,0,0,0\}\,,\qquad \lambda_2\neq0,3\,,\qquad \lambda_3,h_3\neq0\,.
\label{eq:fluxesCY}
\eea
In this case we have:
\bea
\mc{N}_{\rm flux} &=& -\frac{3 \left((\lambda_2 -3) \lambda_2 +3 \right) \lambda_3  h_3^2}{2 (\lambda_2 -3) \lambda_2 }\,, \nn \\
W &=& \frac{1}{2 (\lambda_2 -3) \lambda_2  h_3}\left(h_1+h_3 (\lambda_2  U_1+U_2)\right) \label{eq:WCY} \\
  &\times& \left(2 (\lambda_2 -3) \lambda_2  (f_3-h_3 U_3)+(3-2 \lambda_2 ) \lambda_3  h_1+\lambda_3  h_3 (3 \lambda_2  (U_1-2)+(2 \lambda_2 -3) U_2+9)\right). \nn
\eea
Now, solving $W = \partial_a W=0$, $\forall a=1,2,3$, we see that $U_1$ and $U_2$ depend linearly on $U_3$ with slopes $-1/\lambda_3$ and $\lambda_2/\lambda_3$ respectively. Thus, by requiring $\lambda_2,\lambda_3<0$, we may obtain ${\rm Im}\,U_a$, $\forall a=1,2,3$ to be of the same sign. This keeps $\mc{N}_{\rm flux}$ positive and also ensures that $U_1$ and $U_2$ are in the large complex structure limit when ${\rm Im}\,U_3$ is taken large to be in the weak string coupling regime.

Note the arguments of $f_5(\lambda_2,\lambda_3,h_3)$, $f_6(\lambda_2,\lambda_3,h_3)$, $h_2(\lambda_2,h_3)$ and $\mc{N}_{\rm flux}(\lambda_2,\lambda_3,h_3)$. There are only $488$ triples $(\lambda_2,\lambda_3,h_3)$, $\lambda_2,\lambda_3\in\mathbb{Q}^-$, $h_3\in\mathbb{Z}$, securing $f_5,f_6,h_2\in\mathbb{Z}$ and $\mc{N}_{\rm flux}\in\mathbb{Z}/2$ with $0<\mc{N}_{\rm flux}\leq138$. For $420$ of them there are no $f_3,h_1\in\mathbb{Z}$ that keep all other fluxes in \eqref{eq:fluxesCY} integers. For each of the remaining $68$ triples $(\lambda_2,\lambda_3,h_3)$, we get a subfamily of integer fluxes \eqref{eq:fluxesCY} parameterised by $f_3$ and $h_1$. All the members in any of the aforementioned subfamilies have the same $\mc{N}_{\rm flux}(\lambda_2,\lambda_3,h_3)$ which happens to be an integer. In Tab. \ref{tab:CYpart1} and \ref{tab:CYpart2} we list a representative from each of these $68$ subfamilies. Then, we also discuss one of these subfamilies in detail. Let us stress that among the above $68$ values of $\mc{N}_{\rm flux}(\lambda_2,\lambda_3,h_3)$ only $13$ are distinct.

In all $68$ cases in Tab. \ref{tab:CYpart1} and \ref{tab:CYpart2}, $W$ is a non-homogeneous function of degree $2$. We need to check if these cases are dual to cases where $W$ is homogeneous (as in \cite{Demirtas:2019sip, Broeckel:2021uty}). To do this, we can employ integer shifts of the complex structure moduli $U_1$ and $U_2$ and $SL(2,\mathbb{Z})$ transformations on the axio-dilaton $U_3$.\footnote{See App. \ref{AppA} for the transformation rules.} Note that $h_5,h_6\neq0$ in \eqref{eq:CYmod2} yield a cubic $W$. In all cases in Tab. \ref{tab:CYpart1} and \ref{tab:CYpart2}, $h_5=h_6=0$ and $f_5\neq0$. From \eqref{Adil} we see that an $SL(2,\mathbb{Z})$ duality transformation with non-zero $c$ and $f_5$ leads to a non-zero $h_5$, yielding a non-zero coefficient for $U_2^2 U_3$ in $W$. Thus, for the above check we must keep $c=0$, and only integer shifts of $U_a$ $\forall\, a=1,2,3$ are useful. We find that in $53$ of these $68$ cases, appropriate integer shifts of $U_a$ can transform $W$ into a homogeneous function of degree $2$. Interestingly, after including instanton corrections to the superpotential, it can be checked that only $2$ out of these $53$ solutions feature a weak string coupling and an instanton expansion which is definitely under control, corresponding to the $2$ old vacua already found in \cite{Broeckel:2021uty, Carta:2021kpk} ($1$ of them has been originally discovered in \cite{Demirtas:2019sip}). 

\begin{table}[H]
\begin{center}
\resizebox{\columnwidth}{!}{
  \begin{tabular}{|c|c|c|c|c|c|}
    \hline
    & $(\lambda_2,\lambda_3,h_3)$ & $F$ & $H$ & $\mc{N}_{\rm flux}$ & $W$ \\
    \hline
    1 & (-3,-72,1) & \{-9,-9,-9,0,36,-144\} & \{-2,-3,1,0,0,0\} & 126 & $-(3 U_1 - U_2+2) (18 U_1 + 18 U_2 - U_3-27)$ \\
    \hline
    2 & (-3,-72,-1) & \{-3,-9,15,0,-36,144\} & \{2,3,-1,0,0,0\} & 126 & $(3 U_1-U_2+2) (18 U_1+18 U_2-U_3-33)$ \\
    \hline
    3 & (-3,-8,3) & \{-11,-10,2,0,12,-48\} & \{-5,-9,3,0,0,0\} & 126 & $-(9 U_1-3 U_2+5) (2 (-1+U_1+U_2)-U_3)$ \\
    \hline
    4 & (-3,-8,-3) & \{-14,-13,11,0,-12,48\} & \{7,9,-3,0,0,0\} & 126 & $(9 U_1-3 U_2+7) (2 U_1+2 U_2-U_3-5)$ \\
    \hline
    5 & (-1,-48,1) & \{-33,72,0,0,60,-192\} & \{0,-1,1,0,0,0\} & 126 & $-(U_1-U_2) (18 U_1+30 U_2-U_3-90)$ \\
    \hline
    6 & (-1,-48,-1) & \{-13,-10,-14,0,-60,192\} & \{1,1,-1,0,0,0\} & 126 & $(U_1-U_2+1) (18 U_1+30 U_2-U_3-46)$ \\
    \hline
    7 & $(-1,-\frac{16}{3},3)$ & \{-3,-14,6,0,20,-64\} & \{-6,-3,3,0,0,0\} & 126 & $-(U_1-U_2+2) (6 U_1+10 U_2-3 U_3-4)$ \\
    \hline
    8 & $(-1,-\frac{16}{3},-3)$ & \{-10,-9,1,0,-20,64\} & \{3,3,-3,0,0,0\} & 126 & $(U_1-U_2+1) (6 U_1+10 U_2-3 (7+U_3))$ \\
    \hline
    9 & $(-3,-\frac89,9)$ & \{-9,-14,6,0,4,-16\} & \{-9,-27,9,0,0,0\} & 126 & $-(3 U_1-U_2+1) (2 (1+U_1+U_2)-9 U_3)$ \\
    \hline
    10 & $(-3,-\frac89,-9)$ & \{1,-10,2,0,-4,16\} & \{9,27,-9,0,0,0\} & 126 & $(3 U_1-U_2+1) (2 (-3+U_1+U_2)-9 U_3)$ \\
    \hline
    11 & $(-\frac34,-\frac52,4)$ & \{-14,-13,14,0,16,-50\} & \{-7,-3,4,0,0,0\} & 124 & $-(3 U_1-4 U_2+7) (U_1+2 U_2-U_3+1)$ \\
    \hline
    12 & $(-\frac34,-\frac52,-4)$ & \{-11,-12,6,0,-16,50\} & \{3,3,-4,0,0,0\} & 124 & $(3 U_1-4 U_2+3) (U_1+2 U_2-U_3-6)$ \\
    \hline
    13 & (-12,-80,1) & \{-99,120,0,0,12,-100\} & \{0,-12,1,0,0,0\} & 122 & $-(12 U_1-U_2) (-18+8 U_1+6 U_2-U_3)$ \\
    \hline
    14 & (-12,-80,-1) & \{99,-120,0,0,-12,100\} & \{0,12,-1,0,0,0\} & 122 & $(12 U_1-U_2) (-18+8 U_1+6 U_2-U_3)$ \\
    \hline
    15 & (-6,-72,1) & \{-77,108,0,0,20,-108\} & \{0,-6,1,0,0,0\} & 114 & $-(6 U_1-U_2) (-30+12 U_1+10 U_2-U_3)$ \\
    \hline
    16 & (-6,-72,-1) & \{77,-108,0,0,-20,108\} & \{0,6,-1,0,0,0\} & 114 & $(6 U_1-U_2) (-30+12 U_1+10 U_2-U_3)$ \\
    \hline
    17 & $(-\frac12,-7,2)$ & \{-3,-13,12,0,32,-98\} & \{-4,-1,2,0,0,0\} & 114 & $-(4+U_1-2 U_2) (-2+3 U_1+8 U_2-U_3)$ \\
    \hline
    18 & $(-\frac12,-7,-2)$ & \{-9,-14,0,0,-32,98\} & \{1,1,-2,0,0,0\} & 114 & $(1+U_1-2 U_2) (-20+3 U_1+8 U_2-U_3)$ \\
    \hline
    19 & (-3,-64,1) & \{-10,-11,-7,0,32,-128\} & \{-2,-3,1,0,0,0\} & 112 & $-(2+3 U_1-U_2) (-23+16 U_1+16 U_2-U_3)$ \\
    \hline
    20 & (-3,-64,-1) & \{-4,-10,14,0,-32,128\} & \{2,3,-1,0,0,0\} & 112 & $(2+3 U_1-U_2) (-30+16 U_1+16 U_2-U_3)$ \\
    \hline
    21 & (-3,-16,2) & \{-10,-13,-1,0,16,-64\} & \{-4,-6,2,0,0,0\} & 112 & $-(2+3 U_1-U_2) (-9+8 U_1+8 U_2-2 U_3)$ \\
    \hline
    22 & (-3,-16,-2) & \{-6,-11,9,0,-16,64\} & \{4,6,-2,0,0,0\} & 112 & $(2+3 U_1-U_2) (-17+8 U_1+8 U_2-2 U_3)$ \\
    \hline
    23 & (-3,-4,4) & \{-10,-14,2,0,8,-32\} & \{-8,-12,4,0,0,0\} & 112 & $-2 (2+3 U_1-U_2) (-1+2 U_1+2 U_2-2 U_3)$ \\
    \hline
    24 & (-3,-4,-4) & \{-6,-10,6,0,-8,32\} & \{8,12,-4,0,0,0\} & 112 & $2 (2+3 U_1-U_2) (-5+2 U_1+2 U_2-2 U_3)$ \\
    \hline
    25 & (-3,-1,8) & \{-9,-14,6,0,4,-16\} & \{-8,-24,8,0,0,0\} & 112 & $-2 (1+3 U_1-U_2) (1+U_1+U_2-4 U_3)$ \\
    \hline
    26 & (-3,-1,-8) & \{1,-10,2,0,-4,16\} & \{8,24,-8,0,0,0\} & 112 & $2 (1+3 U_1-U_2) (-3+U_1+U_2-4 U_3)$ \\
    \hline
    27 & (-3,-56,1) & \{-9,-10,-6,0,28,-112\} & \{-2,-3,1,0,0,0\} & 98 & $-(2+3 U_1-U_2) (-20+14 U_1+14 U_2-U_3)$ \\
    \hline
    28 & (-3,-56,-1) & \{-5,-11,13,0,-28,112\} & \{2,3,-1,0,0,0\} & 98 & $(2+3 U_1-U_2) (-27+14 U_1+14 U_2-U_3)$ \\
    \hline
    29 & $(-3,-\frac{8}{7},7)$ & \{-9,-13,3,0,4,-16\} & \{-14,-21,7,0,0,0\} & 98 & $-(2+3 U_1-U_2) (1+2 U_1+2 U_2-7 U_3)$ \\
    \hline
    30 & $(-3,-\frac{8}{7},-7)$ & \{-9,-14,6,0,-4,16\} & \{14,21,-7,0,0,0\} & 98 & $(2+3 U_1-U_2) (2 (-4+U_1+U_2)-7 U_3)$ \\
    \hline
    31 & (-3,-48,1) & \{-10,-12,-4,0,24,-96\} & \{-2,-3,1,0,0,0\} & 84 & $-(2+3 U_1-U_2) (-16+12 U_1+12 U_2-U_3)$ \\
    \hline
    32 & (-3,-48,-1) & \{-4,-9,11,0,-24,96\} & \{2,3,-1,0,0,0\} & 84 & $(3 U_1-U_2+2) (-23+12 U_1+12 U_2-U_3)$ \\
    \hline
    33 & (-3,-12,2) & \{-12,-9,3,0,12,-48\} & \{-3,-6,2,0,0,0\} & 84 & $-(6 U_1-2 U_2+3) (3 (-1+U_1+U_2)-U_3)$ \\
    \hline
    34 & (-3,-12,-2) & \{-14,-9,11,0,-12,48\} & \{5,6,-2,0,0,0\} & 84 & $(6 U_1-2 U_2+5) (-7+3 U_1+3 U_2-U_3)$ \\
    \hline
    35 & (-1,-32,1) & \{-22,48,0,0,40,-128\} & \{0,-1,1,0,0,0\} & 84 & $-(U_1-U_2) (-60+12 U_1+20 U_2-U_3)$ \\
    \hline
    36 & (-1,-32,-1) & \{-12,-10,-6,0,-40,128\} & \{1,1,-1,0,0,0\} & 84 & $(1+U_1-U_2) (-34+12 U_1+20 U_2-U_3)$ \\
    \hline
    37 & (-1,-8,2) & \{-3,-14,6,0,20,-64\} & \{-4,-2,2,0,0,0\} & 84 & $-2 (2+U_1-U_2) (-2+3 U_1+5 U_2-U_3)$ \\
    \hline
    38 & (-1,-8,-2) & \{-10,-9,1,0,-20,64\} & \{2,2,-2,0,0,0\} & 84 & $(1+U_1-U_2) (-21+6 U_1+10 U_2-2 U_3)$ \\
    \hline
    39 & $(-3,-\frac{16}{3},3)$ & \{-10,-14,2,0,8,-32\} & \{-6,-9,3,0,0,0\} & 84 & $-(2+3 U_1-U_2) (-2+4 U_1+4 U_2-3 U_3)$ \\
    \hline
    40 & $(-3,-\frac{16}{3},-3)$ & \{-6,-10,6,0,-8,32\} & \{6,9,-3,0,0,0\} & 84 & $(2+3 U_1-U_2) (-10+4 U_1+4 U_2-3 U_3)$ \\
    \hline
    41 & $(-3,-\frac{4}{3},6)$ & \{-9,-13,3,0,4,-16\} & \{-12,-18,6,0,0,0\} & 84 & $-(2+3 U_1-U_2) (1+2 U_1+2 U_2-6 U_3)$ \\
    \hline
    42 & $(-3,-\frac{4}{3},-6)$ & \{-9,-14,6,0,-4,16\} & \{12,18,-6,0,0,0\} & 84 & $2 (2+3 U_1-U_2) (-4+U_1+U_2-3 U_3)$ \\
    \hline
    43 & (-2,-40,1) & \{3,-14,-3,0,28,-100\} & \{-2,-2,1,0,0,0\} & 78 & $-(2+2 U_1-U_2) (-17+12 U_1+14 U_2-U_3)$ \\
    \hline
    44 & (-2,-40,-1) & \{8,-10,-5,0,-28,100\} & \{1,2,-1,0,0,0\} & 78 & $(1+2 U_1-U_2) (-23+12 U_1+14 U_2-U_3)$ \\
    \hline
    45 & $(-\frac{3}{2},-9,2)$ & \{-9,-9,12,0,16,-54\} & \{-2,-3,2,0,0,0\} & 78 & $-(2+3 U_1-2 U_2) (-2+3 U_1+4 U_2-U_3)$ \\
    \hline
    46 & $(-\frac{3}{2},-9,-2)$ & \{-2,3,-2,0,-16,54\} & \{3,3,-2,0,0,0\} & 78 & $(3+3 U_1-2 U_2) (-5+3 U_1+4 U_2-U_3)$ \\
    \hline
    47 & (-3,-40,1) & \{-9,-11,-3,0,20,-80\} & \{-2,-3,1,0,0,0\} & 70 & $-(2+3 U_1-U_2) (-13+10 U_1+10 U_2-U_3)$ \\
    \hline
    48 & (-3,-40,-1) & \{-5,-10,10,0,-20,80\} & \{2,3,-1,0,0,0\} & 70 & $(2+3 U_1-U_2) (10 (-2+U_1+U_2)-U_3)$ \\
    \hline
    49 & $(-3,-\frac{8}{5},5)$ & \{-9,-13,3,0,4,-16\} & \{-10,-15,5,0,0,0\} & 70 & $-(3 U_1-U_2+3) (2 U_1+2 U_2-5 U_3+1)$ \\
    \hline
    50 & $(-3,-\frac{8}{5},-5)$ & \{-14,-9,7,0,-4,16\} & \{15,15,-5,0,0,0\} & 70 & $(3 U_1-U_2+3) (2 U_1+2 U_2-5 U_3-7)$ \\
		\hline
  \end{tabular}
}
\end{center}
\caption{Representatives of families of integer fluxes for solutions with $W=0$ and $1$ flat direction, Part $1$.}
\label{tab:CYpart1}
\end{table}

\begin{table}[H]
\begin{center}
\resizebox{\columnwidth}{!}{
  \begin{tabular}{|c|c|c|c|c|c|}
    \hline
    & $(\lambda_2,\lambda_3,h_3)$ & $F$ & $H$ & $\mc{N}_{\rm flux}$ & $W$ \\
    \hline
    51 & (-3,-32,1) & \{-10,-13,-1,0,16,-64\} & \{-2,-3,1,0,0,0\} & 56 & $-(2+3 U_1-U_2) (-9+8 U_1+8 U_2-U_3)$ \\
    \hline
    52 & (-3,-32,-1) & \{-6,-11,9,0,-16,64\} & \{2,3,-1,0,0,0\} & 56 & $(2+3 U_1-U_2) (-17+8 U_1+8 U_2-U_3)$ \\
    \hline
    53 & (-3,-8,2) & \{-10,-14,2,0,8,-32\} & \{-4,-6,2,0,0,0\} & 56 & $-2 (2+3 U_1-U_2) (-1+2 U_1+2 U_2-U_3)$ \\
    \hline
    54 & (-3,-8,-2) & \{-6,-10,6,0,-8,32\} & \{4,6,-2,0,0,0\} & 56 & $2 (2+3 U_1-U_2) (-5+2 U_1+2 U_2-U_3)$ \\
    \hline
    55 & (-3,-2,4) & \{-9,-13,3,0,4,-16\} & \{-8,-12,4,0,0,0\} & 56 & $-(2+3 U_1-U_2) (1+2 U_1+2 U_2-4 U_3)$ \\
    \hline
    56 & (-3,-2,-4) & \{-9,-14,6,0,-4,16\} & \{8,12,-4,0,0,0\} & 56 & $2 (2+3 U_1-U_2) (U_1+U_2-2 (2+U_3))$ \\
    \hline
    57 & (-3,-24,1) & \{-9,-12,0,0,12,-48\} & \{-2,-3,1,0,0,0\} & 42 & $-(2+3 U_1-U_2) (6 (-1+U_1+U_2)-U_3)$ \\
    \hline
    58 & (-3,-24,-1) & \{3,3,3,0,-12,48\} & \{2,3,-1,0,0,0\} & 42 & $(2+3 U_1-U_2) (-9+6 U_1+6 U_2-U_3)$ \\
    \hline
    59 & (-1,-16,1) & \{-3,-14,6,0,20,-64\} & \{-2,-1,1,0,0,0\} & 42 & $-(2+U_1-U_2) (-4+6 U_1+10 U_2-U_3)$ \\
    \hline
    60 & (-1,-16,-1) & \{-10,-9,1,0,-20,64\} & \{1,1,-1,0,0,0\} & 42 & $(1+U_1-U_2) (-21+6 U_1+10 U_2-U_3)$ \\
    \hline
    61 & $(-3,-\frac{8}{3},3)$ & \{-9,-13,3,0,4,-16\} & \{-6,-9,3,0,0,0\} & 42 & $-(2+3 U_1-U_2) (1+2 U_1+2 U_2-3 U_3)$ \\
    \hline
    62 & $(-3,-\frac{8}{3},-3)$ & \{-14,-9,7,0,-4,16\} & \{9,9,-3,0,0,0\} & 42 & $(3+3 U_1-U_2) (-7+2 U_1+2 U_2-3 U_3)$ \\
    \hline
    63 & (-3,-16,1) & \{-2,-2,-2,0,8,-32\} & \{-2,-3,1,0,0,0\} & 28 & $-(2+3 U_1-U_2) (-6+4 U_1+4 U_2-U_3)$ \\
    \hline
    64 & (-3,-16,-1) & \{0,-1,3,0,-8,32\} & \{2,3,-1,0,0,0\} & 28 & $(2+3 U_1-U_2) (-7+4 U_1+4 U_2-U_3)$ \\
    \hline
    65 & (-3,-4,2) & \{-9,-13,3,0,4,-16\} & \{-4,-6,2,0,0,0\} & 28 & $-(2+3 U_1-U_2) (1+2 U_1+2 U_2-2 U_3)$ \\
    \hline
    66 & (-3,-4,-2) & \{-14,-9,7,0,-4,16\} & \{6,6,-2,0,0,0\} & 28 & $(3 U_1-U_2+3) (2 U_1+2 U_2-2 U_3-7)$ \\
    \hline
    67 & (-3,-8,1) & \{-1,-1,-1,0,4,-16\} & \{-2,-3,1,0,0,0\} & 14 & $-(3 U_1-U_2+2) (2 U_1+2 U_2-U_3-3)$ \\
    \hline
    68 & (-3,-8,-1) & \{-1,-2,2,0,-4,16\} & \{2,3,-1,0,0,0\} & 14 & $(3 U_1-U_2+2) (2 (-2+U_1+U_2)-U_3)$ \\
    \hline
  \end{tabular}
}
\end{center}
\caption{Representatives of families of integer fluxes for solutions with $W=0$ and $1$ flat direction, Part $2$.}
\label{tab:CYpart2}
\end{table}

On the other hand, $W$ remains non-homogeneous in the remaining $15$ cases, with the following case numbers in Tab. \ref{tab:CYpart1} and \ref{tab:CYpart2}: $3, 4, 21, 22, 23, 24, 25, 26, 33, 34, 38, 41, 55, 65, 66$. These $15$ solutions represent therefore novel perturbatively flat vacua which are qualitatively different from the ones studied in \cite{Demirtas:2019sip, Broeckel:2021uty}. In order to check if these can be solutions with small $g_s$, one should perform a careful study of dilaton stabilisation via instantons which we leave however for future research.

Now, we consider one of the above $68$ triples, $(\lambda_2,\lambda_3,h_3)=(-3,-72,1)$. For this, $\mc{N}_{\rm flux}=126$ and \eqref{eq:fluxesCY} becomes:
\bea
\{f_1,f_2,f_3,f_4,f_5,f_6\} &=& \{-63 + h_1 (f_3 - 18 (3 + h_1)), 108 - 3 f_3 + 72 h_1, f_3,0, 36, -144\}\,, \nn \\
\{h_1,h_2,h_3,h_4,h_5,h_6\} &=& \{h_1, -3, 1,0, 0, 0\}\,,
\eea
and \eqref{eq:WCY} gives:
\be
W = -\left(U_2 - 3 U_1 + h_1\right) \left(U_3 - 18 (-3 + U_1 + U_2) -f_3 + 18 h_1\right).
\ee
Clearly, every $f_3,h_1\in\mathbb{Z}$ retain all the fluxes integers and the solution to $W=\partial_a W=0$, $\forall a=1,2,3$ is given by:
\be
(U_1,U_2,U_3) = \left(\frac{1}{72} \left(U_3-f_3+36 h_1+54\right),\frac{1}{24} \left(U_3-f_3+12 h_1+54\right),U_3\right).
\label{ExplEx}
\ee
Now, choosing $h_1=0$ and $f_3 = 54$, $W$ becomes a homogeneous function of degree $2$. For this subfamily, although we obtain a non-homogeneous $W$ with other choices of $h_1$ and $,f_3$, $W$ can always be made homogeneous by appropriate integer shifts of $U_1$, $U_2$ and $U_3$.

\subsubsection*{Solutions with $W\neq0$}

As already pointed out in Sec. \ref{sec:qual}, supersymmetric solutions require $\partial_a W=-W \partial_a K$ $\forall a=1,2,3$. Thus, if $W\neq 0$, solving the global supersymmetry flatness conditions $\partial_a W=0$ does not lead in general to F-flat solutions in supergravity. However, if the solutions to $\partial_a W=0$ feature a flat direction parametrised by $U_3$, it is easy to realise that along the flat direction $\partial_1 K\sim \partial_2 K \sim \partial_3 K \sim 1/{\rm Im}(U_3) \to 0$ for ${\rm Im}(U_3)= g_s^{-1}\to \infty$. This limit corresponds to weak string couplings and, as can be seen in the explicit solution (\ref{ExplEx}), to large complex structure where the non-perturbative contributions to the prepotential can be ignored. Therefore solving the global supersymmetry flatness conditions can be a useful starting point to construct solutions in a perturbative expansion. When $U_3$ is flat, this approximation can be made exact by taking by hand ${\rm Im}(U_3)$ arbitrarily large, while when $U_3$ is lifted by instanton corrections, one has to make sure that at the minimum $W \partial_a K$ is infinitesimally small. 

Below, we discuss some solutions to the global supersymmetry flatness conditions in the $\mathbb{CP}_{[1,1,1,6,9]}[18]$ example. Given \eqref{eq:CYmod2}, consider the $3$ polynomial equations in $3$ variables: $\partial_a W=0$ $\forall a=1,2,3$. The degree of each of these equations is $3$ or less, depending upon the choices of fluxes. For this system of equations to admit a solution with at least $1$ flat direction, one of them should be dependent on the others. Here we examine cases when this dependence is linear:
\begin{itemize}
\item[($i$)] When $\partial_1 W = \lambda_2\,\partial_2 W+\lambda_3\,\partial_3 W$ with at least one of $\lambda_2$ and $\lambda_3$ which is non-zero and subject to $\mc{N}_{\rm flux} \neq0$, we find only $1$ family of fluxes (details are given below) for which $\partial_a W=0$ $\forall a=1,2,3$ admit complex solutions $\hat{U}_a$, i.e. ${\rm Im}\,\hat{U}_a\neq0$ $\forall a$. With the above family of fluxes we have only $1$ flat direction which is the dilaton;

\item[($ii$)] When $\partial_2 W=\lambda_3\,\partial_3W$ with $\lambda_3\neq0$, the conditions on the fluxes have no solution in keeping with $\mc{N}_{\rm flux}\neq0$.
\end{itemize}

Let us provide the aforesaid family of fluxes which are dependent on $6$ parameters $\lambda_2$, $\lambda_3$, $f_1$, $f_3$, $h_1$ and $h_3$:
\bea
\{f_1,f_2,f_3,f_4,f_5,f_6\} &=& \left\{f_1,\lambda_2  f_3-\lambda_3  h_1-\frac{3 \lambda_3  h_3}{2},f_3,0,\frac{(2 \lambda_2 -3) \lambda_3  h_3}{(\lambda_2 -3) \lambda_2 },\frac{(\lambda_2 -3) \lambda_3  h_3}{\lambda_2 }\right\}\,, \nn \\
\{h_1,h_2,h_3,h_4,h_5,h_6\} &=& \{h_1,\lambda_2  h_3,h_3,0,0,0\}\,,\qquad \lambda_2\neq0,3\,,\qquad \lambda_3,h_3\neq0\,.
\label{eq:fluxesCYWn0}
\eea
In this case we have:
\bea
\mc{N}_{\rm flux} &=& -\frac{3 \left((\lambda_2 -3) \lambda_2 +3\right) \lambda_3  h_3^2}{2 (\lambda_2 -3) \lambda_2 }\,, \nn \\
W &=& \left(f_1+(f_3-h_3 U_3) (\lambda_2  U_1+U_2)-h_1 \left(\lambda_3 U_1+U_3\right)\right)  \label{eq:WCYn0} \\
&+& \frac{\lambda_3 h_3}{4 (\lambda_2 -3) \lambda_2 }\left(2 \lambda_2 ^2 (U_1 (3 U_1+2 U_2-6)-3)+2 \lambda_2  (9 U_1+2 (U_2-3) U_2+1)-6 (U_2-3) U_2-3\right) . \nn
\eea
Now, solving $\partial_a W=0$ $\forall a=1,2,3$, we see that $U_1$ and $U_2$ depend linearly on $U_3$ with slopes $-1/\lambda_3$ and $\lambda_2/\lambda_3$ respectively. Thus, only by requiring $\lambda_2,\lambda_3<0$, we may obtain ${\rm Im}\,U_a$, $\forall a=1,2,3$ to be of the same sign. This also keeps $\mc{N}_{\rm flux}$ positive. At the solution we have:
\be
W = f_1-\frac{f_3 h_1}{h_3}+\frac{\lambda_3  \left((4 \lambda_2 -6) h_1^2+6 (2 \lambda_2 -3) h_1 h_3+\left(-6 \lambda_2 ^2+2 \lambda_2 -3\right) h_3^2\right)}{4 (\lambda_2 -3) \lambda_2  h_3}\,,
\ee
which we require not to vanish. In this case, by integer translations of $U_1$, $U_2$ and $U_3$, $W$ cannot be made a degree-$2$ homogeneous function since a necessary condition for doing so is the same as the vanishing condition of $W$ at the minimum.

Note the arguments of $f_5(\lambda_2,\lambda_3,h_3)$, $f_6(\lambda_2,\lambda_3,h_3)$, $h_2(\lambda_2,h_3)$ and $\mc{N}_{\rm flux}(\lambda_2,\lambda_3,h_3)$. There are only $488$ triples $(\lambda_2,\lambda_3,h_3)$, $\lambda_2,\lambda_3\in\mathbb{Q}^-$, $h_3\in\mathbb{Z}$, securing $f_5,f_6,h_2\in\mathbb{Z}$ and $\mc{N}_{\rm flux}\in\mathbb{Z}/2$ with $0<\mc{N}_{\rm flux}\leq138$. For each of the triples $(\lambda_2,\lambda_3,h_3)$, there exist $f_1,f_3,h_1\in\mathbb{Z}$ that keep all other fluxes in \eqref{eq:fluxesCYWn0} integers, as well as $W\neq0$. In fact, for each of the triples, we get a subfamily of integer fluxes \eqref{eq:fluxesCYWn0} parameterised by $f_1$, $f_3$ and $h_1$. All the members in any of the aforementioned subfamilies have the same $\mc{N}_{\rm flux}(\lambda_2,\lambda_3,h_3)$. Let us point out that among these $488$ values of $\mc{N}_{\rm flux}(\lambda_2,\lambda_3,h_3)$ only $64$ are distinct. Below, we discuss one of these subfamilies in detail.

We consider one of the $488$ triples given by $(\lambda_2,\lambda_3,h_3)=(-4,-56,-1)$. For this, $\mc{N}_{\rm flux}=93$ and \eqref{eq:fluxesCYWn0} becomes:
\bea
\{f_1,f_2,f_3,f_4,f_5,f_6\} &=& \{f_1,-4 (f_3-14 h_1+21),f_3,0,-22,98\}\,, \nn \\
\{h_1,h_2,h_3,h_4,h_5,h_6\} &=& \{h_1, 4, -1, 0, 0, 0\}\,,
\eea
and \eqref{eq:WCYn0} gives:
\be
W = f_1-4 U_1 (f_3-14 h_1-8 U_2+U_3+33)+U_2 (f_3-11 U_2+U_3+33)-h_1 U_3+48 U_1^2-\frac{107}{2}\,. \nn
\ee
Clearly, every $f_1,f_3,h_1\in\mathbb{Z}$ retain all fluxes integers and the solution to $\partial_a W=0$, $\forall a=1,2,3$ is given by:
\be
(U_1,U_2,U_3) = \left(\frac{1}{56} \left(U_3-22 h_1+33+f_3\right),\frac{1}{14} \left(U_3-8 h_1+33+f_3\right),U_3\right).
\ee
At the solution we have:
\be
W = f_1+h_1 \left(f_3-11 h_1+33\right)-\frac{107}{2}\,,
\ee
which is non-zero since $f_1,f_3,h_1\in\mathbb{Z}$.

\subsection{Flat directions in a Calabi-Yau with 4 moduli}

In this section we search for flat directions using the CY discussed in \cite{Cicoli:2013cha} which features effectively $3$ complex structure moduli at the $\mc{G}$-symmetric locus. We begin by quoting the large complex structure expansion of the prepotential (denoting $(U^1,U^2,U^4,\phi)$ by $(U_1,U_2,U_3,U_4)$):
\bea
\mc{F}(U_a) &=& -3 U_1 U_2 U_4-3 U_1 U_3 U_4-3 U_2 U_3 U_4-3 U_2 U_4^2-3 U_3 U_4^2-\frac32 \,U_1^2 U_4 -\frac92 \,U_1 U_4^2 -\frac52\,U_4^3 \nn \\
 &+& 3 U_1 U_4 +\frac32\, U_2 U_4 +\frac32\, U_3 U_4 +\frac{15}{4}\,U_4^2 + \frac32\,U_1 + U_2 + U_3 + \frac{11}{4}\,U_4 + \xi\,,
\eea
where $\xi$ (that involves the CY Euler number) is imaginary with irrational imaginary part. The period vector $\Pi$ is given by (\ref{per}) and the superpotential can be written explicitly using (\ref{eq:wis}), together with the fluxes $F=(f_1\ f_2\ f_3\ f_4\ f_5\ f_6\ f_7\ f_8\ f_9\ f_{10})^t$ and $H=(h_1\ h_2\ h_3\ h_4\ h_5\ h_6\ h_7\ h_8\ h_9\ h_{10})^t$ that are integer-valued. For simplicity we set $f_6=h_6=0$ which eliminates the $\xi$ dependence in $W$. Furthermore, due to the $\mc{G}$-symmetry, we identify:
\be
f_4 = f_3\,,\qquad h_4 = h_3\,,\qquad f_9 = f_8\,,\qquad h_9 = h_8\,,\qquad U_2=U_3\,.
\ee
Moreover the orientifold and brane setup discussed in \cite{Cicoli:2013cha} give a tadpole bound $\mc{N}_{\rm flux}=-\frac12 H^t\cdot\Sigma\cdot F\leq \frac12\left(20+3(1+2n_b)^2\right)$ with $n_b\in\mathbb{Z}$. For definiteness, we shall choose $n_b=-2$ which yields $\mc{N}_{\rm flux}\leq 47/2$. With the above, we have:
\bea
\mc{N}_{\rm flux} &=& \frac12 (f_5 h_{10} - f_7 h_2 - 2 f_8 h_3 - f_{10} h_5 + f_2 h_7 + 2 f_3 h_8)\,, \nn \\
W &=& (h_{10} U_4-f_{10}) \left[\frac{11}{4}+3(U_1+U_2)+\frac{15}{2} U_3-\frac32 \left((U_1+2 U_2+3 U_3)^2-(U_2+2U_3)^2-U_2^2\right)\right] \nn \\
&+& U_1 (f_2-h_2 U_4)+2 U_2 (f_3-h_3 U_4)+U_3 (f_5-h_5 U_4) \nn \\
&+&\frac32 (f_7-h_7 U_4) (U_3 (2 U_1+4 U_2+3 U_3-2)-1) +f_1 \nn \\
&+& (f_8-h_8 U_4) (3 U_3 (2 U_1+2 U_2+2 U_3-1)-2)-h_1 U_4\,.
\label{eq:CYmod3}
\eea
In this case we will not perform a systematic search for supersymmetric solutions with approximate flat directions. As a preliminary analysis, we note however that, given \eqref{eq:CYmod3} and considering $\partial_1 W = \lambda_2\,\partial_2W$, $\lambda_2\neq0$, $\mc{N}_{\rm flux}\neq0$, there exists a class of fluxes for which $\partial_a W=0$, $\forall a=1,\dots,4$ admit complex solutions $\hat{U}_a$, i.e. ${\rm Im}\,\hat{U}_a\neq0$ $\forall a$ with $2$ flat directions. The aforesaid class of fluxes and corresponding $\mc{N}_{\rm flux}$ are given by:
\bea
&&\{f_1, f_2, f_3, f_5, f_7, f_8, f_{10}, h_1, h_2, h_3, h_5, h_7, h_8, h_{10}\}= \nn \\
&&\left\{f_1,-2 f_3,f_3,-\frac{f_8 (h_1+h_3)}{h_3},-\frac43\,f_8,f_8,0,h_1,-2 h_3,h_3,0,0,0,0\right\}\,, \nn \\
&& \mc{N}_{\rm flux} =-\frac73\, f_8 h_3\,,
\eea
where $f_8,h_3\neq0$ and $\lambda_2=-1$. Let us present an explicit example. For $f_8=-3$ and $h_3=3$, $\mc{N}_{\rm flux}=21$. In this case the fluxes become:
\bea
&&\{f_1, f_2, f_3, f_5, f_7, f_8, f_{10}, h_1, h_2, h_3, h_5, h_7, h_8, h_{10}\} = \nn \\
&& \left\{f_1, -2 f_3, f_3, 3 + h_1, 4, -3, 0, h_1, -6, 3, 0, 0, 0, 0\right\}\,,
\eea
and the superpotential reads:
\be
W = f_1+2 f_3 \left(U_2-U_1\right)+\left(U_3-U_4\right) \left(h_1-6U_1+6 U_2\right).
\ee
Clearly, $f_1,f_3,h_1\in\mathbb{Z}$ retain all fluxes integer and the solution to $\partial_a W=0$, $\forall a=1,\dots,4$ is given by:
\be
(U_1,U_2,U_3,U_4) = \left(U_2+\frac{h_1}{6},U_2,U_4-\frac{f_3}{3},U_4\right).
\ee
Notice that when the flat directions $U_2$ and $U_4$ are in the large complex structure limit, the same is necessarily true for $U_1$ and $U_3$. Moreover, the superpotential at the supersymmetric minimum is:
\be
W = f_1-\frac{f_3 h_1}{3}\,.
\ee
As in the $\mathbb{CP}_{[1,1,1,6,9]}[18]$ case, by integer translations of $U_1$, $U_2$, $U_3$ and $U_4$, a necessary condition for making $W$ a degree-$2$ homogeneous function is the same as the vanishing condition of $W$ at the minimum. Hence, for solutions with $W\neq 0$ at the minimum, the superpotential is a non-homogeneous polynomial of degree $2$, while for solutions with vanishing $W$ at the minimum, the superpotential in some cases can be brought to a homogeneous function of degree $2$. An example where it can be done is the case with $f_1 = 6$, $f_3 = 3$ and $h_1 = 6$. However, in the case with $f_1 = -14$, $f_3 = -6$ and $h_1 = 7$, $W$ cannot be brought to a degree-$2$ homogeneous function by integer shifts of $U_a$, although it vanishes at the minimum. Detailed explorations in various CYs will be carried out in the future.

\section{Lifting flat directions: phenomenology and applications}
\label{sec:pheno}

The flat directions studied in this paper have interesting phenomenological implications. Before mentioning some of them, let us stress that these flat directions are approximate since they are expected to be lifted by subleading effects at either perturbative or non-perturbative level. In the $T^6/\mathbb{Z}_2$ case, a non-zero $W$ should be generated by non-perturbative effects which depend on the K\"ahler moduli. A non-zero scalar potential for the leading order flat directions is then generated by the $U$-dependence of the prefactor of non-perturbative effects and the coefficients of $\alpha'$ and string loop corrections to the K\"ahler potential. Moreover one should carefully check potential modifications of the primitivity condition by quantum corrections.

For the CY cases, the flat direction of perturbative flat vacua with $W=0$ should be lifted by instantons along the lines of \cite{Demirtas:2019sip}. On the other hand, for solutions with $W\neq 0$, the imaginary part of the approximate flat direction would be lifted already at perturbative level by including $U$-dependent effects that arise from the supergravity contribution to the K\"ahler covariant derivative $W K_U$.

Let us now briefly discuss several potential applications to phenomenology of approximate flat directions:
\begin{enumerate}
\item \textbf{K\"ahler moduli stabilisation:} There are various mechanisms for stabilising the K\"ahler moduli in type IIB (see for example \cite{Kachru:2003aw, Balasubramanian:2005zx, Berg:2005yu, Westphal:2006tn, Cicoli:2008va, Cicoli:2012fh, Cicoli:2016chb, Gallego:2017dvd, Antoniadis:2018hqy, AbdusSalam:2020ywo}). In particular, an exponentially low $W_0$ is crucial for KKLT constructions \cite{Kachru:2003aw}. Flux vacua with $W=0$ and flat directions have been shown to be a promising starting point to realise these scenarios \cite{Demirtas:2019sip, Demirtas:2020ffz,Alvarez-Garcia:2020pxd, Demirtas:2021ote, Demirtas:2021nlu, Honma:2021klo,Broeckel:2021uty}. 

Even if not strictly required, very low values of $W_0$ might be needed also in some dS LVS constructions where the visible sector lives on D3-branes at singularities \cite{Cicoli:2021dhg}. In these models consistency conditions, like D7-tadpole and Freed-Witten anomaly cancellation, in general induce a T-brane background which yields a positive contribution to the scalar potential in the presence of background $3$-form fluxes \cite{Cicoli:2015ylx}. In \cite{Cicoli:2021dhg}, this contribution has been shown to be able to give a dS minimum for exponentially small values of $W_0$. On the other hand, \cite{Cicoli:2017shd} presented a different global model with D3-branes at singularities where dS moduli stabilisation with T-brane uplifting can be achieved also for $W_0\sim \mc{O}(1)$. 

\item \textbf{Winding uplift:} Solutions with $W\neq 0$ could be a promising starting point for explicit realisations of dS uplifting via exponentially small F-terms of the complex structure moduli \cite{Hebecker:2020ejb}. The idea is to have at leading order supersymmetric solutions in the large complex structure limit with $W\neq 0$ and $1$ axionic flat direction. In turn, this axion is lifted by instantons which induce exponentially suppressed but non-zero F-terms for the complex structure moduli, so leading to a tunable (via flux choices) and positive uplifting contribution to the scalar potential. 

To be more explicit, the $W\neq 0$ solutions discussed in Sec. \ref{cy11169} for the $\mathbb{CP}_{[1,1,1,6,9]}[18]$ case, feature $\partial_1 W = \lambda_2 \partial_2 W + \lambda_3 \partial_3 W$. Thus solving the full supergravity F-term equations $\partial_a W + W \partial_a K =0$ $\forall a=1,2,3$, is equivalent to solving:
\bea
\partial_1 W &=& 0-W \partial_1 K\,, \label{eq1} \\
\partial_2 W &=& 0-W \partial_2 K\,, \label{eq2} \\
\partial_1 K &=& \lambda_2 \partial_2 K + \lambda_3 \partial_3 K \,.
\label{eq3}
\eea
Eq. (\ref{eq1}) and (\ref{eq2}) are $2$ complex equations, and so would fix the $2$ complex moduli $U_1$ and $U_2$ in terms of $U_3$. Moreover their solutions would very well be approximated by the solutions to $\partial_1 W=\partial_2 W=0$ already found in Sec. \ref{cy11169}, if the minimum is such that the supergravity corrections are infinitesimally small in the large complex structure limit. Finally (\ref{eq3}) is a real equation since $K$ is just a function of the imaginary parts of the complex structure moduli and the axio-dilaton in the large complex structure limit (denoting ${\rm Im}(U_a)\equiv u_a$ $\forall a=1,2,3$):
\be
K =  -\ln\left[4 u_1\left(3 u_1^2 + 3 u_1 u_2 + u_2^2\right)-4{\rm Im}(\xi)\right] -\ln\left(2 u_3\right).
\ee
Thus (\ref{eq3}) should fix only ${\rm Im}(U_3)$, leaving ${\rm Re}(U_3)$ as the only axionic flat direction that is expected to be lifted by instanton corrections to the prepotential. In the large complex structure limit these contributions would be exponentially suppressed by $e^{-{\rm Im}(U_1)}\ll 1$ and $e^{-{\rm Im}(U_2)}\ll 1$.

\item \textbf{Cosmology:} Approximate flat directions have natural applications to cosmology where inflaton fields are required to be lighter than the Hubble scale during inflation to be in the slow-roll regime. In fact, flat directions in the type IIB flux superpotential have already been used in \cite{Hebecker:2017lxm} to enlarge the inflaton field range, and more recently in \cite{Kallosh:2021vcf} to build models of sequestered inflation. Approximate flat directions could be promising candidates also to drive the present day accelerated expansion of our universe since quintessence fields need to be very light to reproduce the observed cosmological constant scale. Moreover, leading order flat directions can help to avoid any destabilisation problem coming from contributions to the dark energy potential due to the large inflationary energy scale \cite{Cicoli:2021fsd, Cicoli:2021skd}. 

\item \textbf{Supersymmetry breaking:} Leading order flat directions can also play a relevant role in any model of supersymmetry breaking if $W=0$ at classical level. In fact, in this case the F-terms of the K\"ahler moduli are vanishing at leading order and the effective field theory after integrating out the heavy complex structure moduli has to include the K\"ahler moduli and all the complex structure moduli, including the axio-dilaton, which are massless at leading order \cite{Choi:2005ge}. The dynamics which stabilises the K\"ahler moduli and the leading order flat directions is expected to break supersymmetry and to develop non-zero F-terms for all these fields which will play an important role in generating soft supersymmetry breaking terms. The F-term of the dilaton would be particularly important in D3-brane models with sequestered supersymmetry breaking where it is the main source for generating non-zero gaugino masses \cite{Blumenhagen:2009gk, Aparicio:2014wxa}.

\item \textbf{Statistics in the landscape:} The statistical approach to string phenomenology has received a lot of attention during the last two decades (see e.g. \cite{Douglas:2003um, Ashok:2003gk, Denef:2004ze, Denef:2004dm, Denef:2004cf, Broeckel:2020fdz,Broeckel:2021dpz, Halverson:2019cmy, Susskind:2004uv, Douglas:2004qg, Dine:2004is, Arkani-Hamed:2005zuc, Kallosh:2004yh, Dine:2005yq, Sun:2022xdl}). Trying to achieve a complete classification of flux vacua with exponentially small $W_0$ is crucial to understand the statistical significance of these vacua. The analysis of \cite{Broeckel:2020fdz, Broeckel:2021dpz} implies that if $W_0$ is uniformly distributed at very small values, then the scale of supersymmetry breaking has a power-law distribution, while if $W_0$ is exponentially small in the dilaton, as in the models of \cite{Demirtas:2019sip}, then the gravitino mass has a logarithmic distribution. Preliminary steps in understanding the statistical significance of perturbatively flat vacua were taken in \cite{Broeckel:2021uty} which found that they represent a small fraction of the full set of vacua at low $W_0$ as estimated in \cite{Denef:2004ze}. Our paper goes in the direction to explore novel classes of vacua at low $W_0$ to enrich their knowledge.

\item \textbf{CRG Conjecture:} The solutions found should be interesting in the context of studies on the consistency conditions of $4$ graviton scattering in the classical limit (see \cite{Camanho:2014apa, Chowdhury:2019kaq}). The solutions obtained are warped Minkowski compactifications in which the string coupling can be tuned to arbitrarily small values. The solutions are in the supergravity approximation. Developing a precise understanding of the fate of the solutions beyond the supergravity approximations, i.e. checking if there can be a solution where the flat direction survives to all orders in $\alpha'$,\footnote{Warping dependent corrections would also have to be incorporated, see e.g. \cite{Giddings:2005ff, Frey:2006wv, Shiu:2008ry, Martucci:2016pzt}.} (with the solution remaining Minkowski) and the study of $4$ graviton scattering in these backgrounds is relevant in the context of the classical Regge growth conjecture. 
\end{enumerate}

\section{Conclusions}
\label{sec:conc}

In this paper we have presented a novel method to obtain type IIB flux vacua with flat directions at tree level. The key idea is to make choices for flux quanta so that there are relations between the flux superpotential and its derivatives. These relations ensure that the equations of motion are satisfied. We implemented this method in toroidal and Calabi-Yau compactifications in the large complex structure limit. Explicit solutions were obtained and classified on the basis of duality equivalences. In the toroidal setting we presented solutions with both $N=1$ and $N=2$ supersymmetry. For the $\mathbb{CP}_{[1,1,1,6,9]}[18]$ CY, on top of solutions which were already known in the literature, we found $15$ novel perturbatively flat vacua with approximate flat directions where the superpotential is not a homogeneous function of degree $2$. We also presented solutions with $W\neq 0$ which might lead to an explicit realisation of winding dS uplift. We also performed a preliminary analysis of flux vacua for the CY considered in \cite{Cicoli:2013cha} finding supersymmetric solutions with $2$ approximate flat directions.

We also discussed the lifting of these solutions by higher order effects (both perturbative and non-perturbative) and applications in a wide variety of settings such as K\"ahler moduli stabilisation, explicit dS uplifting contributions from non-zero F-terms of the complex structure moduli, cosmology (in the context of inflation and quintessence), statistical studies in the landscape, classical Regge growth conjecture and supersymmetry breaking.

There are many interesting directions to pursue in the future. We have considered the simplest possible relations between the superpotential and its derivatives -- linear relationships. It will be interesting to consider non-linear relations and relations involving moduli-dependent coefficients. These are likely to provide new classes of flux vacua. The solutions obtained are also important in the context of developing a more precise understanding of flux vacua. In this context, the solutions with extended supersymmetry and arbitrarily weak coupling are particularly interesting. One can attempt to describe them by worldsheet methods, thereby going away from the large radius limit. Future phenomenological applications have been outlined Sec. \ref{sec:pheno}. We hope to return to these questions in the near future.

\section*{Acknowledgments}

We would like to thank Anirban Basu, Andrew Frey, Arthur Hebecker, Liam McAllister, Nicola Pedron, Fernando Quevedo and Roberto Valandro for useful discussions. During the initial stages of this work RM has been supported in part by the INFOSYS scholarship for senior students (HRI). AM is supported in part by the SERB, DST, Government of India by the grant MTR/2019/000267.

\vspace{0.5cm}

\appendix

\section{Duality transformations in type IIB}
\label{AppA}

\noindent In this section we briefly summarise the dualities relevant for our discussion and record
our conventions.

\vspace{0.4 cm}

\noindent \underline{$SL(2, \mathbb{Z})$ symmetry}: The type IIB theory enjoys an $SL(2,\mathbb{Z})$ symmetry. Under this, the $3$-form flux and the axio-dilaton transform as:
\be
\begin{pmatrix}
H_3 \cr F_3 
\end{pmatrix}
\to
\begin{pmatrix}
d & c \cr
b & a
\end{pmatrix}
\begin{pmatrix}
 H_3 \cr F_3
\end{pmatrix},
\qquad \phi \to  \phi'={{a \phi + b} \over {c\phi + d}}\,,
\label{Adil}
\ee
where:
\be
\begin{pmatrix}
a & b \cr
c & d
\end{pmatrix} \in SL(2,\mathbb{Z})\,.
\ee
Note that $\int F_3 \wedge H_3$ is invariant under this transformation, implying that the D3-charge of a flux configuration is invariant. However, the superpotential transforms as:
\be
W[U^a,\phi]\qquad\to\qquad W'[U^a,\phi']=\frac{W[U^a,\phi(\phi')]}{c\phi(\phi')+d}\,,
\ee
where $U^a$ are the complex structure moduli and $\phi(\phi')$ is obtained inverting \eqref{Adil}.

\vspace{0.4 cm}

\noindent \underline{$SL(6, \mathbb{Z})$ symmetry for $T^6$}: $T^6$ is obtained as the quotient of $\mathbb{R}^6$ by a 6D lattice, and $SL(6, \mathbb{Z})$ matrices relate the different choices of basis of the same lattice. An $SL(6, \mathbb{Z})$ action transforms the fluxes as well the period matrix. $2$ flux configurations are equivalent (or dual) if the fluxes and the solution for the complex structure moduli are related by an $SL(6, \mathbb{Z})$ transformation. In our solutions the $T^6$ is factorised into $T^2 \times T^2 \times T^2$. The relevant $SL(6, \mathbb{Z})$ transformations are the ones that permute the $3$ $2$-tori and the $SL(2, \mathbb{Z}) \times SL(2, \mathbb{Z}) \times SL(2, \mathbb{Z})$ subgroup that acts on each of the $3$ tori. The action of each of these $SL(2, \mathbb{Z})$ on their respective tori is as follows. The coordinates on the $2$-torus transform as:\footnote{The transformation of the fluxes follows from this via the usual transformation rule of $3$-forms. One can check that under the action of $SL(6,\mathbb{Z})$ the transformed flux quanta are even integers as long as the original ones are.}
\be
\label{eq:action2tori}
\begin{pmatrix}
x \cr y 
\end{pmatrix}
=
\begin{pmatrix}
a & b \cr
c & d
\end{pmatrix}
\begin{pmatrix}
x' \cr y' 
\end{pmatrix}\,,
\qquad
\begin{pmatrix}
a & b \cr
c & d
\end{pmatrix} \in SL(2,\mathbb{Z})\,,
\ee
where we can think of the primed coordinates as the new coordinates and the unprimed ones as the old ones. For the complex structure of the $2$-torus we have:\footnote{To be consistent with our notations, here we denote the $\tau$ parameter of a $2$-torus by $U$.}
\be
U' = { {d U+ b}  \over {c U + a} }\,.
\ee
An $SL(2,\mathbb{Z})$ transformation can be generated by successive action of $\mathcal{T}$- and $\mathcal{S}$-transformations given by:
\bea
\mathcal{T} &=& 
\begin{pmatrix}
1 & 1 \\
0 & 1
\end{pmatrix}\,, \qquad \mathcal{T}\,:\,U\to U+1\,, \\
\mathcal{S} &=&
\begin{pmatrix}
0 & 1 \\
-1 & 0
\end{pmatrix}\,, \qquad \mathcal{S}\,:\,U\to -\frac{1}{U}\,.
\eea
In what follows we often use a product of $n$ $\mathcal{T}$-transformations given by:
\be
\mathcal{T}^n = 
\begin{pmatrix}
1 & n \\
0 & 1
\end{pmatrix}\,,\qquad \mathcal{T}^n\,:\,U\to U+n\,.
\ee
Note that configurations $\{a_i,b_i,c_i,d_i\}$ and $\{-a_i,-b_i,-c_i,-d_i\}$ are dual by an action of $\mathcal{S}^2 \times \mathcal{S}^2 \times \mathcal{S}^2$ on $T^2 \times T^2 \times T^2$, which helps to classify inequivalent solutions. This action preserves $N_{\rm flux}$ and the solution to $W=\partial_a W=0$ $\forall a=1,\dots,4$, since $N_{\rm flux}\,,W$ are respectively quadratic and linear in $\{a_i,b_i,c_i,d_i\}$.

\vspace{0.4 cm}

\noindent \underline{$Sp(2h^{2,1}_- +2, \mathbb{Z})$ symmetry for Calabi Yaus}: The perturbative K\"ahler potential \eqref{eq:wis2} for CY compactifications is independent of the axions ${\rm Re}(U^a)$, $a=1,\dots,h^{2,1}_-$. Due to this, the discrete gauge symmetries of the theory are the integer shifts of the complex structure moduli:
\be
U^a \to U^a+n^a\,,\qquad n^a\in\mathbb{Z},\qquad a=1,\dots,h^{2,1}_-\,,
\label{Ushift}
\ee
causing the period and flux vectors to undergo a monodromy transformation:
\be
\{\Pi,H,F\} \to M_{\{n^a\}}\{\Pi,H,F\}\,,\qquad M_{\{n^a\}}\in Sp(2h^{2,1}_- +2, \mathbb{Z})\,.
\ee
Furthermore, the monodromy matrix is required to be unipotent:
\be
\left(M_{\{n^a\}}-I\right)^p\neq0,\qquad \left(M_{\{n^a\}}-I\right)^{p+1}=0,\qquad 1\leq p\leq3\,.
\ee
We can compute the monodromy matrix $M_{\{n^a\}}$ as follows. Notice that
\be
\Pi^i(U^a+n^a)={\sum}_j\left(M_{\{n^a\}}\right)^i_j\Pi^j(U^a),\ i=1,\dots,2h^{2,1}_- +2\,,
\ee
are a set of functional relations. Using the definition of the period vector \eqref{per}, the above relations can be evaluated at multiple values $\hat{U}^a$ to generate independent linear equations in the elements of the monodromy matrix. Inverting the latter we obtain the matrix elements uniquely. For example, in the $\mathbb{CP}_{[1,1,1,6,9]}[18]$ case (discussed in Sec. \ref{cy11169}) we get:
\be
\resizebox{1.00\hsize}{!}{
$M_{\{n_1,n_2\}}=
\left(
\begin{array}{cccccc}
 1 & -n_1 & -n_2 & 3 n_2+\frac{1}{2} n_1 \left(3 n_1^2+3 n_2 n_1+n_2^2+17\right) & \frac{1}{2} \left(3 n_1+n_2\right) \left(3 n_1+n_2+3\right) & \frac{3}{2} n_1 \left(n_1+1\right)+n_1 n_2 \\
 0 & 1 & 0 & -\frac{1}{2} \left(3 n_1+n_2-3\right) \left(3 n_1+n_2\right) & -3 \left(3 n_1+n_2\right) & -3 n_1-n_2 \\
 0 & 0 & 1 & -\frac{1}{2} n_1 \left(3 n_1+2 n_2-3\right) & -3 n_1-n_2 & -n_1 \\
 0 & 0 & 0 & 1 & 0 & 0 \\
 0 & 0 & 0 & n_1 & 1 & 0 \\
 0 & 0 & 0 & n_2 & 0 & 1 \\
\end{array}
\right)\,.$
}
\ee
It is easy to see that the above matrix belongs to $Sp(6,\mathbb{Z})$, i.e. with $\Sigma$ as given in \eqref{SimplMatrix} we obtain: $M_{\{n_1,n_2\}}^T\cdot\Sigma\cdot M_{\{n_1,n_2\}}=\Sigma$. Also, it is unipotent as per requirement. Moreover note that the shift (\ref{Ushift}) keeps $\mc{N}_{\rm flux}=-\frac12\,H^t\cdot\Sigma\cdot F$ invariant.

\section{Duality in toroidal solutions}
\label{AppB}

In this appendix we discuss the duality relations among the solutions with flat directions of the toroidal compactification case.

\subsection{Solutions with $1$ flat direction}

Let us now discuss in detail the duality among the solutions \eqref{Sol1flat} with $1$ flat direction. They are parametrised by an integer $p$, and $N_{\rm flux}=24$ irrespective of $p$. Below we show that the $p=0$ case is dual to any $p\neq 0$ case via an $SL(6,\mathbb{Z})$ transformation.

Let us use unprimed and primed coordinates for $p=0$ and $p\neq0$ respectively. We act with an $SL(6,\mathbb{Z})$ matrix $M$ on the coordinates of $T^2 \times T^2 \times T^2$ in accordance with \eqref{eq:action2tori}, where $M$ is given by:
\be
M = 
\begin{pmatrix}
M_1 & 0 & 0 \\
0 & M_2 & 0 \\
0 & 0 & M_2
\end{pmatrix}\,, \qquad
M_1 = 
\begin{pmatrix}
1 & 2p \\
0 & 1
\end{pmatrix}\,, \qquad	
M_2 = 
\begin{pmatrix}
		1 & p \\
		0 & 1
\end{pmatrix}\,.
\ee
This transforms the period matrix as:
\be
M\,:\, {\rm diag}\{U_1,U_2,U_3\} \to {\rm diag}\{U_1+2p,U_2+p,U_3+p\}\,.
\ee
Under this, the solution \eqref{Sol1flat} with $p=0$ is clearly mapped to a solution with $p\neq0$. Now we need to show that the fluxes \eqref{eq:fluxesEx1flat} map between the $p=0$ and $p\neq0$ cases. Indeed, using \eqref{threebasis} and \eqref{eq:action2tori}, we have:\footnote{$\alpha'$ and $\beta'$ denote the basis of $3$-forms \eqref{threebasis} with respect to the primed coordinates $(x'^i,y'^i)$.}
\bea
F_3 &=& 4\alpha_{11}-2\alpha_{22}-2\alpha_{33} \quad \to \quad 4 \alpha' _{11}-2 \alpha' _{22}-2 \alpha' _{33}+4 p \beta'^{11}-4 p^2 \beta'^0=F'_3\,, \nn \\
H_3 &=& -4\beta^{11}+2\beta^{22}+2\beta^{33} \quad \to \quad -4 \beta'^{11}+2 \beta'^{22}+2 \beta'^{33}+4 p \beta'^0=H'_3\,.
\eea

\subsection{Solutions with $2$ flat directions}

\subsubsection*{Dualities of family $\mc{A}$}

First we show that $\mc{A}_1$, $\mc{A}_2$ and $\mc{A}_3$ are dual via permutations of the 3 2-tori. Then the question to classify the inequivalent solutions in family $\mc{A}$ essentially boils down to that of subfamily $\mc{A}_1$, which we address subsequently.

\noindent \underline{Duality between $\mc{A}_1$, $\mc{A}_2$ and $\mc{A}_3$}: The fluxes in subfamilies $\mc{A}_1$ and $\mc{A}_2$, given respectively by \eqref{a1flux} and \eqref{eq:fluxesA2}, depend on the 6 parameters $\lambda_1,\lambda_2,\lambda_3,b_3,d_0,d_3$, while those of $\mc{A}_3$, given in \eqref{eq:fluxesA3}, depend on the 6 parameters $\lambda_1,\lambda_2,\lambda_3,b_2,d_0,d_2$. Under the permutation between the first and the second tori of $T^2 \times T^2 \times T^2$, the fluxes of $\mc{A}_1$ map to those of $\mc{A}_2$ when we identify $\{\lambda_1,\lambda_2,\lambda_3,b_3,d_0,d_3\}$ of $\mc{A}_1$ with $\{\lambda_2,\lambda_1,\lambda_3,b_3,d_0,d_3\}$ of $\mc{A}_2$. Moreover the respective transformation of the period matrix, ${\rm diag}\{U_1,U_2,U_3\} \to {\rm diag}\{U_2,U_1,U_3\}$, along with the above identification, relate their solutions. Similarly, under the permutation between the second and the third tori of $T^2 \times T^2 \times T^2$, the fluxes of $\mc{A}_1$ map to those of $\mc{A}_3$ when we identify $\{\lambda_1,\lambda_2,\lambda_3,b_3,d_0,d_3\}$ of $\mc{A}_1$ with $\{\lambda_1,\lambda_3,\lambda_2,b_2,d_0,d_2\}$ of $\mc{A}_3$. The respective transformation of the period matrix, ${\rm diag}\{U_1,U_2,U_3\} \to {\rm diag}\{U_1,U_3,U_2\}$, along with the above identification, relate their solutions as well.

\noindent \underline{Inequivalent solutions in $\mc{A}_1$}: The requirement that $a_1$, $b_2$ and $b_3$ in \eqref{a1flux} be even integers results in the parametrisation shown below:
\bea
&&b_3=2p\,,\qquad d_0=2q\lambda_2\,,\qquad d_3=2r\lambda_2\,, \qquad r\neq0,\quad p,q,r\in\mathbb{Z}\,, \nn \\
&&N_{\rm flux}(r,\lambda_2,\frac{\lambda_3}{\lambda_1})=\frac{8r^2\lambda_2\lambda_3}{\lambda_1}\,.
\eea
The dependence of the fluxes \eqref{a1flux} on $\lambda_1$ and $\lambda_3$ are only through the ratio $\lambda_3/\lambda_1$. For the present analysis we confine to integer values of $\lambda_2$ and $\lambda_3/\lambda_1$. It can be shown that whenever it is possible to find a triple $(r,\lambda_2,\lambda_3/\lambda_1)$ with $\frac{8r^2\lambda_2\lambda_3}{\lambda_1},\frac{\lambda _3 r}{\lambda _1},\frac{\lambda _2 \lambda _3 r}{\lambda _1}\in\mathbb{Z}$\footnote{These respectively ensure that $N_{\rm flux}$ takes integer values and $a_3,d_1$ are even integers.} and $0<N_{\rm flux}\leq32$, there exist infinitely many pairs $(p,q)$ so that all the fluxes \eqref{a1flux} are even integers. For example $q=r$ and any $p\in\mathbb{Z}$ always work. Therefore we first need to find all possible integer triples $(r,\lambda_2,\lambda_3/\lambda_1)$. This will provide all allowed values of $N_{\rm flux}$. Then, among the different flux configurations corresponding to each of those triples (i.e. given an $N_{\rm flux}$) we need to find the distinct equivalence classes (using duality).

Denoting the integer $\lambda_3/\lambda_1$ by $s$ ($\neq0$), we have $N_{\rm flux}=8 r^2 s \lambda _2$. Clearly $N_{\rm flux}$ takes values in $\{8,16,24,32\}$. The possible values of $r$ are $\pm1,\pm2$. The requirement that all the fluxes \eqref{a1flux} be even integers results in:
\bea
&&\text{when}\ r=\pm1\,, \qquad p,q\in\mathbb{Z}\,; \nn \\
&&\text{when}\ r=\pm2\,, \qquad \{p\in2\mathbb{Z},q\in\mathbb{Z}\} \quad \text{or} \quad \{p\in\mathbb{Z},q\in2\mathbb{Z}\}\,.
\eea
Replacing $(r,p,q)$ by $(-r,-p,-q)$ maps the fluxes to minus themselves. Hence, in order to obtain the inequivalent solutions, it would be sufficient to consider $r>0$. Now there are only 4 classes whose respective parametrisations, $N_{\rm flux}$ and the solutions are as follows.

\noindent \underline{\it Class 1}:
\bea
  \frac{\lambda_3}{\lambda_1}&=&s\,,\qquad b_3=2p\,,\qquad d_0=2q\lambda_2\,,\qquad d_3=2\lambda_2\,, \nn \\
  s&=&1,\dots,4\,, \qquad \lambda_2=1,\dots,\left[\frac{4}{s}\right], \quad p,q\in\mathbb{Z}\,, \nn \\
  N_{\rm flux}&=&8 s\lambda_2\,, \qquad (U_1,U_2,U_3,U_4) =\left(\frac{q}{s}+\frac{U_3}{s},\lambda _2 U_4-p,U_3,U_4\right),
\label{eq:A1subcat1}
\eea
where $[n]$ denotes the greatest integer $\leq n$ and $N_{\rm flux}$ takes values in $\{8,16,24,32\}$.

\noindent \underline{\it Class 2}:
\bea
  \frac{\lambda_3}{\lambda_1}&=&1\,,\qquad \lambda_2=1\,,\qquad b_3=2p\,,\qquad d_0=2q\,,\qquad d_3=4\,, \nn \\
  &&\left\{p\in2\mathbb{Z},q\in\mathbb{Z}\right\} \qquad \text{or} \qquad \left\{p\in\mathbb{Z},q\in2\mathbb{Z}\right\}, \nn \\
  N_{\rm flux}&=&32\,, \qquad (U_1,U_2,U_3,U_4) =\left(\frac{q}{2}+U_3,U_4-\frac{p}{2},U_3, U_4\right).
\label{eq:A1subcat2}
\eea

\noindent \underline{\it Class 3}:
\bea
  \frac{\lambda_3}{\lambda_1}&=&s\,,\qquad b_3=2p\,,\qquad d_0=2q\lambda_2\,,\qquad d_3=2\lambda_2,\quad s,\lambda_2<0\,, \nn \\
  |s|&=&1,\dots,4\,, \qquad |\lambda_2|=1,\dots,\left[\frac{4}{|s|}\right], \quad p,q\in\mathbb{Z}\,, \nn \\
  N_{\rm flux}&=&8 s\lambda_2\,, \qquad (U_1,U_2,U_3,U_4) =\left(\frac{q}{s}+\frac{U_3}{s},\lambda _2 U_4-p,U_3,U_4\right),
\label{eq:A1subcat3}
\eea
where $N_{\rm flux}$ takes values in $\{8,16,24,32\}$.

\noindent \underline{\it Class 4}:
\bea
  \frac{\lambda_3}{\lambda_1}&=&-1\,,\qquad \lambda_2=-1\,,\qquad b_3=2p\,,\qquad d_0=-2q\,,\qquad d_3=-4\,, \nn \\
  &&\left\{p\in2\mathbb{Z},q\in\mathbb{Z}\right\} \qquad \text{or} \qquad \left\{p\in\mathbb{Z},q\in2\mathbb{Z}\right\}, \nn \\
  N_{\rm flux}&=&32\,, \qquad (U_1,U_2,U_3,U_4) =\left(-\frac{q}{2}-U_3,-\frac{p}{2}-U_4,U_3, U_4\right).
\label{eq:A1subcat4}
\eea

\noindent A duality may exist between 2 flux configurations with the same $N_{\rm flux}$. After incorporating such dualities, we find that each of the 4 classes has only a finite number of physically distinct flux configurations. To check aforesaid dualities, the solution space for the moduli in all the 4 classes suggests that only $SL(2,Z)$-actions on the first and the second tori of $T^2 \times T^2 \times T^2$ may help. Thus the $SL(6,\mathbb{Z})$ matrix in our considerations will be:
\be
M = 
\begin{pmatrix}
M_1 & 0 & 0 \\
0 & M_2 & 0 \\
0 & 0 & I
\end{pmatrix}\,, \quad
M_1 = 
\begin{pmatrix}
1 & k \\
0 & 1
\end{pmatrix}\,, \quad
M_2 = 
\begin{pmatrix}
		1 & l \\
		0 & 1
\end{pmatrix}\,, \quad
k,l\in\mathbb{Z}\,.
\ee
For all 4 classes the action of $M$ transforms the fluxes keeping $N_{\rm flux}$ unaltered. The following details depend on the class in consideration.

\noindent \underline{\it For the case of Class 1}, the new solution with the transformed fluxes is:
\be
(U_1,U_2,U_3,U_4) =\left(k+\frac{q}{s}+\frac{U_3}{s},,\lambda _2 U_4-p+l,U_3,U_4\right).
\ee
When $q=m$ modulo $s$ (i.e. $m-q$ is a multiple of $s$) with the choices:
\be
k=\frac{m-q}{s},\ l=p\,,
\ee
the transformed fluxes as well as the new solution respectively coincide with the fluxes and solution of the case with $p = 0$ and $q = m$ for each $\lambda_2=1,\dots,[4/s]$. In the later case $N_{\rm flux}$ and the solution are given by:
\be
N_{\rm flux}=8s\lambda_2\,, \quad (U_1,U_2,U_3,U_4) =\left(\frac{m}{s}+\frac{U_3}{s},\lambda _2 U_4,U_3,U_4\right), \quad m=0,\dots,s-1\,.
\ee

\noindent \underline{\it For the case of Class 2}, the new solution with the transformed fluxes is:
\be
(U_1,U_2,U_3,U_4) =\left(k+\frac{q}{2}+U_3,U_4-\frac{p}{2}+l,U_3, U_4\right).
\ee
When $p=m,q=n$ modulo $2$ (i.e. $m-p$ and $n-q$ are multiples of $2$) with the choices:
\be
k=\frac{n-q}{2}\,,\qquad l=\frac{p-m}{2}\,,
\ee
the transformed fluxes as well as the new solution respectively coincide with the fluxes and solution of the case with $p = m$ and $q = n$. In the later case $N_{\rm flux}$ and the solution are given by:
\bea
&&N_{\rm flux}=32\,, \qquad (U_1,U_2,U_3,U_4) =\left(\frac{n}{2}+U_3,U_4-\frac{m}{2},U_3, U_4\right), \nn \\
&&\{m=0,\ n=0,1\}\qquad \text{or}\qquad \{m=0,1,\ n=0\}\,.
\eea

\noindent \underline{\it For Classes 3 and 4}, the analysis is similar to that for classes 1 and 2 respectively.

\subsubsection*{Dualities of family $\mc{B}$}

First we show that $\mc{B}_1$ is dual to $\mc{B}_2$ via an $SL(6,\mathbb{Z})$ transformation. Then the question to classify the inequivalent solutions in family $\mc{B}$ essentially boils down to that of subfamily $\mc{B}_1$, which we address subsequently.

\noindent \underline{Duality between $\mc{B}_1$ and $\mc{B}_2$}: To prove the duality between $\mc{B}_1$ and $\mc{B}_2$, we act with an $\mathcal{S}$-transformation only on the first 2-torus of $T^2 \times T^2 \times T^2$, transforming the period matrix as:
\bea
&&M\,:\, {\rm diag}\{U_1,U_2,U_3\} \quad\to\quad {\rm diag}\{-\frac1U_1,U_2,U_3\}\,, \nn \\[5pt]
&&M = 
\begin{pmatrix}
M_1 & 0 & 0 \\
0 & I & 0 \\
0 & 0 & I
\end{pmatrix}\,, \qquad
M_1 = 
\begin{pmatrix}
0 & 1 \\
-1 & 0
\end{pmatrix}\,, \qquad	
I = 
\begin{pmatrix}
		1 & 0 \\
		0 & 1
\end{pmatrix}\,.
\label{eq:SonU1}
\eea
This transforms the fluxes \eqref{diagflux} as:
\bea
\{a_0, a_1, a_2, a_3\} &\to& \{-a_1, a_0 , b_3 , b_2\}\,,\qquad \{b_0, b_1, b_2, b_3\}\to\{-b_1 , b_0 , -a_3 , -a_2 \}\,, \nn \\
\{c_0, c_1, c_2, c_3\} &\to& \{-c_1 , c_0 , d_3 , d_2\}\,,\qquad \{d_0, d_1, d_2, d_3\}\to\{-d_1 , d_0 , -c_3 , -c_2\}\,.
\eea
It is straightforward to check that, under the above action, the fluxes of $\mc{B}_1$, given by \eqref{eq:fluxesB1}, map to those of $\mc{B}_2$, given by \eqref{eq:fluxesB2}, when we identify $\{b_2,d_2,d_1,d_0,c_3\}$ of $\mc{B}_1$ with $\{a_3, c_3, -d_0, d_1,-d_2\}$ of $\mc{B}_2$. Such identification relates the crucial condition $d_2\neq0$ of \eqref{eq:fluxesB1} to the condition $c_3\neq0$ of \eqref{eq:fluxesB2}, and leaves $N_{\rm flux} = \frac{2}{\lambda_3}\left(c_3d_0-d_1d_2\right)$ invariant. With this identification now \eqref{eq:SonU1} maps the solution \eqref{eq:solgenB1} to \eqref{eq:solgenB2}, establishing the duality.

\noindent \underline{Inequivalent solutions in $\mc{B}_1$}: The fluxes \eqref{eq:fluxesB1} depend on $\lambda_3,b_2,c_3,d_0,d_1,d_2$. For the present analysis we confine to integer values of $\lambda_3$. There are only 4 classes consistent with even integer fluxes and $0<N_{\rm flux}\leq32$. Their respective parametrisations, $N_{\rm flux}$ and the solutions are as follows.

\noindent \underline{\it Class 1}:
\bea
  b_2&=&2 k s\,,\qquad c_3=2p\lambda_3\,,\qquad d_0=2q\lambda_3\,,\qquad d_1=2r\lambda_3\,,\qquad d_2=2s\lambda_3\,, \nn \\
  pq-rs&=&1,\dots,4\,,\qquad \lambda_3=1,\dots,\left[\frac{4}{p q - r s}\right]\,, \qquad k,p,q,r,s\in\mathbb{Z}\,,  \\
  N_{\rm flux}&=&8 (p q - r s)\lambda_3\,,\qquad (U_1,U_2,U_3,U_4) =\left(-\frac{s U_2+q}{p U_2+r},U_2,-k + \lambda_3 U_4, U_4\right), \nn
\label{eq:B1subcat1}
\eea
where $N_{\rm flux}$ takes values in $\{8,16,24,32\}$.

\noindent \underline{\it Class 2}:
\bea
  \lambda_3&=&1\,,\qquad b_2=2 k s\,,\qquad c_3=4p\,,\qquad d_0=4q\,,\qquad d_1=4r\,,\qquad d_2=4s\,, \nn \\
  pq-rs&=&1\,,\qquad k\in2\mathbb{Z}+1\,,\qquad p,q,r,s\in\mathbb{Z}\,, \nn \\
  N_{\rm flux}&=&32\,,\qquad (U_1,U_2,U_3,U_4) =\left(-\frac{s U_2+q}{p U_2+r},U_2,-\frac{k}{2} + U_4, U_4\right).
\label{eq:B1subcat2}
\eea

\noindent \underline{\it Class 3}:
\bea
  b_2 &=& 2 k s\,,\quad c_3=2p\lambda_3\,,\quad d_0=2q\lambda_3\,,\quad d_1=2r\lambda_3\,,\quad d_2=2s\lambda_3\,,\quad pq-rs,\lambda_3<0\,, \nn \\
  |pq-rs|&=&1,\dots,4\,,\qquad |\lambda_3|=1,\dots,\left[\frac{4}{|p q - r s|}\right]\,,\qquad k,p,q,r,s\in\mathbb{Z}\,, \nn \\
  N_{\rm flux}&=&8 (p q - r s)\lambda_3\,,\qquad (U_1,U_2,U_3,U_4) =\left(-\frac{s U_2+q}{p U_2+r},U_2,-k + \lambda_3 U_4, U_4\right).
\label{eq:B1subcat3}
\eea

\noindent \underline{\it Class 4}:
\bea
  \lambda_3 &=& -1\,,\qquad b_2=2 k s\,,\qquad c_3=-4p\,,\qquad d_0=-4q\,,\qquad d_1=-4r\,,\qquad d_2=-4s\,, \nn \\
  pq-rs&=&-1\,,\qquad k\in2\mathbb{Z}+1\,,\qquad p,q,r,s\in\mathbb{Z}\,, \nn \\
  N_{\rm flux}&=&32 \,,\qquad (U_1,U_2,U_3,U_4) =\left(-\frac{s U_2+q}{p U_2+r},U_2,-\frac{k}{2} - U_4, U_4\right).
\label{eq:B1subcat4}
\eea

\noindent A duality may exist between 2 flux configurations with the same $N_{\rm flux}$. After incorporating such dualities, we find that each of the 4 classes has only a finite number of physically distinct flux configurations. To check aforesaid dualities, the solution space for the moduli in all the 4 classes suggests that only $SL(2,Z)$-actions on the first and the third tori of $T^2 \times T^2 \times T^2$ may help. Thus the $SL(6,\mathbb{Z})$ matrix in our considerations will be:
\bea
&&M = 
\begin{pmatrix}
M_1 & 0 & 0 \\
0 & I & 0 \\
0 & 0 & M_3
\end{pmatrix}\,, \qquad
M_1 = 
\begin{pmatrix}
g & h \\
i & j
\end{pmatrix}\,, \qquad	
M_3 = 
\begin{pmatrix}
		1 & l \\
		0 & 1
\end{pmatrix}\,, \nn \\
&&gj-hi=1\,, \quad g,h,i,j,l\in\mathbb{Z}\,.
\eea
For all 4 classes the action of $M$ transforms the fluxes keeping $N_{\rm flux}$ unaltered. The following details depend on the class in consideration.

\noindent \underline{\it For the case of Class 1}, the new solution with the transformed fluxes is:
\be
(U_1,U_2,U_3,U_4) =\left(\frac{(h p - j s) U_2+ (h r-j q) }{(g p - i s) U_2 + (g r-i q)},U_2,-k + l+\lambda_3 U_4, U_4\right).
\ee
When $pq-rs=1$, with the choices:
\be
g = -s\,,\qquad h = q\,,\qquad i = -p\,,\qquad j = r\,,\qquad l = k\,,
\ee
the transformed fluxes as well as the new solution respectively coincide with the fluxes and solution of the case with $p = 0$, $q = 0$, $r = 1$, $s = -1$ and $k = 0$ $\forall \,\lambda_3=1,2,3,4$. In the later case $N_{\rm flux}$ and the solution are given by:
\be
N_{\rm flux}=8\lambda_3\,, \quad (U_1,U_2,U_3,U_4) =\left(U_2,U_2,\lambda_3 U_4, U_4\right).
\ee
When $pq-rs=2$, depending on each of $p,q,r,s$ even (e) or odd (o), the transformed fluxes and solution coincide with those of some specific configuration. In keeping with $pq-rs=2$, $p,q,r,s$ can only be:
\be
{\rm eeeo},\ {\rm eeoe},\ {\rm eoee},\ {\rm eoeo},\ {\rm eooe},\ {\rm oeee},\ {\rm oeeo},\ {\rm oeoe},\ {\rm oooo}\,.
\ee
For $p,q,r,s={\rm eoeo},\ {\rm oeoe},\ {\rm oooo}$, with the choices:
\be
g=q\,,\qquad h=\frac{s-q}{2}\,,\qquad i=r\,,\qquad j=\frac{p-r}{2}\,,\qquad l=k\,,
\ee
the transformed fluxes as well as the new solution respectively coincide with the fluxes and solution of the case with $p=2$, $q=1$, $r=0$, $s=1$ and $k=0$ for each $\lambda_3=1,2$. In the later case $N_{\rm flux}$ and the solution are given by:
\be
N_{\rm flux}=16\lambda_3\,, \quad (U_1,U_2,U_3,U_4) =\left(\frac{-U_2-1}{2 U_2},U_2,\lambda_3 U_4, U_4\right).
\ee
For $p,q,r,s={\rm eeoe},\ {\rm eoee},\ {\rm eooe}$, with the choices:
\be
g=q\,,\qquad h=\frac{s}{2}\,,\qquad i=r\,,\qquad j=\frac{p}{2}\,,\qquad l=k\,,
\ee
the transformed fluxes as well as the new solution respectively coincide with the fluxes and solution of the case with $p=2$, $q=1$, $r=0$, $s=1$ and $k=0$ for each $\lambda_3=1,2$. In the later case $N_{\rm flux}$ and the solution are given by:
\be
N_{\rm flux}=16\lambda_3\,, \quad (U_1,U_2,U_3,U_4) =\left(-\frac{1}{2 U_2},U_2,\lambda_3 U_4, U_4\right).
\ee
For $p,q,r,s={\rm eeeo},\ {\rm oeee},\ {\rm oeeo}$, with the choices:
\be
g=-s\,,\qquad h=\frac{q}{2}\,,\qquad i=-p\,,\qquad j=\frac{r}{2}\,,\qquad l=k\,,
\ee
the transformed fluxes as well as the new solution respectively coincide with the fluxes and solution of the case with $p=0$, $q=0$, $r=2$, $s=-1$ and $k=0$ for each $\lambda_3=1,2$. In the later case $N_{\rm flux}$ and the solution are given by:
\be
N_{\rm flux}=16\lambda_3\,, \quad (U_1,U_2,U_3,U_4) =\left(\frac{U_2}{2},U_2,\lambda_3 U_4, U_4\right).
\ee
When $pq-rs=3$, we need to analyse cases where each of $p,q,r,s=0,1,2$ modulo $3$.\footnote{2 integers $n_1$ and $n_2$ are equal modulo $3$ if there exists an integer $n_3$ such that $n_1=3n_3+n_2$. For example, note that $-2=1$ and $-1=2$ modulo $3$.} Out of $3^4$ possibilities, only $32$ cases are consistent with $pq-rs=3$ where $p,q,r,s$ can be:
\bea
0 0 0 1,\ 0 0 0 2,\ 0 0 1 0,\ 0 0 2 0,\ 0 1 0 0,\ 0 1 0 1,\ 0 1 0 2,\ 0 1 1 0\,, \nn \\
0 1 2 0,\ 0 2 0 0,\ 0 2 0 1,\ 0 2 0 2,\ 0 2 1 0,\ 0 2 2 0,\ 1 0 0 0,\ 1 0 0 1\,, \nn \\
1 0 0 2,\ 1 0 1 0,\ 1 0 2 0,\ 1 1 1 1,\ 1 1 2 2,\ 1 2 1 2,\ 1 2 2 1,\ 2 0 0 0\,, \nn \\
2 0 0 1,\ 2 0 0 2,\ 2 0 1 0,\ 2 0 2 0,\ 2 1 1 2,\ 2 1 2 1,\ 2 2 1 1,\ 2 2 2 2\,.
\eea
For $p,q,r,s=0001,\ 0002,\ 1000,\ 1001,\ 1002,\ 2000,\ 2001,\ 2002$, with the choices:
\be
g=-s\,,\qquad h=\frac{q}{3}\,,\qquad i=-p\,,\qquad j=\frac{r}{3}\,,\qquad l=k\,,
\ee
the transformed fluxes as well as the new solution respectively coincide with the fluxes and solution of the case with $p=0$, $q=0$, $r=3$, $s=-1$, $k=0$ and $\lambda_3=1$. In the later case $N_{\rm flux}$ and the solution are given by:
\be
N_{\rm flux}=24\,, \quad (U_1,U_2,U_3,U_4) =\left(\frac{U_2}{3},U_2, U_4, U_4\right).
\ee
For $p,q,r,s=0 0 1 0,\ 0 0 2 0,\ 0 1 0 0,\ 0 1 1 0,\ 0 1 2 0,\ 0 2 0 0,\ 0 2 1 0,\ 0 2 2 0$, with the choices:
\be
g=-\frac{s}{3}\,,\qquad h=q\,,\qquad i=-\frac{p}{3}\,,\qquad j=r\,,\qquad l=k\,,
\ee
the transformed fluxes as well as the new solution respectively coincide with the fluxes and solution of the case with $p=0$, $q=0$, $r=1$, $s=-3$, $k=0$ and $\lambda_3=1$. In the later case $N_{\rm flux}$ and the solution are given by:
\be
N_{\rm flux}=24\,, \quad (U_1,U_2,U_3,U_4) =\left(3U_2,U_2, U_4, U_4\right).
\ee
For $p,q,r,s=0101,\ 0202,\ 1010,\ 1111,\ 1212,\ 2020,\ 2121,\ 2222$, with the choices:
\be
g=q\,,\qquad h=\frac{s-q}{3}\,,\qquad i=r\,,\qquad j=\frac{p-r}{3}\,,\qquad l=k\,,
\ee
the transformed fluxes as well as the new solution respectively coincide with the fluxes and solution of the case with $p=3$, $q=1$, $r=0$, $s=1$, $k=0$ and $\lambda_3=1$. In the later case $N_{\rm flux}$ and the solution are given by:
\be
N_{\rm flux}=24\,, \quad (U_1,U_2,U_3,U_4) =\left(\frac{-U_2-1}{3 U_2},U_2, U_4, U_4\right).
\ee
For $p,q,r,s=0102,\ 0201,\ 1020,\ 1122,\ 1221,\ 2010,\ 2112,\ 2211$, with the choices:
\be
g=q\,,\qquad h=\frac13 (s-2 q)\,,\qquad i=r\,,\qquad j=\frac13 (p-2 r)\,,\qquad l=k\,,
\ee
the transformed fluxes as well as the new solution respectively coincide with the fluxes and solution of the case with $p=3$, $q=1$, $r=0$, $s=2$, $k=0$ and $\lambda_3=1$. In the later case $N_{\rm flux}$ and the solution are given by:
\be
N_{\rm flux}=24\,, \quad (U_1,U_2,U_3,U_4) =\left(\frac{-2 U_2-1}{3 U_2},U_2, U_4, U_4\right).
\ee
When $pq-rs=4$, a similar analysis can be done.

\noindent \underline{\it For the case of Class 2}, the new solution with the transformed fluxes is:
\be
(U_1,U_2,U_3,U_4) =\left(\frac{(h p - j s) U_2+ (h r-j q) }{(g p - i s) U_2 + (g r-i q)},U_2,-\frac{k}{2} + l+U_4, U_4\right).
\ee
Now with the choices:
\be
g = -s\,,\qquad h = q\,,\qquad i = -p\,,\qquad j = r\,,\qquad l = \frac{k-1}{2}\,,
\ee
the transformed fluxes as well as the new solution respectively coincide with the fluxes and solution of the case $p = 0,\ q = 0,\ r = 1,\ s = -1,\ k = 1$. In the later case $N_{\rm flux}$ and the solution are given by:
\be
N_{\rm flux}=32\,, \quad (U_1,U_2,U_3,U_4) =\left(U_2,U_2,U_4-\frac12, U_4\right).
\ee

\noindent \underline{\it For Classes 3 and 4}, the analysis is similar to that for classes 1 and 2 respectively.

\subsubsection*{Dualities of family $\mc{C}$}

As per \eqref{eq:fluxesC1}, \eqref{eq:fluxesC2} and \eqref{eq:fluxesC3}, the fluxes of $\mc{C}_1$ and $\mc{C}_2$ have $5$ independent parameters, whereas it is $6$ in case of $\mc{C}_3$. Despite this, we are able to prove that $SL(6,\mathbb{Z})$ transformations relate $\mc{C}_3$ to $\mc{C}_2$, while $\mc{C}_1$ to a subset of $\mc{C}_2$. Below we provide the details. Then the question to classify the inequivalent solutions in family $\mc{C}$ essentially boils down to that of $\mc{C}_2$, which we address subsequently.

\noindent \underline{Duality between $\mc{C}_2$ and $\mc{C}_3$}: The fluxes of $\mc{C}_3$, given by \eqref{eq:fluxesC3}, depend on the $6$ parameters $\lambda_2,b_2,b_3,d_0,d_2$ and $d_3$ with $\lambda_2,d_0,d_2\neq0$ and $b_2d_3\neq b_3d_2$. We divide $\mc{C}_3$ in $2$ complementary subsets with $d_3=0$ and $d_3\neq0$ respectively. Each of these is shown to be dual to $\mc{C}_2$.

To prove the duality between the subset of $\mc{C}_3$ with $d_3=0$ and $\mc{C}_2$, we act with an $\mathcal{S}$-transformation only on the third 2-torus of $T^2 \times T^2 \times T^2$ transforming the period matrix as:
\bea
&&M\,:\, {\rm diag}\{U_1,U_2,U_3\} \quad\to\quad {\rm diag}\{U_1,U_2,-\frac1U_3\}\,, \nn \\[5pt]
&&M = 
\begin{pmatrix}
I & 0 & 0 \\
0 & I & 0 \\
0 & 0 & M_3
\end{pmatrix}\,, \qquad
I = 
\begin{pmatrix}
1 & 0 \\
0 & 1
\end{pmatrix}\,, \qquad	
M_3 = 
\begin{pmatrix}
		0 & 1 \\
		-1 & 0
\end{pmatrix}\,.
\label{eq:SonU3}
\eea
This transforms the fluxes \eqref{diagflux} as:
\bea
\{a_0, a_1, a_2, a_3\} &\to& \{-a_3, b_2 , b_1 , a_0\}\,,\quad \{b_0, b_1, b_2, b_3\}\to\{-b_3 , -a_2 , -a_1 , b_0 \}\,, \nn \\
\{c_0, c_1, c_2, c_3\} &\to& \{-c_3, d_2, d_1, c_0\}\,,\quad \{d_0, d_1, d_2, d_3\}\to\{-d_3, -c_2, -c_1, d_0\}\,.
\eea
It is straightforward to check that, under the above action, the fluxes of $\mc{C}_3$, given by \eqref{eq:fluxesC3} with $d_3=0$, map to those of $\mc{C}_2$, given by \eqref{eq:fluxesC2}, when we identify $\{-\frac{b_3 d_2}{d_0},\frac{b_2 d_0}{d_2},d_2 \lambda _2,d_0\}$ of $\mc{C}_3$ with $\{b_2,b_3,c_2,d_3\}$ of $\mc{C}_2$. Such identification relates the crucial conditions $b_3,d_0,d_2\neq0$ of \eqref{eq:fluxesC3} (when $d_3=0$) to the conditions $b_2,c_2,d_3\neq0$ of \eqref{eq:fluxesC2}. With this identification now \eqref{eq:SonU3} maps the solution \eqref{eq:solgenC3} with $d_3=0$ to \eqref{eq:solgenC2}, establishing the duality.

To prove the duality between the subset of $\mc{C}_3$ with $d_3\neq0$ and $\mc{C}_2$, we act with an $SL(2,\mathbb{Z})$-transformation only on the third 2-torus of $T^2 \times T^2 \times T^2$ transforming the period matrix as:
\bea
&&M\,:\, {\rm diag}\{U_1,U_2,U_3\} \quad\to\quad {\rm diag}\{U_1,U_2,\frac{jU_3+h}{iU_3+g}\}\,, \nn \\[5pt]
&&M = 
\begin{pmatrix}
I & 0 & 0 \\
0 & I & 0 \\
0 & 0 & M_3
\end{pmatrix}\,, \qquad
I = 
\begin{pmatrix}
1 & 0 \\
0 & 1
\end{pmatrix}\,, \qquad	
M_3 = 
\begin{pmatrix}
		g & h \\
		i & j
\end{pmatrix}\,, \nn \\
&&gj-hi=1\,, \quad g,h,i,j\in\mathbb{Z}\,.
\label{eq:SL2ZonU3}
\eea
This transforms the fluxes \eqref{diagflux} as:
\bea
&&\{a_0, a_1, a_2, a_3\} \quad\to\quad \{a_0 g + a_3 i,\ a_1 g - b_2 i,\ a_2 g - b_1 i,\ a_0 h + a_3 j\}\,, \nn \\ 
&&\{b_0, b_1, b_2, b_3\} \quad\to\quad \{-b_3 h + b_0 j, -a_2 h + b_1 j, -a_1 h + b_2 j,\ b_3 g - b_0 i \}\,, \nn \\
&&\{c_0, c_1, c_2, c_3\} \quad\to\quad \{c_0 g + c_3 i,\ c_1 g - d_2 i,\ c_2 g - d_1 i,\ c_0 h + c_3 j\}\,, \nn \\
&&\{d_0, d_1, d_2, d_3\} \quad\to\quad \{-d_3 h + d_0 j, -c_2 h + d_1 j, -c_1 h + d_2 j,\ d_3 g - d_0 i\}\,.
\eea
It is straightforward to check that, under the above action, the fluxes of $\mc{C}_3$, given by \eqref{eq:fluxesC3} with $d_3\neq0$, map to those of $\mc{C}_2$, given by \eqref{eq:fluxesC2}, when we implement the following steps:
\begin{enumerate}
\item Given non-zero even integer fluxes $d_0$ and $d_3$ in $\mc{C}_3$ find $4$ integers $g,h,i,j$ satisfying:
\be
h=\frac{d_0 j}{d_3}\,,\qquad i=\frac{d_3 (g j-1)}{d_0 j}\,,\quad j\neq0\,.
\ee
This can be done if the following holds. Given $2$ integers $(p,q)\neq(0,0)$ (i.e., taking $d_0=2p$ and $d_3=2q$) one can always find other $2$ integers $(g,j),\ j\neq0$ such that $(\frac{j p}{q},\frac{q (g j-1)}{j p})$ are integers. We have verified this numerically for $p,q=-1000,\dots,1000$.

\item Identify $\left\{j (b_2-\frac{b_3 d_2}{d_3}),\frac{b_2 d_3 (1-g j)}{d_2 j}+b_3 g,\frac{d_2 d_3 \lambda _2}{d_0 j},\frac{d_3}{j}\right\}$ of $\mc{C}_3$ with $\{b_2,b_3,c_2,d_3\}$ of $\mc{C}_2$. As \eqref{eq:fluxesC3} are even integer fluxes and $g,h,i,j$ are chosen to be integers, clearly $j (b_2-\frac{b_3 d_2}{d_3})=-a_1 h + b_2 j,\ \frac{b_2 d_3 (1-g j)}{d_2 j}+b_3 g=b_3 g - b_0 i,\ \frac{d_2 d_3 \lambda _2}{d_0 j}=c_2 g - d_1 i,\ \frac{d_3}{j}=d_3 g - d_0 i$ are even integers. Alternatively, in the transformed fluxes of $\mc{C}_3$ one can substitute $b_2,b_3,d_2,d_3$ in terms of $b_2,b_3,c_2,d_3$ of $\mc{C}_2$ and $d_0$ of $\mc{C}_3$ (obtained by inverting the above identification map) to get the fluxes of $\mc{C}_2$, i.e. the explicit dependence on $d_0$ of $\mc{C}_3$ goes away. The above identification also relates the crucial conditions $d_0,d_2\neq0,\ b_2d_3\neq b_3d_2$ of \eqref{eq:fluxesC3} (with $d_3\neq0$) to the conditions $b_2,c_2,d_3\neq0$ of \eqref{eq:fluxesC2}. 
\end{enumerate}
Now \eqref{eq:SL2ZonU3} maps the solution \eqref{eq:solgenC3} with $d_3\neq0$ to \eqref{eq:solgenC2}, establishing the duality.

\noindent \underline{Duality between $\mc{C}_1$ and a subset of $\mc{C}_2$}: Consider a $\mathcal{T}^l$- and an $SL(2,\mathbb{Z})$-action respectively on the first and the third 2-tori of $T^2 \times T^2 \times T^2$, i.e. the $SL(6,\mathbb{Z})$ matrix is:
\bea
&&M = 
\begin{pmatrix}
M_1 & 0 & 0 \\
0 & I & 0 \\
0 & 0 & M_3
\end{pmatrix}\,, \qquad
M_1 = 
\begin{pmatrix}
1 & l \\
0 & 1
\end{pmatrix}\,, \qquad	
M_3 = 
\begin{pmatrix}
		g & h \\
		i & j
\end{pmatrix}\,, \nn \\
&&gj-hi=1\,, \quad g,h,i,j,l\in\mathbb{Z}\,.
\eea
This action, together with an appropriate choice for $g,h,i,j$, transforms the fluxes of $\mc{C}_1$, given by \eqref{eq:fluxesC1}, to those of $\mc{C}_2$, given by \eqref{eq:fluxesC2}, with $\frac{d_3}{c_2}=-l$ (i.e. integer) only. The appropriate choices depend on the flux quanta \eqref{eq:fluxesC1} as follows:
\bea
g&=&1\,,\qquad h=0\,,\qquad j=1\,,\qquad \text{when}\quad d_2=0\,, \nn \\
h&=&1\,,\qquad i=-1\,,\qquad j=0\,,\qquad \text{when}\quad d_2\neq0\,,\,c_2=0\,, \nn \\
j&=&\frac{c_2h}{d_2\lambda_2}\,,\qquad g=\frac{d_2\lambda_2(1+hi)}{c_2h}\,,\qquad h\neq0\,,\qquad \text{when}\quad d_2\,,\,c_2\neq0\,.
\eea
For non-zero even integer fluxes $c_1=\frac{c_2}{\lambda}=2p$ and $d_2=2q$ in $\mc{C}_1$, the last choice can always be made (which we checked numerically when $p,q=-1000,\dots,1000$). The period matrix transforms in a way that in all the above cases the solution for the moduli in $\mc{C}_1$ maps to that of the corresponding subset of $\mc{C}_2$.

Clearly, there are flux configurations in $\mc{C}_2$ for which $\frac{d_3}{c_2}$ is non-integer. For example $b_2=4,b_3=2,c_2=4,d_3=2$ with $N_{\rm flux}=32$ is not dual to any flux configuration in $\mc{C}_1$.

\noindent \underline{Inequivalent solutions in $\mc{C}_2$}: The fluxes \eqref{eq:fluxesC2} depend on $\lambda_2,b_2,b_3,c_2,d_3$. The requirement that $b_2,b_3,c_1,d_3$ be even integers results in the parametrisation shown below:
\bea
b_2&=&2p\,,\quad b_3=2q\,,\quad c_2=2r\lambda_2\,,\quad d_3=2s\,, \quad p,r,s\neq0\,,\quad p,q,r,s\in\mathbb{Z}\,, \nn \\
N_{\rm flux}&=&8pr\lambda_2\,.
\eea
For the present analysis we confine to integer values of $\lambda_2$. This allows $N_{\rm flux}$ to take values in $\{8,16,24,32\}$ and one can show that, whenever we find a triple $(p,r,\lambda_2)$ corresponding to a given $N_{\rm flux}$ value, there exist infinitely many pairs $(q,s)$ so that all the fluxes \eqref{eq:fluxesC2} are even integers. For example $s=r$ and any $q\in\mathbb{Z}$ always work. Therefore, given an $N_{\rm flux}$, we first need to find all possible integer triples $(p,r,\lambda_2)$. Then, among the different flux configurations corresponding to each of those triples, we need to find the distinct equivalence classes (using duality). The number of possible triples is $4$ when $N_{\rm flux}=8$, $12$ for both cases with $N_{\rm flux}=16$ and $N_{\rm flux}=24$, and $24$ when $N_{\rm flux}=32$. To demonstrate the aforesaid dualities, we consider below only the $N_{\rm flux}=8$ case with $p=-1$, $r=1$ and $\lambda_2=-1$.

With $(p,r,\lambda_2)=(-1,1,-1)$ more generally one can take $q=ks,\ k\in\mathbb{Z}$ that leads to even integer fluxes \eqref{eq:fluxesC2}. In this case the solution \eqref{eq:solgenC2} reads:
\be
(U_1,U_2,U_3,U_4) =\left(s+U_2,U_2,\frac{1}{k-U_4},U_4\right).
\ee
Now the above fluxes and solution with $(s,k)\neq(0,1)$ can be mapped to those with $(s,k)=(0,1)$ by acting with $\mathcal{T}^{1-s}$ and $\mathcal{S}\mathcal{T}^k\mathcal{S}$ respectively on the first and the third 2-tori of $T^2 \times T^2 \times T^2$.

\subsubsection*{Dualities between families}

The linear relation that the derivatives of the superpotential satisfy differs across the families $\mc{A},\mc{B},\mc{C}$, see \eqref{eq:linrelA}, \eqref{eq:linrelB} and \eqref{crel}. Despite this, below we find certain dualities among them. In summary, we show that $\mc{B}_1$ contains $\mc{A}_3$. Also, we know from the previous subsection that $\mc{C}_3$ contains 2 copies of $\mc{C}_2$, one of which is shown here to be dual to $\mc{B}_1$.

\noindent \underline{Duality between $\mc{A}_3$ and a subset of $\mc{B}_1$}: The fluxes of $\mc{B}_1$, given by \eqref{eq:fluxesB1}, with $c_3=0$ map to those of $\mc{A}_3$, given by \eqref{eq:fluxesA3}, when we identify $\{\lambda_3,b_2,d_0,d_1,d_2\}$ of $\mc{B}_1$ with $\{\lambda_3,b_2,d_0,-\frac{d_2\lambda_2}{\lambda_1},d_2\}$ of $\mc{A}_3$.\footnote{Note that the fluxes of $\mc{A}_3$ depend on $\lambda_1,\lambda_2$ via the ratio $\frac{\lambda_2}{\lambda_1}$.} Such identification relates the crucial conditions $\lambda_3,d_1,d_2\neq0$ of \eqref{eq:fluxesB1} (when $c_3=0$) to the conditions $\frac{\lambda_2}{\lambda_1},\lambda_3,d_2\neq0$ of \eqref{eq:fluxesA3}. Furthermore, with this identification, the solution \eqref{eq:solgenB1} with $c_3=0$ is same as the solution \eqref{eq:solgenA3}, establishing the duality.

\noindent \underline{Duality between $\mc{B}_1$ and a subset of $\mc{C}_3$}: The fluxes of $\mc{C}_3$, given by \eqref{eq:fluxesC3}, depend on the parameters $\lambda_2,b_2,b_3,d_0,d_2$ and $d_3$ with $\lambda_2,d_0,d_2\neq0$ and $b_2d_3\neq b_3d_2$. We take the subset of $\mc{C}_3$ for which $d_3=0$ and show that it is dual to $\mc{B}_1$. To prove this, we act with an $SL(2,\mathbb{Z})$-transformation only on the first 2-torus of $T^2 \times T^2 \times T^2$ transforming the period matrix as:
\bea
&&M\,:\, {\rm diag}\{U_1,U_2,U_3\} \quad\to\quad {\rm diag}\{\frac{jU_1+h}{iU_1+g},U_2,U_3\}\,, \nn \\[5pt]
&&M = 
\begin{pmatrix}
M_1 & 0 & 0 \\
0 & I & 0 \\
0 & 0 & I
\end{pmatrix}\,, \qquad	
M_1 = 
\begin{pmatrix}
		g & h \\
		i & j
\end{pmatrix}\,, \qquad
I = 
\begin{pmatrix}
1 & 0 \\
0 & 1
\end{pmatrix}\,, \nn \\
&&gj-hi=1\,, \quad g,h,i,j\in\mathbb{Z}\,.
\label{eq:SL2ZonU1}
\eea
This action, together with an appropriate choice for $g,h,i,j$, transforms the fluxes of $\mc{C}_3$, given by \eqref{eq:fluxesC3}, with $d_3=0$ to those of $\mc{B}_1$, given by \eqref{eq:fluxesB1}. The appropriate choices depend on the flux quanta \eqref{eq:fluxesB1} as follows:\footnote{To distinguish, here we use prime for the flux quanta \eqref{eq:fluxesB1}.}
\bea
j&=&\frac{d_2\lambda_2h}{d_0}\,,\qquad g=\frac{d_0(1+hi)}{d_2\lambda_2h}\,,\qquad h\neq0\,,\qquad \text{when}\quad d'_0=0\,, \nn \\
g&=&j=1\,,\qquad h=i=0\,,\qquad \text{when}\quad d'_0\neq0\,,\,c'_3=0\,, \nn \\
g&=&i=j=1\,,\qquad h=0\,,\qquad \text{when}\quad d'_2\,,\,c'_3\neq0\,.
\eea
For non-zero even integer fluxes $d_0=2p$ and $d_1=d_2\lambda_2=2q$ in $\mc{C}_3$, the last choice can always be made (which we checked numerically when $p,q=-1000,\dots,1000$). The period matrix transforms in a way that in all the above cases the solution for the moduli in $\mc{C}_3$ (when $d_3=0$) maps to that of the corresponding subset of $\mc{B}_1$. 

Thus, we conclude that $\mc{B}_1$ is the master family which contains all distinct solutions.

\end{document}